\documentclass{jfm}
\usepackage{graphicx}
\usepackage{epstopdf, epsfig}

\usepackage{ulem}
\usepackage{color}
\usepackage{comment,amsmath}
\usepackage{microtype}
\def\U{{\boldsymbol{U}}}

\def\ij{\mathrm{i}}

\def\C{\mathsfbi{C}}
\def\Q{\mathsfbi{Q}}
\def\A{\mathsfbi{A}}
\def\Del{\mathrm{\Delta}}

\def\S{\mathsfbi{S}}
\def\P{{\bf P}}

\newcommand{\h}{\eta}

\newcounter{saveeqn}%

\newcommand{\be}{\begin{equation}}
\newcommand{\ee}{\end{equation}}
\newcommand{\bdm}{\begin{equation*}}
\newcommand{\edm}{\end{equation*}}
\newcommand{\bea}{\begin{eqnarray}}
\newcommand{\eea}{\end{eqnarray}}

\newcommand{\partialf}[2]
{
 \ifthenelse{\equal{#1}{}}{\frac{\partial}{\partial #2}}{\frac{\partial #1}{\partial #2}}
}

\renewcommand{\(}{\left(}
\renewcommand{\)}{\right)}
\renewcommand{\[}{\left[}
\renewcommand{\]}{\right]}
\newcommand{\<}{\left\langle}
\renewcommand{\>}{\right\rangle}

\newcommand{\df}{\mathrm{d}}

\newcommand{\la}{\lambda}

\renewcommand{\b}{\beta}
\renewcommand{\a}{\alpha}

\newcommand{\z}{\zeta}

\providecommand\bnabla{\boldsymbol{\nabla}}
\providecommand\bcdot{\boldsymbol{\cdot}}

\def\bit{\vphantom{\dot{W}}}

\def\bt{\tilde{\beta}}

\def\st{\sin{\vartheta}}

\def\x{\xv }

\def\xv{\boldsymbol{x}}

\def\nv{\boldsymbol{n}}
\def\kv{\boldsymbol{k}}

\newcommand{\defn}{\ensuremath{\stackrel{\mathrm{def}}{=}}}
\renewcommand{\equiv}{\defn}
\shorttitle{Is spontaneous generation of coherent baroclinic flows possible?}
\shortauthor{N. A. Bakas and P. J. Ioannou}

\title{Is spontaneous generation of coherent baroclinic flows possible?}

\author{Nikolaos A. Bakas\aff{1}
\corresp{\email{nbakas@uoi.gr}}
\and Petros J. Ioannou\aff{2}}
\affiliation{\aff{1}Laboratory of Meteorology and Climatology, Department of Physics, University of Ioannina, Ioannina, Greece
\aff{2}Department of Physics, National and Kapodistrian University of Athens, Athens, Greece}

\begin{document}

\maketitle

\begin{abstract}
Geophysical turbulence is observed to self-organize into large-scale flows such as zonal jets
and coherent vortices. Previous studies on barotropic beta-plane turbulence, 
a simple model that retains the basic self-organization dynamics, have shown that coherent flows emerge 
out of a background of homogeneous turbulence as a bifurcation when
the turbulence intensity increases and the emergence of large scale flows has been attributed to a new type
of collective, symmetry breaking instability of the statistical state dynamics of the turbulent
flow. In this work we extend the analysis to stratified flows and
investigate turbulent self-organization in a two-layer fluid with no imposed mean north-south
thermal gradient and turbulence supported by an external random stirring. We use a second
order closure of the statistical state dynamics (S3T) with an appropriate averaging ansatz that allows the
identification of statistical turbulent equilibria and their structural stability.
The bifurcation of the statistically homogeneous equilibrium state to inhomogeneous
equilibrium states comprising of zonal jets and/or large scale waves when the energy input rate
of the excitation passes a critical threshold is analytically studied. The theory predicts that
when the flow transitions to a statistical state with large-scale structures, these states are barotropic
if the scale of excitation is larger than the deformation radius.  Mixed barotropic-baroclinic states with jets and/or waves  arise
when the excitation is at scales shorter than the deformation radius with the baroclinic component of the flow being
subdominant for low energy input rates and non-significant for higher energy input rates.  The results of the S3T theory are compared
to nonlinear simulations. The theory is found to accurately predict both the critical transition parameters and the
scales of the emergent structures but underestimates their amplitude.
\end{abstract}


\section{Introduction}

Large scale flows like the zonal jets and the large-scale vortices of  the gaseous planets
are a common feature  of  the turbulent state in planetary atmospheres
\citep{Ingersoll-90, Maximenko-etal-2005,Galperin-etal-2014}.
There are two type of theories that have been advanced for the emergence and maintenance of these
flows: the deep origin theories that  propose that the origin of the jets is due to   Reynolds stresses arising from convection from the interior of the
planet
(cf. \cite{Busse-1976,Kaspi-etal-2009})
and theories that posit that the jets have a shallow origin and emerge when large-scale atmospheric turbulence
evolves  in the presence of latitudinal variations of the Coriolis parameter
on the surface of the planet (the planetary  $\beta$ effect)
\citep{Williams-78,Williams-03,Cho-Polvani-1996pof,Kaspi-Flierl-2007,Showman-2007,Scott-Polvani-2007,Liu-Schneider-2010,Thomson-McIntyre-2016}.
Comprehensive review of these  theories can be found in \cite{Vasavada-and-Showman-05} and in \cite{Lavega-etal-2018-book}.

%In order to understand the
%process of turbulence self-organization into these large scale structures, simplified models
%containing the relevant dynamics have been studied for some years now.
%
One of the simplest
models that has been widely studied is  the organization of large scale structures in barotropic $\beta$-plane turbulence
(cf. \cite{Rhines-1975,Vallis-Maltrud-93}).  Recently, following theoretical predictions \citep{Farrell-Ioannou-2007-structure}, direct numerical simulations of forced-dissipative
barotropic turbulence on a beta-plane indicated that large-scale coherent flows   could emerge
through a symmetry breaking bifurcation of the turbulent flow:
as dissipation or turbulence intensity varies, the homogeneity of the flow is broken by the spontaneous emergence of zonal jets and
large-scale waves and the large-scale structures are  supported at finite amplitude
by the turbulent eddies through shear straining \citep{Srinivasan-Young-2012,Bakas-Ioannou-2013-prl,Constantinou-etal-2014}.
Similar numerical calculations on a stratified two-layer $\beta$-plane channel with no temperature gradient across the channel have
indicated that $\beta$-plane turbulence can maintain large scale jets with vary small baroclinic component
\citep{Farrell-Ioannou-2017-saturn}.
In this work we extend the analysis to stratified two-layer  flows
and address the following question: in the absence of temperature gradients can externally forced turbulence in  a
two-layer stratified flow  support turbulent equilibria in which large scale baroclinic coherent flows are  sustained at finite amplitude? Or
are the large scale flows that emerge necessarily barotropic?

To address the emergence of large scale structure in turbulence, a new point of view was recently advanced. Through analysis
of the Statistical State Dynamics (SSD) of the flow, that is of the dynamics that governs the evolution of the flow statistics, a new type of
collective instability was revealed and it was proposed that emergence of large scale structure resulted from this symmetry
breaking instability \citep{Farrell-Ioannou-2003-structural,Farrell-Ioannou-2007-structure,Srinivasan-Young-2012,
Parker-Krommes-2013,Bakas-Ioannou-2013-prl,Parker-Krommes-2014-generation}.
%For example,  second order statistical state dynamics of a homogeneously
%forced barotropic fluid on a $\beta$-plane channel predicts that the homogeneous equilibrium state bifurcates at a critical parameter
%to an inhomogeneous equilibrium state with the emergence of large scale flows and/or large scale waves supported against dissipation
%by the Reynolds stresses of the fluctuations from the mean.
%can transition from a homogeneous turbulent state to a state
Analysis of the SSD and the resulting instability is only possible through a closure assumption for the dynamics, as a
straightforward calculation leads
to an infinite hierarchy of equations for the moments and is therefore intractable \citep{Hopf-1952,Kraichnan-1964, Frisch-1995}. There
is now a large number
of studies of barotropic turbulence \citep{Farrell-Ioannou-2007-structure,Marston-2010,Srinivasan-Young-2012}, shallow-water turbulence
\citep{Farrell-Ioannou-2009-equatorial}, baroclinic turbulence
\citep{DelSole-96,Farrell-Ioannou-2008-baroclinic,Farrell-Ioannou-2015-book,Marston-etal-2016}, turbulence in pipe flows
\citep{Constantinou-etal-Madrid-2014,Farrell-etal-2016-VLSM,Farrell-etal-2016-PTRSA}, turbulence in a convectively unstable flows
\citep{Herring-1963,Fitzgerald-Farrell-2014,Ait-Chaalal-etal-2016} and turbulence in plasma and astrophysical
flows \citep{Farrell-Ioannou-2009-plasmas,Tobias-etal-2011,Parker-Krommes-2013} providing evidence that whenever there is a
coherent flow coexisting with the turbulent field, the SSD can be accurately captured by a second-order closure. Such
closures of the SSD are either termed Stochastic Structural Stability Theory (S3T) \citep{Farrell-Ioannou-2003-structural} or
second-order cumulant expansion (CE2) \citep{Marston-etal-2008}.

%({\bf maybe move the following discussion in the conclusions?})

At first sight the success of a second order closure is surprising. Consider for example the case of
zonal jets supported by barotropic turbulence. The question is whether the quasi-linear eddy--mean flow interaction, which represents
shearing of the eddies by the jet,  dominates  the  disruptive  effects of the eddy--eddy interactions  that are represented in the  higher order
cumulants.
Close to the bifurcation point of jet emergence, the mean flow shear is infinitesimal and presumably the eddy-eddy interactions dominate
and the quasi-linear interaction is inadequate to form  jets  (cf. \cite{Srinivasan-Young-2012,Frishman-Herbert-2018}).
 Also, in highly supercritical regimes and  in the limit of high turbulence intensity
 the eddy-eddy interactions may dominate and the turbulence  may become   strongly diffusive   \citep{Held-Larichev-1996}.
 %the eddy--eddy interactions mix  potential vorticity and may eventually dominate the dynamics  \citep{McIntyre-2008,Scott-Dritschel-2012}.
However, recent studies have  provided evidence that the S3T dynamics are dominant
both at the first bifurcation point and at reasonably high supercriticalities.

Close
to the bifurcation point of jet emergence, a second-order closure was shown to be extremely accurate if the dynamics of coherent
large-scale waves
that significantly influence the flow in this regime are suppressed \citep{Constantinou-etal-2014} or are accounted for
\citep{Bakas-Ioannou-2013-prl}. In addition, \citet{Bakas-Ioannou-2013-jas} and \citet{Bakas-etal-2015} studied in detail the eddy--mean flow
dynamics underlying jet formation. They showed that an infinitesimal jet perturbation can influence the stochastically forced eddies and induce
eddy vorticity fluxes that can be interpreted as resulting from shearing of wave-packets by the infinitesimal jet. They also found in
this regime in which the eddies are dissipated before they are sheared over by the weak jet, that the induced fluxes are
up-gradient leading to a positive feedback loop and to exponential growth of the jet. It is the existence of this exponential instability
that underlies the success of the S3T predictions at the bifurcation points.

Several studies have identified the strongly non-linear regime using a single non-dimensional parameter: the zonostrophic parameter $R_\beta$
that measures the ratio of the Rhines scale over the scale where the isotropic energy cascade gets anisotropized by $\beta$
\citep{Galperin-etal-2006}.
% have shown that in the zonostrophic regime with $R_\beta\geq2$, there are strong and robust zonal jets
%supported in the turbulent flow.
The Earth's atmosphere is estimated to have $R_\beta=1.6$ while  Jupiter's and Saturn's jets are estimated to have $R_\beta=5$
\citep{Galperin-etal-2014}.  It has been confirmed  that the S3T approximations extends at least up to $R_{\beta}=2.5$.
\citet{Bakas-Ioannou-2015-book} performed numerical experiments within the zonostrophic regime at $R_\beta=2.5$ and
demonstrated that the scale of the
emergent structures is well captured by a second order closure while there are some  quantitative differences  in  the intensity of the
flows.  This demonstrated  that even though the eddy--eddy interactions are not small and may contribute to added eddy diffusion in the mean dynamics
they do not affect the S3T collective instability mechanism  that  supports the large-scale  flows.
 \cite{Scott-Dritschel-2012} were able to perform simulations at even higher $R_\beta$ and have demonstrated that
 well formed potential vorticity (PV) staircases form when  $R_\beta \geq 10$.
The validity
of the S3T dynamics in the PV staircase regime at these values of $R_\beta$ has not been checked yet.

S3T has also another important theoretical advantage. The turbulent equilibria and more importantly their instability that can lead to transitions
in the turbulent flow have analytic expression only in the SSD. The reason is that in turbulent simulations the Reynolds stresses have large
fluctuations and the mean drift in the Reynolds stresses which is responsible for the emergence of large-scale structure is not
discernible. As a result, only a framework such as S3T allows for the analytic
treatment of these turbulence-mean flow instabilities and can lead to a more comprehensive understanding of turbulent bifurcations.
For barotropic beta-plane turbulence it was shown that jets \citep{Farrell-Ioannou-2007-structure,Srinivasan-Young-2012} and
large-scale waves \citep{Bakas-Ioannou-2014-jfm}
emerge as the statistical equilibrium of homogeneous turbulence becomes unstable. This instability, was shown to be the generalization
of modulational instability in stochastically forced dissipative flows \citep{Parker-Krommes-2014-book,Bakas-etal-2015}.

In this work, we undertake the task  of presenting the analytical S3T theory for a stratified fluid on a $\beta$-plane
addressing the
following: \emph{(i)}  what type of large scale flows  emerge and are sustained at finite amplitude in a turbulent stratified atmosphere that is
homogeneous both in the horizontal and the vertical? \emph{(ii)} Since in this setting, the turbulence injection scale is not
necessarily the Rossby radius of deformation as is for example the classical case of baroclinic instability, how does the energy flow depend on
the ratio of the turbulence injection scale to the deformation radius?

We use a two-layer model
on a $\beta$-plane that allows both analytic treatment of the dynamics and numerical integration of the high dimensional
SSD of the flow. We choose to maintain this turbulent field in the simplest manner by introducing a homogeneous
and isotropic stochastic excitation that drives the turbulence and dissipation is represented as   linear dissipation of potential
vorticity. The structure of the stochastic excitations is not important so long as it maintains the
observed amplitude of turbulence given that  the anisotropy of the turbulence is induced by its interaction with the mean flow. We
first present numerical simulations of the turbulent flow that demonstrate the regime transitions in the turbulent flow and identify
the scales and characteristics of the emergent structures. We then derive the SSD of the flow under a second-order closure with
the assumption that the average for defining the statistical moments is coarse-graining over fast time scales. We address
the instability of the homogeneous equilibrium of the SSD for a wide range of parameters and the properties of the unstable modes that
represent the emerging structures. We then study the equilibration of the structure forming instability and the coherent structures
supported at finite amplitude. Finally, we compare the predictions of the S3T theory against the direct numerical simulations of the turbulent
flow and showcase its validity.

\section{Numerical simulations of the turbulent flow}

Consider a quasi-geostrophic baroclinic two--layer fluid on an infinite  beta-plane.
The upper and lower layers are denoted with subscripts~1 and~2, have equal depth, $H/2$, and densities $\rho_1$
and $\rho_2$ with $\rho_2>\rho_1$. The quasi-geostrophic dynamics governing the evolution of the barotropic
$\tilde{\psi} = (\psi_1+\psi_2)/2$ and the baroclinic $\tilde{\theta}= (\psi_1-\psi_2)/2$ streamfunction is:
\begin{eqnarray}
\partial_{\tilde{t}} \tilde{\zeta} + J(\tilde{\psi},\tilde{\zeta}) + J(\tilde{\theta},\tilde{\Del}\tilde{\theta}) +
\tilde{\beta}\partial_{\tilde{x}}\tilde{\psi} &=& -\tilde{r}\tilde{\zeta}+\tilde{\xi}^\psi\ ,\label{eq:brtr}\\
\partial_{\tilde{t}} \tilde{\eta} + J(\tilde{\psi},\tilde{\eta}) +
J(\tilde{\theta},\tilde{\zeta}) + \tilde{\beta}\partial_{\tilde{x}}\tilde{\theta} &=& -\tilde{r}\tilde{\eta}+\tilde{\xi}^\theta\ ,\label{eq:brcl}
\end{eqnarray}
where $\tilde{\Del} \equiv \partial^2_{\tilde{x}}+\partial^2_{\tilde{y}}$ is the horizontal Laplacian, $J(f,g)\equiv(\partial_{\tilde{x}} f)(\partial_{\tilde{y}} g)-
(\partial_{\tilde{y}} f)(\partial_{\tilde{x}} g)$ is the Jacobian, $\tilde{\Del}_\la\equiv\tilde{\Del} -2\tilde{\la}^2$,
$\tilde{\zeta}\equiv\tilde{\Del}\tilde{\psi}$  is the barotropic vorticity, $\tilde{\eta}\equiv\tilde{\Del}_\lambda\tilde{\psi}$,
$\tilde{x}$ and $\tilde{y}$ are the coordinates in the zonal and the meridional direction respectively and the tilde denotes
dimensional quantities (cf.~\cite{Cehelsky-Tung-1991}). The deformation radius is
$1/\tilde{\la}= \sqrt{g' (H/2)}/f_0$, where $g'=2g(\varrho_1-\varrho_2)/(\varrho_1+\varrho_2)$ is the reduced gravity and $f_0$ is
the Coriolis parameter at the center of the plane, $\tilde{\beta}$ is the planetary vorticity gradient and $\tilde{r}$ is the coefficient
of linear damping of potential vorticity.

We do not impose a temperature gradient across the channel, which through thermal wind balance would impose a mean
baroclinic shear. Instead of this large scale forcing, we impose random baroclinic and barotropic potential vorticity sources
and sinks, denoted  $\tilde{\xi}^\psi$ and $\tilde{\xi}^\theta$ respectively. These sources and sinks represent barotropic and baroclinic
excitation of the fluid by sub-scale processes or by processes not included in the quasi-geostrophic dynamics such as convection.
The random excitation is assumed to have zero mean, to be temporally delta correlated and to be statistically homogeneous in
the horizontal. Under these assumptions, the excitations in the upper layer, $\tilde{\xi}_1(\tilde{\xv},\tilde{t})$, and in the lower layer,
$\tilde{\xi}_2(\tilde{\xv},\tilde{t})$, have the two-point, two-time covariances:
\begin{equation}
\langle {\tilde{\xi}_i(\tilde{\xv} _a, \tilde{t}_a)\tilde{\xi}_j(\tilde{\xv} _b, \tilde{t}_b)}\rangle =\langle {\tilde{\xi}_j(
\tilde{\xv} _a, \tilde{t}_a)\tilde{\xi}_i(\tilde{\xv}_b, \tilde{t}_b)} \rangle= \delta(\tilde{t}_a-\tilde{t}_b)\tilde{\alpha}_{ij}
(\tilde{\xv} _a-\tilde{\xv} _b), ~~i,j=1,2~~,
\label{eq:forc_hom}
\end{equation}
where $\langle \;\bcdot\;\rangle$ denotes an ensemble average over forcing realizations, $\tilde{\xv}=(\tilde{x},\tilde{y})$, and the subscript
denotes two different points $a$ and $b$. Statistical homogeneity requires that the covariances are symmetric
to the exchange of points $a$ and $b$ and to the exchange of the excitation of the two layers. Therefore the functions
$\tilde{\alpha}_{ij}$ satisfy:
\begin{equation}
\tilde{\alpha}_{11}=\tilde{\alpha}_{22}=2 \tilde{\Xi}~~,~~\tilde{\alpha}_{12}=\tilde{\alpha}_{21}=2  \tilde{\Xi}_{12}~.\label{eq:hom_elem}
\end{equation}
An important consequence of this symmetry is that the baroclinic and barotropic components of the excitation
\begin{equation}
\tilde{\xi}^\psi = \frac{\tilde{\xi}_1+\tilde{\xi}_2}{2}~~,~~\tilde{\xi}^\theta = \frac{\tilde{\xi}_1-\tilde{\xi}_2}{2}~~,
\end{equation}
are uncorrelated:
\begin{equation}
\langle {\tilde{\xi}^\psi(\tilde{\xv} _a, \tilde{t}_a)\tilde{\xi}^\theta(\tilde{\xv} _b, \tilde{t}_b)} \rangle=0.
\end{equation}
The barotropic and baroclinic component covariances of the forcing are:
\begin{eqnarray}
\langle {\tilde{\xi}^\psi(\tilde{\xv} _a, \tilde{t}_a)\tilde{\xi}^\psi(\tilde{\xv} _b, \tilde{t}_b)} \rangle &=& \delta(\tilde{t}_a-\tilde{t}_b)\left[\bit
\tilde{\Xi}(\tilde{\xv} _a-\tilde{\xv} _b) + \tilde{\Xi}_{12}(\tilde{\xv}_a-\tilde{\xv}_b) \right]\ ,\\
\langle \tilde{\xi}^\theta(\tilde{\xv}_a, \tilde{t}_a)\tilde{\xi}^\theta(\tilde{\xv}_b, \tilde{t}_b) \rangle  &=&
\delta(\tilde{t}_a-\tilde{t}_b)\left[\bit \tilde{\Xi}(\tilde{\xv}_a-\tilde{\xv}_b) - \tilde{\Xi}_{12}(\tilde{\xv}_a-\tilde{\xv}_b) \right]\ .
\end{eqnarray}
If we assume a correlation between the layers $\tilde{\Xi}_{12}=p\tilde{\Xi}$, positive definiteness of the covariances
implies that $|p|\leq 1$. We consider three values for $p$ that span this range and exemplify opposite limits.
We consider the value $p=1$ which corresponds to the case of imposing the same excitation on the two layers
($\tilde{\xi}_1=\tilde{\xi}_2$). For
this forcing, the baroclinic forcing covariance is zero and represents at first sight the worst (best) case scenario
for the emergence of mean flows with strong baroclinic (barotropic) components. In the second case we take the opposite limit of exciting
only the baroclinic part of the flow ($p=-1$) by imposing an anti--correlated excitation in the two layers ($\tilde{\xi}_2=-\tilde{\xi}_1$).
This represents the best (worst) case scenario for the emergence of mean flows with strong baroclinic (barotropic) components.
Finally we take $p=0$ and consider an independent excitation of the two layers $\tilde{\Xi}_{12}=0$. In that case, the barotropic and
baroclinic forcing covariances are equal:
\begin{equation}
\langle {\tilde{\xi}^\psi(\tilde{\xv} _a, \tilde{t}_a)\tilde{\xi}^\psi(\tilde{\xv}_b, \tilde{t}_2b)} \rangle=
\langle {\tilde{\xi}^\theta(\tilde{\xv}_a, \tilde{t}_a)\tilde{\xi}^\theta(\tilde{\xv}_b, \tilde{t}_b)} \rangle =
\delta(\tilde{t}_a-\tilde{t}_b)\tilde{\Xi}(\tilde{\xv}_a-\tilde{\xv}_b)\ .\label{eq:forc_prop}
\end{equation}

In this work we address the following questions:\\
{\it a)} For a given spatial structure of the homogeneous forcing $\Xi$, do
any coherent structures with scales different than the ones we directly excite emerge in the flow and what are their characteristics?
We are particularly interested in whether the structures that appear are baroclinic or necessarily barotropic.\\
{\it b)} Can we develop a theory that is able to explain the emergence of coherent structures and predict their characteristics?

To address the first question, we consider (\ref{eq:brtr})-(\ref{eq:brcl}) in a
doubly periodic channel of size $2 \upi \times 2 \upi$ and integrate the equations using a pseudo-spectral code and a
fourth-order Runge-Kutta time stepping scheme. For the spatial structure of the excitation, we assume that the forcing
injects energy in a thin ring in wavenumber space that has radius $\tilde{k}_f$ and width $\Delta \tilde{k}_f$ \citep{Lilly-1969}. The power
spectrum of the spatial covariance of  the forcing is therefore:
\begin{equation}
\hat{\Xi}(\tilde{\kv})=\sum_{\tilde{k}_x}\sum_{\tilde{k}_y} \tilde{\Xi}(\tilde{\xv}_a-\tilde{\xv}_b)e^{-\ij \tilde{\kv}
\bcdot(\tilde{\xv}_a-\tilde{\xv}_b)}=\left\{\begin{array}{ll} a_f,~\mbox{for~}|\tilde{k}-\tilde{k}_f|\leq \Del
\tilde{k}_f\\0,~\mbox{for~}|\tilde{k}-\tilde{k}_f|>\Del \tilde{k}_f\end{array}\right.
,\label{eq:power}
\end{equation}
where $\tilde{\kv} =(\tilde{k}_x, \tilde{k}_y)$ is the wavevector with total wavenumber $\tilde{k}=|\tilde{\kv}|$ and
$\tilde{k}_x$, $\tilde{k}_y$ assume integer
values. The amplitude
\begin{equation}
a_f=\frac{\tilde{\varepsilon}\tilde{k}^2(\tilde{k}^2+2\tilde{\la}^2)^2}{4(\tilde{k}^4+(2+p)\tilde{\la}^4+3\tilde{k}^2\tilde{\la}^2)\Del \tilde{k}_f},
\end{equation}
is chosen so that the forcing injects energy at a rate $\tilde{\varepsilon}$ in the flow.

We use the modest $M=32\times 32$ resolution and the rather low value $\tilde{k}_f=6$ ($\Del \tilde{k}_f=1$). The reason
is that we would like to explain the phenomena observed in the direct numerical simulations denoted as NL,
with the statistical theory to be developed in sections 3-5 that requires the integration of the highly
dimensional covariance matrix (for a grid with $M$ points the covariance has dimension $2M^2$). However,
the low resolution results were compared to higher resolution simulations and there were no significant
differences found. In one layer flows with the present geometry
strong and persistent large scale flows were produced when the non-dimensional planetary vorticity
gradient $\beta=\tilde{\beta}/\tilde{k}_f\tilde{r}$ is large \citep{Bakas-Ioannou-2014-jfm}. Taking into
consideration these results we chose $\tilde{\beta}=60$ and $\tilde{r}=0.1$ yielding the non-dimensional
value of $\beta=100$, which also roughly corresponds to that in the Jovian atmosphere.

Previous studies on the emergence of large scale structures in forced-dissipative turbulence have
used two indices to quantify the energy in the emerging large scale structures. The first is the
zonal mean flow (zmf)  index
\begin{equation}
\mbox{zmf}=\frac{\sum_{\tilde{k}_y:\tilde{k}_y<\tilde{k}_f-\Del \tilde{k}_f} \hat{E}(\tilde{k}_x=0,\tilde{k}_y)}{\sum_{\tilde{k}_x\tilde{k}_y}\hat{E}(\tilde{k}_x,\tilde{k}_y)},\label{eq:zmf}
\end{equation}
with
\begin{equation}
\hat{E}(\tilde{k}_x, \tilde{k}_y)=\lim_{T\rightarrow \infty} \frac{1}{T}\int_{0}^{T}\tilde{k}^2
\left(|\hat\psi|^2+|\hat\theta|^2\right)\,\df \tilde{t}\ ,
\end{equation}
the time averaged kinetic energy of the flow at wavenumbers $(\tilde{k}_x, \tilde{k}_y)$. The zmf index  determines
the ratio  of the energy of the zonal component of the flow ($\tilde{k}_x=0$) to the energy of the perturbation field and was
used by \cite{Srinivasan-Young-2012} to study the emergence of jets in barotropic turbulence. The second is the non-zonal mean
flow (nzmf) index:
\begin{equation}
\mbox{nzmf} =\frac{\sum_{\tilde{k}_x\tilde{k}_y: \tilde{k}<\tilde{k}_f-\Del \tilde{k}_f}\hat{E}(\tilde{k}_x\ne 0, \tilde{k}_y)}
{\sum_{\tilde{k}_x\tilde{k}_y}\hat{E}(\tilde{k}_x, \tilde{k}_y)}\ ,\label{eq:nzmf}
\end{equation}
that measures the energy in waves with scales larger than the scale of the forcing ($\tilde{k}<\tilde{k}_f$) and was
used by \citet{Bakas-Ioannou-2013-prl} to study the emergence of large scale waves. If the large scale structures
that emerge are coherent, then these indices quantify their amplitude.

\begin{figure}
\centerline{\includegraphics[width=.75\textwidth]{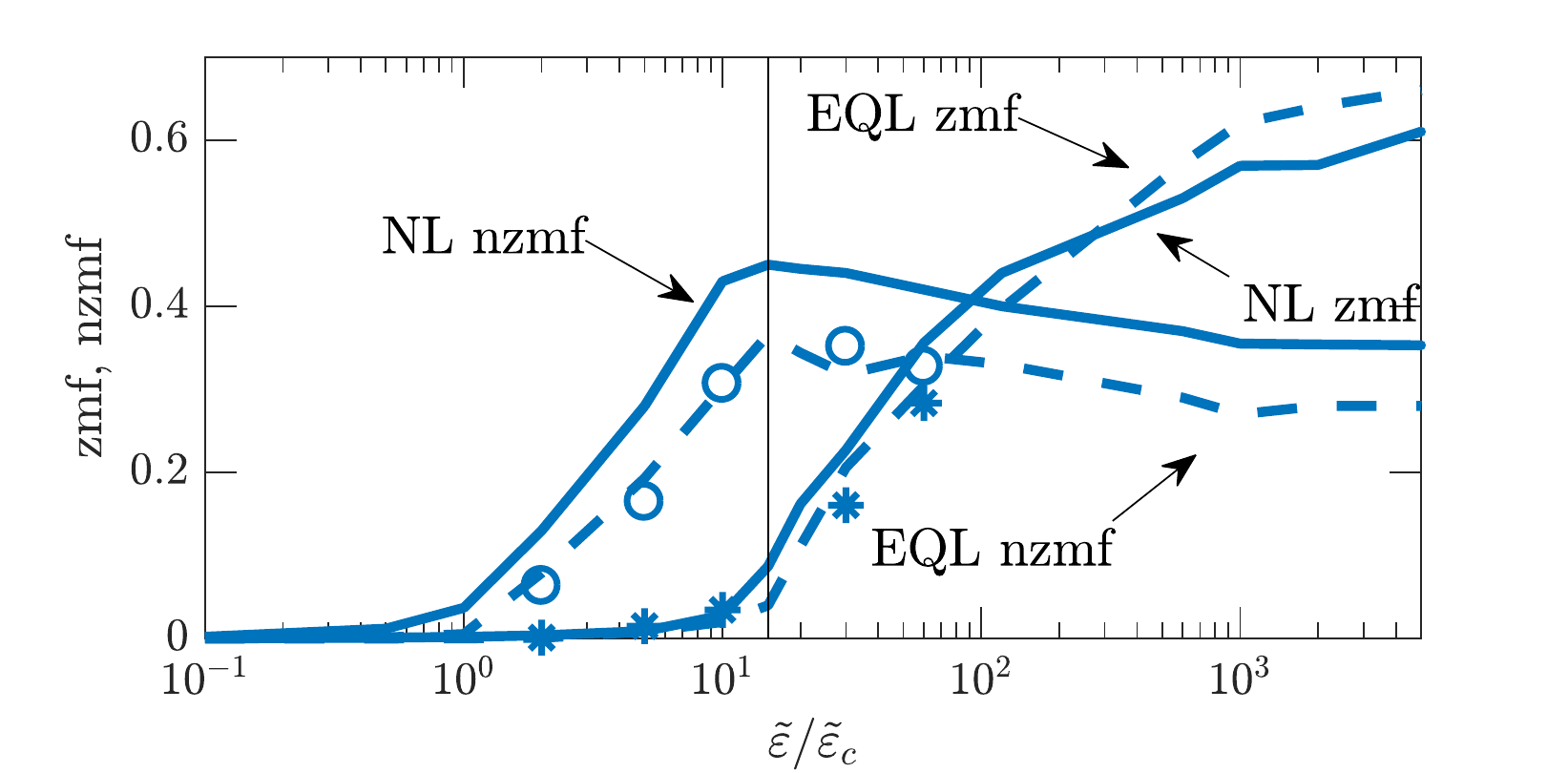}}
\caption{The zmf and nzmf indices defined in (\ref{eq:zmf}) and
(\ref{eq:nzmf}) respectively, as a function of the energy input rate $\tilde{\varepsilon}$ for the
the fully nonlinear (NL) integrations (solid lines), the ensemble quasi-linear (EQL) integrations
(dashed lines) and the S3T simulations (open circles for the nzmf and stars for the zmf index). The later two are discussed in
sections 5-6. The critical energy
input rate $\tilde{\varepsilon}_c$ for which the homogeneous state becomes unstable
is calculated from the S3T stability analysis (in section 4) and the critical energy input rate $\tilde{\varepsilon}_{nl}$ for which
the traveling wave states become unstable with respect to zonal perturbations is shown by the vertical thin line (see section 5).
The parameters values are $\tilde{\beta}=60$, $\tilde{k}_f=6$, $\Delta\tilde{k}_f=1$, $\tilde{r}=0.1$, $\tilde{\lambda}=\tilde{k}_f$ and
the forcing between the two layers is
uncorrelated ($p=0$).}\label{fig:NL_L6_bif}
\end{figure}

\begin{figure}
\centerline{\includegraphics[width=.75\textwidth]{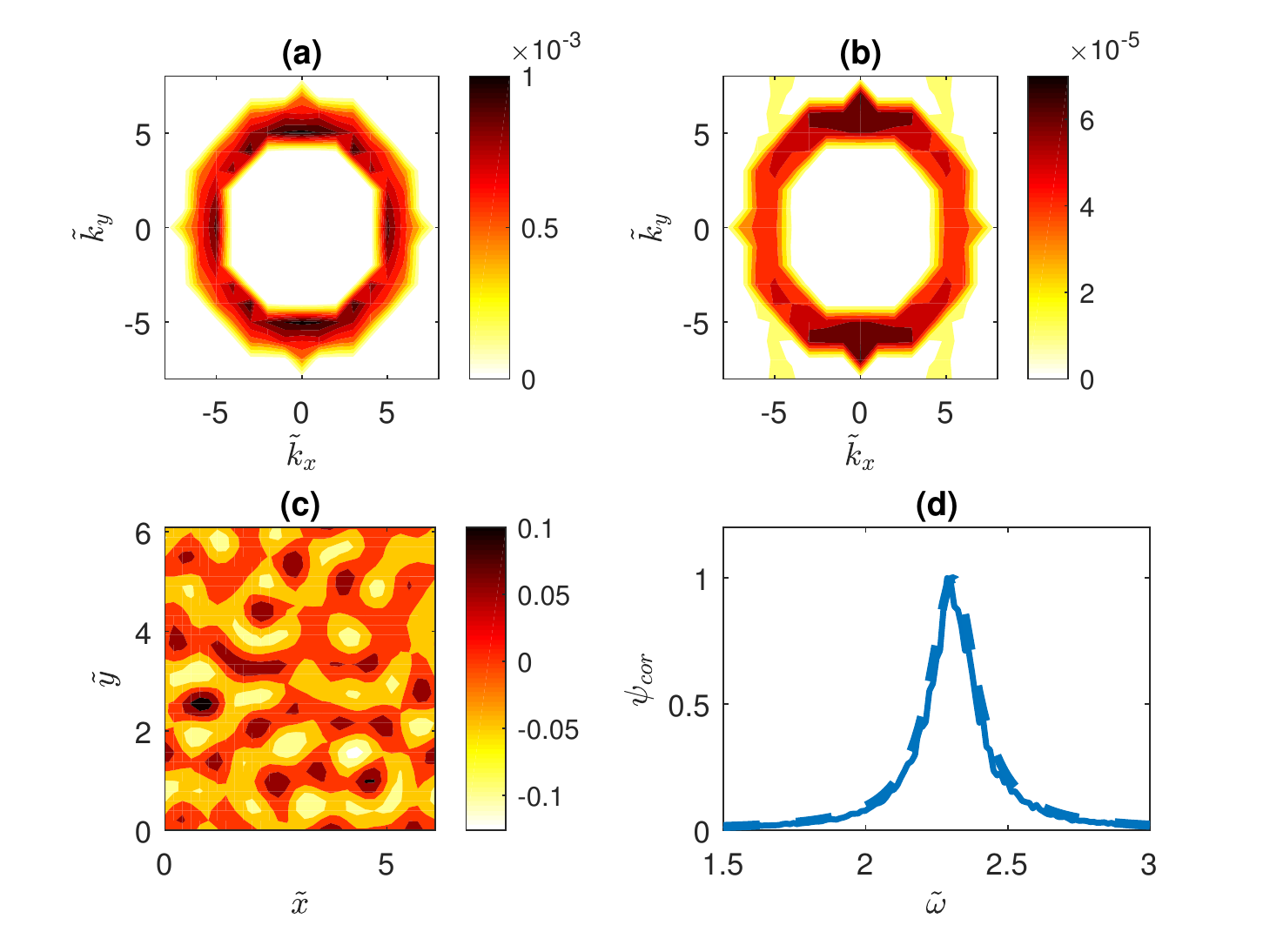}}
\caption{(a)-(b) Long time average of the kinetic energy power spectra of the (a) barotropic and (b) baroclinic part
of the turbulent flow at statistical equilibrium obtained from an NL simulation at $\tilde{\varepsilon}=\tilde{\varepsilon}_c/2$.
(c) Snapshot of the barotropic streamfunction at statistical equilibrium. (d) The ensemble mean frequency power spectrum
$\psi_{cor}$ (defined in (\ref{eq:freq_ps})) for the $\tilde{\kv}=(1,5)$ non-zonal barotropic structure. The frequency power
spectrum is normalized to unit maximum amplitude for illustration purposes. The corresponding normalized $\psi_{cor}$ for
linear Rossby waves with the same wavenumber and stochastically forced by white noise is also shown (dashed lines) for comparison.
The rest of the parameter values are the same as in figure~\ref{fig:NL_L6_bif}.}\label{fig:homog}
\end{figure}

Figure~\ref{fig:NL_L6_bif} shows the dependence of the  nzmf and zmf indices on the energy input rate
$\tilde{\varepsilon}$ for NL simulations with a Rossby radius of deformation comparable to the forcing scale
$\tilde{\lambda}=\tilde{k}_f$ and an uncorrelated forcing between the two layers ($p=0$). We observe that
for $\tilde{\varepsilon}$ lower than the critical value $\tilde{\varepsilon}_c$, which will be theoretically predicted
in section 4, there are no large scale structures in
the flow. This can also be verified by the kinetic energy power spectra shown in figure~\ref{fig:homog}a-b, as
the spectra contain only the ring of forced wavenumbers with little energy spill-over to other scales for the baroclinic
part of the flow. Therefore the flow is dominated by the directly forced waves shown in
figure~\ref{fig:homog}c that obey the Rossby wave dispersion but are phase incoherent. To quantify this we calculate
the ensemble mean of the frequency power spectrum of the streamfunction field at wavenumber $(\tilde{k}_x, \tilde{k}_y)$:
\begin{equation}
\psi_{cor}=\<|\hat{\psi}(\tilde{k}_x,\tilde{k}_y,\tilde{\omega})|^2\>,\label{eq:freq_ps}
\end{equation}
where
\begin{equation}
\hat{\psi}(\tilde{k}_x,\tilde{k}_y,\tilde{\omega})=\int\sum_{\tilde{x},\tilde{y}}\psi(\tilde{x},\tilde{y},\tilde{t})
e^{-\ij\tilde{\kv} \bcdot\tilde{\xv} -\ij\tilde{\omega}\tilde{t}}\,\df \tilde{t}\ ,
\end{equation}
for the barotropic component of the flow. Similarly we calculate $\theta_{cor}$ for the baroclinic part. Travelling waves
manifest as peaks at specific frequencies with phase coherence over times proportional to the
inverse of the half-width of  their resonant peak. We consider that the structure that emerges in the
NL is phase coherent when its coherence time exceeds the dissipation time
scale, which is the time over which a linear wave stochastically excited by white noise remains coherent. Figure~\ref{fig:homog}d
shows the barotropic frequency power spectrum $\psi_{cor}$ normalized to unit maximum amplitude for one of the waves in the
forced ring with wavenumbers $(\tilde{k}_{xr}, \tilde{k}_{yr})=(1,5)$. The frequency power spectrum peaks at the Rossby wave frequency
$\omega_r=\tilde{\beta}\tilde{k}_{xr}/(\tilde{k}_{xr}^2+\tilde{k}_{yr}^2)$, while comparison of its half-width with the
corresponding half-width of the power spectrum of stochastically forced linear Rossby waves with the same wavenumbers
shows that the coherence time for these waves is the dissipation time scale.

\begin{figure}
\centerline{\includegraphics[width=.75\textwidth]{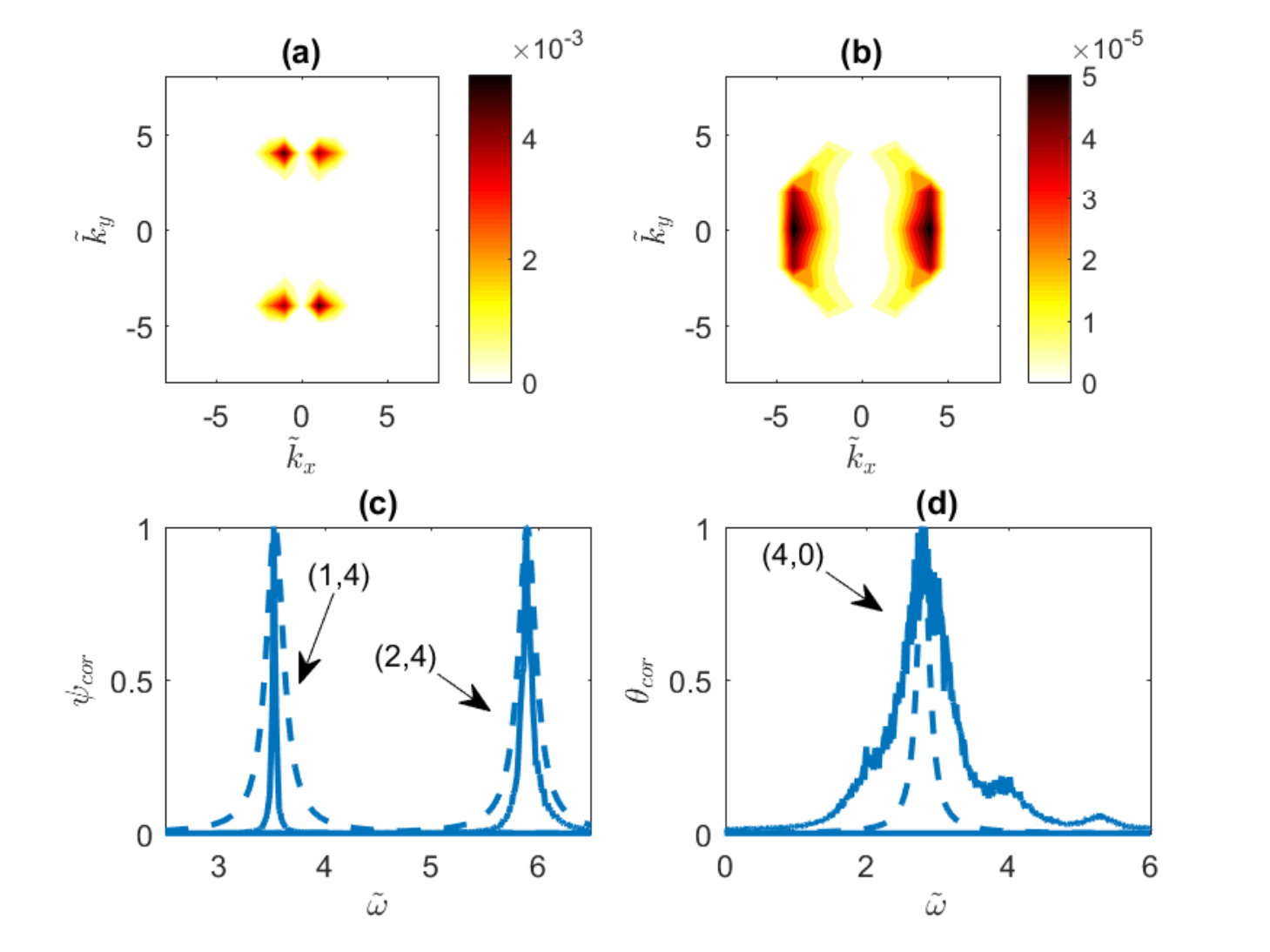}}
\caption{(a)-(b) Long time average of the kinetic energy power spectra of the (a) barotropic and (b)
baroclinic part of the turbulent flow at statistical equilibrium, obtained from an NL simulation at
$\tilde{\varepsilon}=2\tilde{\varepsilon}_c$. Shown are the spectra in the region in wavenumber space inside the forced
ring. (c) The ensemble mean frequency power spectrum $\psi_{cor}$
 for the two $\tilde{\kv}=(1,4)$ and $\tilde{\kv}=(2,4)$ non-zonal barotropic structures
dominating the kinetic energy power spectra in (a).
The frequency power spectrum is normalized to unit maximum amplitude for illustration purposes (as observed in (a)
the $\tilde{\kv}=(1,4)$ structure has about twice the energy of the $\tilde{\kv}=(2,4)$ structure). (d) The ensemble
mean frequency power spectrum $\theta_{cor}$ for the $\tilde{\kv}=(4,0)$ non-zonal baroclinic
structure dominating the kinetic energy power spectra in (b). In (c) and (d) the corresponding normalized
$\psi_{cor}$ and $\theta_{cor}$ for linear Rossby waves with the same wavenumbers and stochastically forced by white
noise is also shown (dashed lines) for comparison. The
rest of the parameter values are the same as in figure~\ref{fig:NL_L6_bif}.}\label{fig:NL_L6_spectra}
\end{figure}

The rapid increase in the nzmf index for $\tilde{\varepsilon}>\tilde{\varepsilon}_c$ shown in figure \ref{fig:NL_L6_bif}, signifies the emergence
of large scale waves in the flow. The scales of the waves are shown in figure~\ref{fig:NL_L6_spectra}
which illustrates the kinetic energy power spectra of the NL simulations for $\tilde{\varepsilon}=2\tilde{\varepsilon}_c$. We
observe that the kinetic energy of the barotropic part of the flow (cf. figure~\ref{fig:NL_L6_spectra}a) peaks at
wavenumbers $(|\tilde{k}_x|, |\tilde{k}_y|)=(1,4)$ and $(|\tilde{k}_x|, |\tilde{k}_y|)=(2,4)$. In Figure~\ref{fig:NL_L6_spectra}c  we
plot the barotropic
frequency power spectrum $\psi_{cor}$ for the two $\tilde{\kv}=(1,4)$ and $\tilde{\kv}=(2,4)$ dominant barotropic
structures and compare them with the corresponding power
spectrum of stochastically forced linear Rossby waves with the same wavenumbers. We observe that both the $(1,4)$
and the $(2,4)$ waves remain phase coherent (over six and two dissipation time scales respectively) and also satisfy to a good
approximation  the Rossby wave dispersion as $\psi_{cor}$ peaks at the corresponding Rossby wave frequencies for
both wavenumbers. In contrast, the baroclinic component of the flow is very weak compared to the barotropic part (cf. figure~\ref{fig:NL_L6_spectra}b)
and the emergent structures have an incoherent baroclinic part. For example, the wave with $\tilde{{\bf k}}=(4,0)$
that achieves maximum kinetic energy for the baroclinic component remains phase coherent for very short times (shorter
than $1/\tilde{r}$) as revealed by its frequency power spectrum $\theta_{cor}$ that is shown in figure~\ref{fig:NL_L6_spectra}d. We can
therefore conclude that large scale, phase coherent, barotropic Rossby waves emerge in the flow for $\tilde{\varepsilon}>\tilde{\varepsilon}_c$.
For larger energy input rates the energy in these large scale waves increases (increasing nzmf index in figure \ref{fig:NL_L6_bif}) as
well as the scales of the dominant waves.

\begin{figure}
\centerline{\includegraphics[width=.75\textwidth]{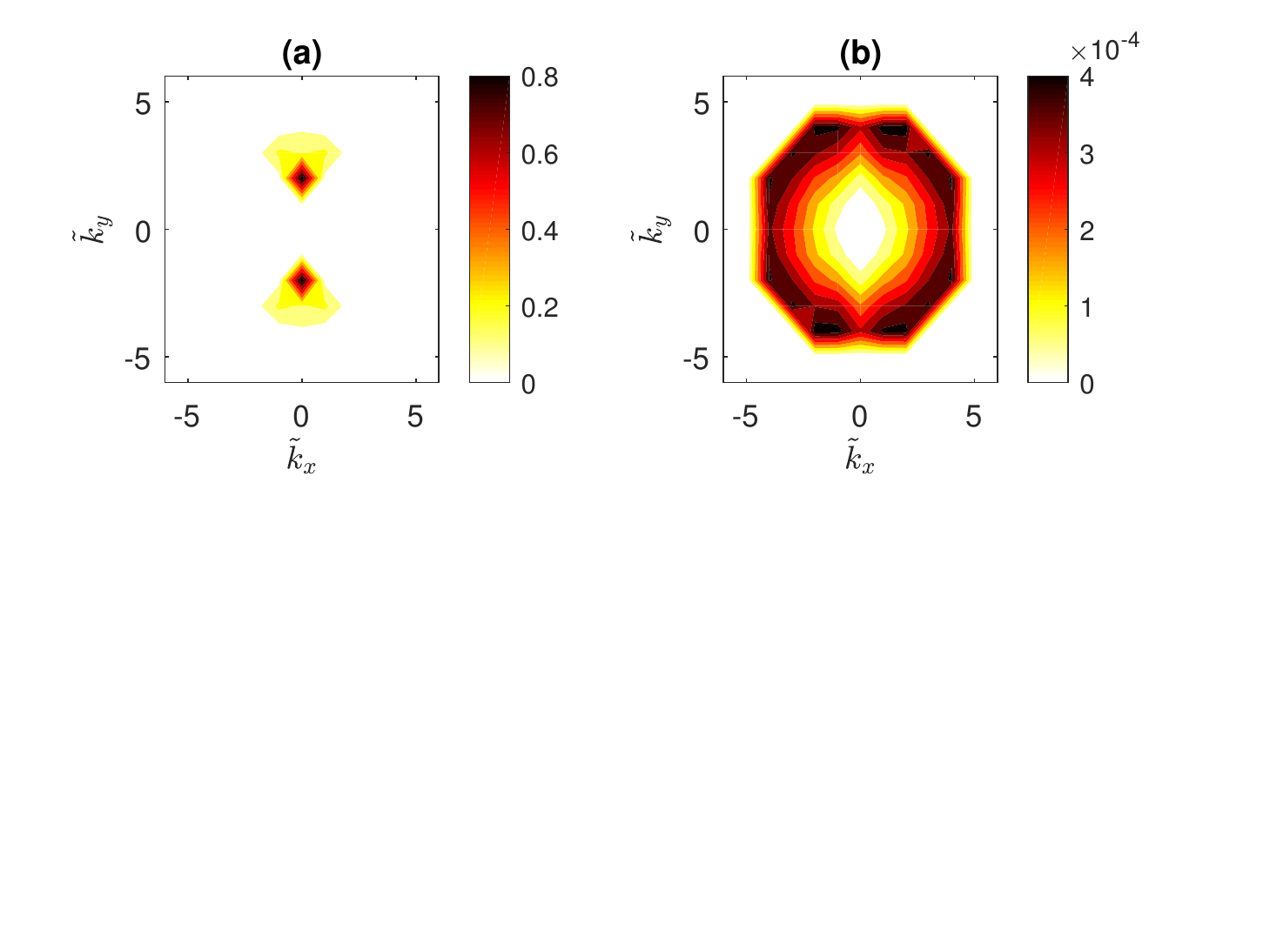}}
\caption{Long time average of the kinetic energy power spectra of the (a) barotropic and (b)
baroclinic part of the turbulent flow at statistical equilibrium obtained for a NL simulation with
$\tilde{\varepsilon}=60\tilde{\varepsilon}_c$. The
rest of the parameter values are the same as in figure~\ref{fig:NL_L6_bif}.}\label{fig:NL_L6_spectra2}
\end{figure}

When the energy input rate passes a second threshold $\tilde{\varepsilon}_{nl}$, which will be calculated in
section 5, large scale zonal jets emerge as signified by the increase of the zmf index for
$\tilde{\varepsilon}>\tilde{\varepsilon}_{nl}$ shown in figure \ref{fig:NL_L6_bif}. The barotropic kinetic energy spectra for
$\tilde{\varepsilon}=60\tilde{\varepsilon}_c$ shown in figure~\ref{fig:NL_L6_spectra2}a,
reveal that the zonal jet with $(|\tilde{k}_x|, |\tilde{k}_y|)=(0,2)$ dominates the flow, while there is
a secondary peak at the wave with $(|\tilde{k}_x|, |\tilde{k}_y|)=(1,3)$ that is phase coherent (not shown).
The baroclinic part of the spectrum is isotropic and fills the whole area inside the ring of forced wavenumbers but
is five orders of magnitude smaller. As the energy input rate further increases the energy being pumped into the barotropic
zonal jets increases, while the energy of the large scale barotropic waves decreases (cf. figure \ref{fig:NL_L6_bif}; The
scales of both the jets and the
waves also increase. In summary, there are two regime
transitions in the flow as the energy input rate of the
forcing increases. In the first transition, phase coherent, large scale Rossby waves emerge and break the homogeneity of the
turbulent flow and in the second transition large scale barotropic zonal jets emerge.

\begin{figure}
\centerline{\includegraphics[width=.75\textwidth]{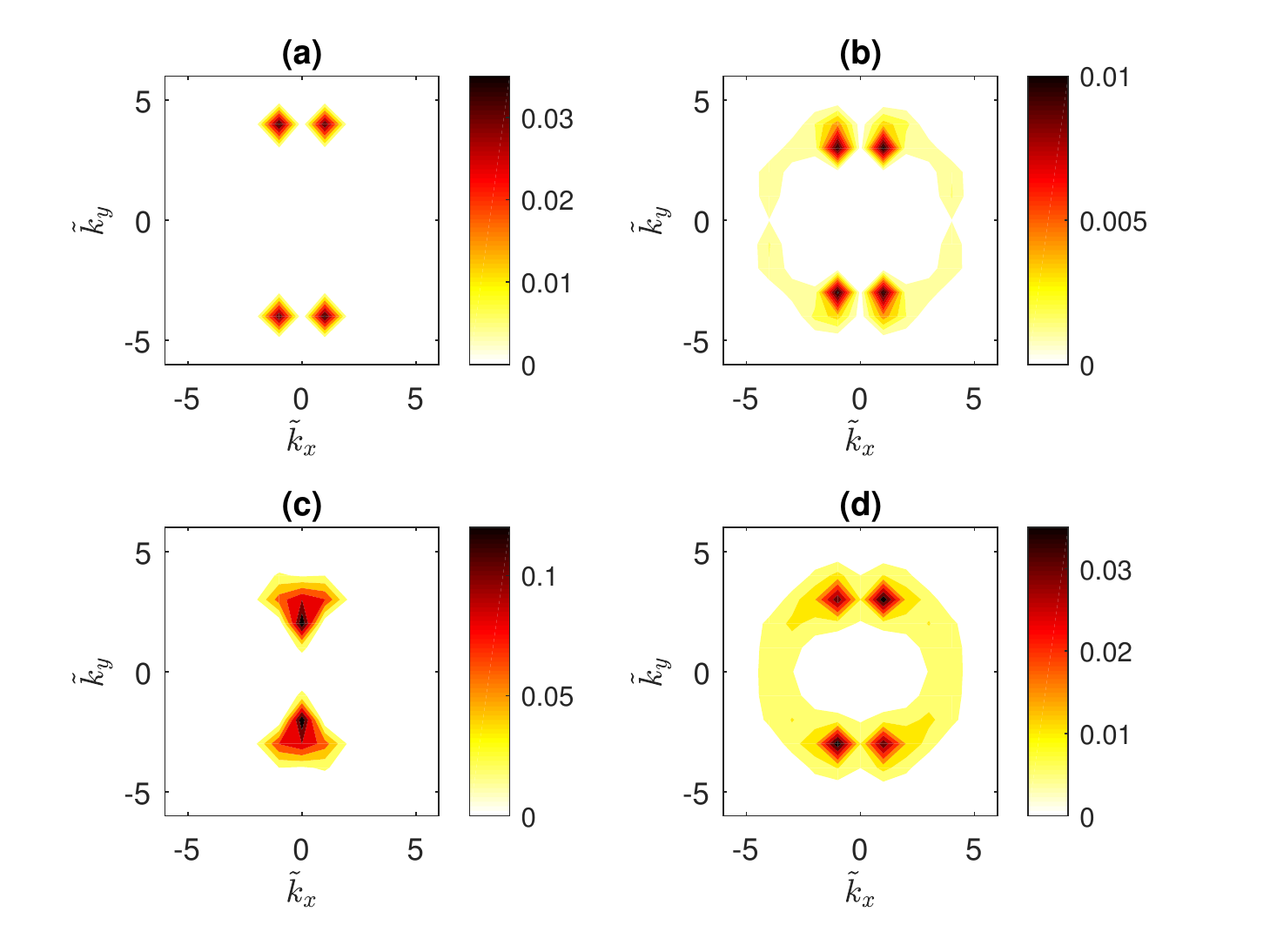}}
\caption{(a)-(b) Long-time average of the kinetic energy power spectra of the (a) barotropic and (b) baroclinic part of the
turbulent flow at statistical equilibrium for a NL simulation with
$\tilde{\varepsilon}=10\tilde{\varepsilon}_c$. (c)-(d) The same as (a)-(b) but for $\tilde{\varepsilon}=40\tilde{\varepsilon}_c$. The
deformation radius is $\tilde{\lambda}=\tilde{k}_f/6$ and the rest of the parameter values are the same as in
figure~\ref{fig:NL_L6_bif}.}\label{fig:NL_L1_spectra}
\end{figure}

For lower values of the Rossby radius of deformation $\tilde{\lambda}$, the bifurcation diagram is similar
with the two regimes of large scales waves and large scale jets evident when the energy input rate passes a
critical value. However, the structures that dominate the flow have a baroclinic component as well. Figure~\ref{fig:NL_L1_spectra}
shows the kinetic energy power spectra for $\tilde{\lambda}=\tilde{k}_f/6$ and two values of the energy input
rate above the critical threshold $\tilde{\varepsilon}_c$. For low energy input rates (c.f. figure~\ref{fig:NL_L1_spectra}a,b), the
flow is dominated by a barotropic wave with $(|\tilde{k}_x|, |\tilde{k}_y|)=(1,4)$ and a baroclinic wave
with $(|\tilde{k}_x|, |\tilde{k}_y|)=(1,3)$ and an amplitude for the baroclinic streamfunction about half of the amplitude
of the barotropic streamfunction. Calculation of the frequency power spectra $\psi_{cor}$ and $\theta_{cor}$ reveals
that both of these waves follow the Rossby wave dispersion and remain phase coherent over times longer
than the dissipation time scale (not shown).

\begin{figure}
\centerline{\includegraphics[width=.75\textwidth]{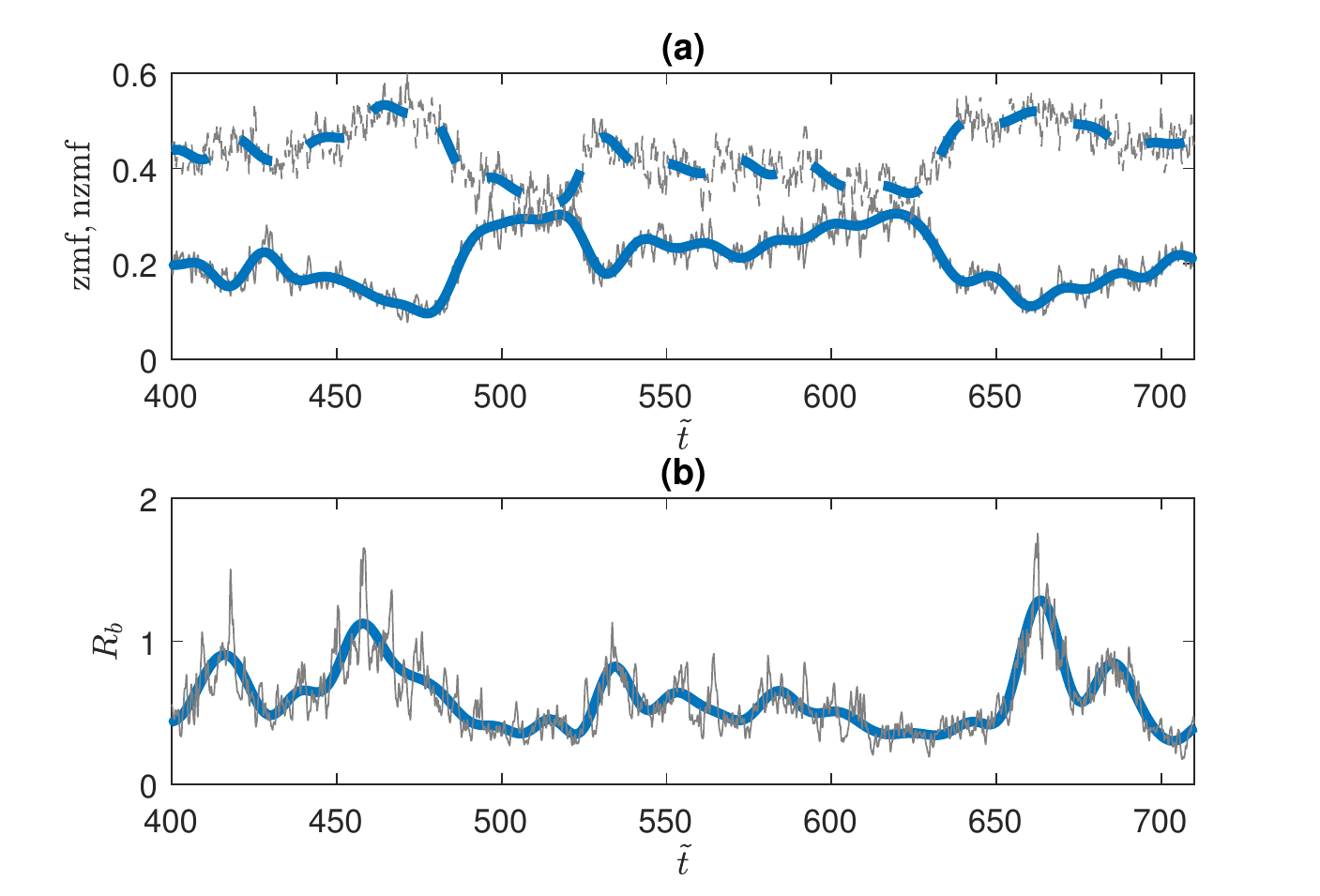}}
\caption{(a) Evolution of the zmf (solid) and nzmf (dashed) indices (thin lines). The thick lines show the low pass
filtered time series of the indices. (b) Evolution of the baroclinicity ratio (thin line). The thick line shows a low pass
filtered time series of $R_b$. The energy input rate is $\tilde{\varepsilon}=40\tilde{\varepsilon}_c$, the deformation radius
is $\tilde{\lambda}=\tilde{k}_f/6$ and
the rest of the parameters are as in figure~\ref{fig:NL_L6_bif}.}\label{fig:NL_L1_40ec_trans}
\end{figure}

At larger
energy input rates (c.f. figure~\ref{fig:NL_L1_spectra}c,d), the flow is dominated by a barotropic jet accompanied by a
baroclinic wave. This state was found to exhibit significant variability at long time scales of the order of $10/\tilde{r}$.
Figure \ref{fig:NL_L1_40ec_trans}a shows the evolution of the zmf and nzmf indices. We observe a low frequency variability
of the two indices that are anti-correlated, with time periods of stronger jets/weaker waves and time periods of weaker jets/stronger waves.
Similar variability is observed in the baroclinicity measure:
\begin{equation}
R_b=\frac{\sum_{\tilde{k}_x,\tilde{k}_y} (\tilde{k}^2+2\tilde{\la}^2)|\hat\theta(\tilde{\kv})|^2}
{\sum_{\tilde{k}_x,\tilde{k}_y} \tilde{k}^2|\hat\psi(\tilde{\kv})|^2},\label{eq:brcl_ratio}
\end{equation}
that is shown in figure \ref{fig:NL_L1_40ec_trans}b, where the periods of stronger waves are accompanied by
higher values for the flow baroclinicity. Therefore, we observe an energy exchange between the barotropic zonal jet and
the large scale baroclinic wave. For larger energy input rates, the baroclinic part of the flow weakens along with
the amplitude of this low frequency variability. As
a result, for geophysical flows that are highly supercritical with respect to the structure forming instability such
as the Jovian atmosphere, the baroclinic component of the flow is expected to be very weak.

We now test the sensitivity of the obtained results to the forcing correlation between the two layers. The
barotropic forcing ($p = 1$) presents at first sight the worst case scenario for symmetry breaking by baroclinic
structures since only barotropic eddies are injected in the flow. However, the regime transitions observed for
uncorrelated forcing remain the same. That is, when $\tilde{\lambda}=\tilde{k}_f$ barotropic structures break the
homogeneity of the turbulent flow when the energy input rate passes a critical threshold. When the
energy is injected at scales smaller than the Rossby radius of deformation ($\tilde{\lambda}=\tilde{k}_f/6$) both
barotropic and baroclinic large scale, phase coherent waves emerge in the flow. For higher energy input rates,
the flow gets zonated and barotropized. Therefore, the regime transitions in this case are not shown.

\begin{figure}
\centerline{\includegraphics[width=.75\textwidth]{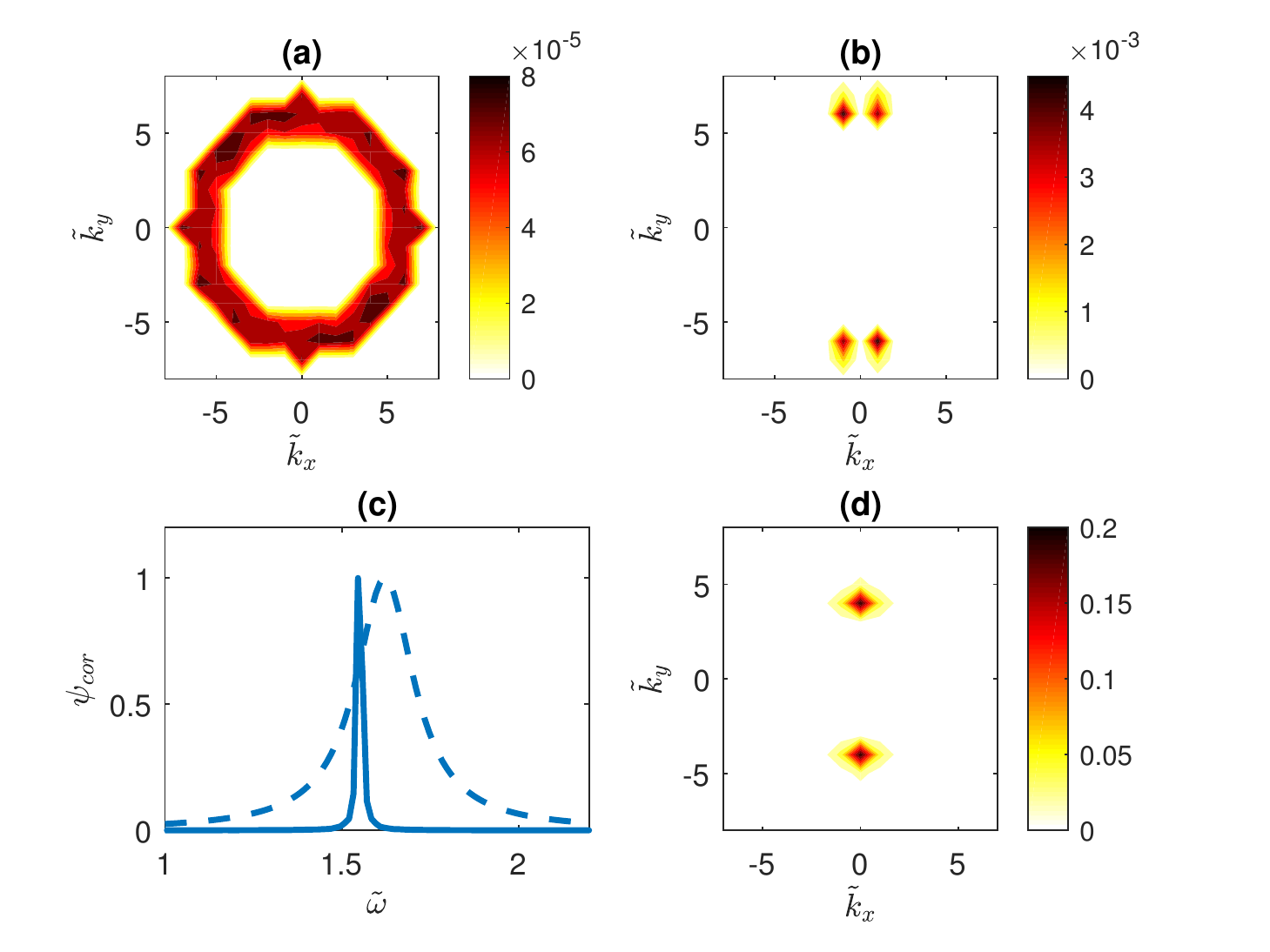}}
\caption{(a) Long time average of the kinetic energy power spectra of the baroclinic part of the turbulent flow at
statistical equilibrium for $\tilde{\varepsilon}=\tilde{\varepsilon}_c/2$. (b) Long time average of the kinetic energy power
spectra of the barotropic part of the flow for $\tilde{\varepsilon}=4\tilde{\varepsilon}_c$. (c) The ensemble mean frequency power
spectrum $\psi_{cor}$ for the $\tilde{\kv}=(1,6)$ barotropic wave dominating the power spectra in (b). (d) Long time average of the
kinetic energy power spectra of the barotropic part of the flow for $\tilde{\varepsilon}=60\tilde{\varepsilon}_c$. The forcing
is baroclinic ($p=-1$), $\tilde{\lambda}=\tilde{k}_f$ and the rest of the parameter values are the same as in
figure~\ref{fig:NL_L6_bif}. These simulations were performed with a $64\times 64$ resolution.}\label{fig:NL_pm1_L6_spectra}
\end{figure}

In contrast, the baroclinic forcing ($p=-1$) presents at first sight the worst case scenario for the emergence of
barotropic flows. For $\tilde{\lambda}=\tilde{k}_f/6$ the regime transitions are the same as in the case of uncorrelated forcing
($p=0$) and are not shown. The case of $\tilde{\lambda}=\tilde{k}_f$ is shown in figure \ref{fig:NL_pm1_L6_spectra}. For energy
input rates below a critical threshold the flow remains homogeneous. As shown in figure \ref{fig:NL_pm1_L6_spectra}a, the kinetic
energy power spectrum of the baroclinic component shows significant power only in the directly forced ring of wavenumbers, while
the power spectrum of the barotropic part of the flow is more than two orders of magnitude smaller and is not shown. This is no
surprise since we inject only baroclinic eddies in the flow. When the energy input rate passes a critical threshold, barotropic waves
emerge in the flow with scales comparable to the forcing scale $\tilde{k}_f$ even though we do not
directly force barotropic eddies. This is revealed by the kinetic energy power spectrum
of the barotropic part of the flow that is shown for $\tilde{\varepsilon}=4\tilde{\varepsilon}_c$ in figure \ref{fig:NL_pm1_L6_spectra}b
and peaks at $(|\tilde{k}_x|, |\tilde{k}_y|)=(1,6)$. The correlation power spectrum of this dominant wave that is shown in figure \ref{fig:NL_pm1_L6_spectra}c
reveals that these are phase coherent waves. For larger input rates
($\tilde{\varepsilon}=60\tilde{\varepsilon}_c$) shown in figure \ref{fig:NL_pm1_L6_spectra}, the coherent part of the flow is zonated as
in all cases previously discussed.

\begin{figure}
\centerline{\includegraphics[width=.75\textwidth]{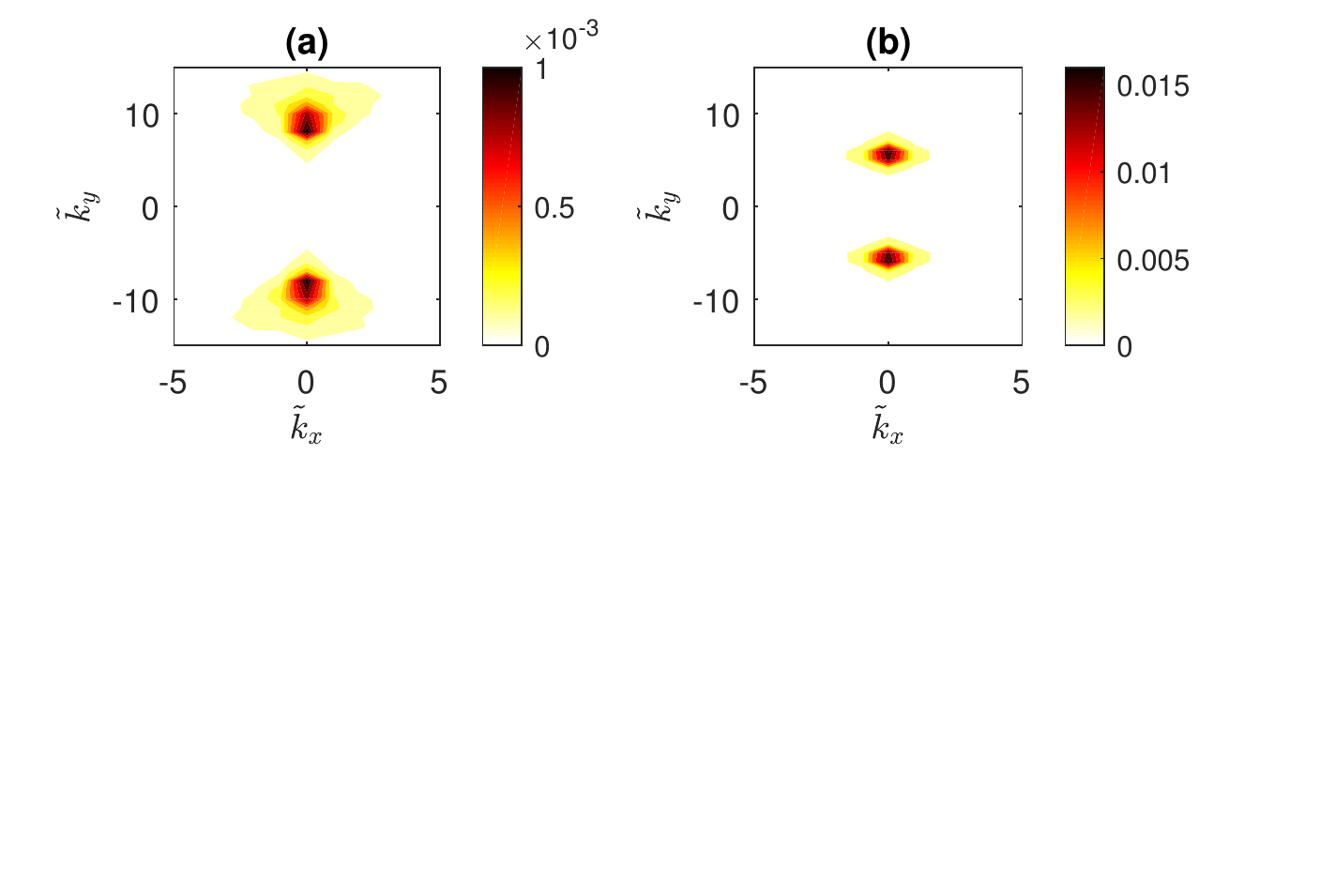}}
\caption{Long time average of the kinetic energy power spectra of the barotropic part of the turbulent flow at
statistical equilibrium for (a) $\tilde{\varepsilon}=4\tilde{\varepsilon}_c/2$ and (b) $\tilde{\varepsilon}=60\tilde{\varepsilon}_c$.
The forcing is baroclinic ($p=-1$), $\tilde{\lambda}=2\tilde{k}_f$ and the rest of the parameter values are the same as in
figure~\ref{fig:NL_L6_bif}. These simulations were performed with a $64\times 64$ resolution.}\label{fig:NL_pm1_L12_spectra}
\end{figure}

A slightly different picture
emerges when energy is injected at scales larger than the deformation radius as the wave regime is absent. For example, when
$\tilde{\lambda}=2\tilde{k}_f$ there is only one regime transition in the turbulent flow as the energy input
rate of the forcing is increased with the emergence of zonal jets breaking the translational symmetry in the flow. Figure
\ref{fig:NL_pm1_L12_spectra} illustrates the kinetic energy power spectrum of the barotropic part for two supercritical values of the
energy input rate. For low energy input rates shown in figure \ref{fig:NL_pm1_L12_spectra}a ($\tilde{\varepsilon}=4\tilde{\varepsilon}_c$),
the jets have scales comparable to the deformation scale as the power spectra peak at $|\tilde{k}_y|=3\tilde{\lambda}/4$. For larger
energy input rates shown in figure \ref{fig:NL_pm1_L12_spectra}b the jets obtain larger scales and as the energy input rate
is further increased the energy is pumped into larger and larger scale jets.

To summarize, when
the energy injection scale is comparable to the Rossby radius of deformation, the homogeneity of the flow
is broken by the emergence of large scale barotropic waves and large scale zonal jets when the energy input rate
passes certain thresholds. When the energy injection scale is smaller than the Rossby radius of deformation,
both barotropic and baroclinic large scale waves emerge in the flow and for large energy input rates the barotropic part of
the flow is zonated and becomes dominant. These results are in general independent of the correlation of the excitation
between the two layers with minor exceptions (for example the absence of the wave regime when the energy injection scale
is larger than the Rossby radius of deformation). In the following sections we develop a theory that explains
the emergence of large scale flows as a bifurcation in the turbulent flow and is also able to predict
the characteristics and amplitudes of the emergent structures.

\section{Stochastic Structural Stability Theory}

In order to analyze the transition from a statistically homogeneous
to a statistically inhomogeneous turbulent state, we consider the Statistical State Dynamics (SSD) of the two-layer model~(\ref{eq:brtr})-(\ref{eq:brcl})
which comprises the dynamics of the cumulants of the flow. Due to the infinite hierarchy of the resulting equations,
this system has to be closed at a certain order through an approximation or assumption. We follow previous studies
and consider a second order closure that is called  S3T  or CE2 \citep{Farrell-Ioannou-2003-structural,Marston-etal-2008}.
The cumulants are defined by using
a proper averaging operator  (cf. \cite{Monin-Yaglom-1971})
that captures  the emergent large-scale structure (e.g.,~zonal jet or planetary-scale wave)
and additionally satisfies the ergodic property
that the second-order cumulants are equal
to the ensemble average over forcing realizations  under the same first cumulant. In order to address
the emergence of both zonal jets and coherent waves in this work, we employ the ensemble average and 
interpret it as an average over an intermediate time scale which is long compared to the time scale of the evolution
of the incoherent flow but short compared to the time scale for the evolution of the mean. Similar assumptions
for the average separating fast from slow motions or large scale from small
scale motions were also made in previous studies \citep{Bernstein-Farrell-2010,Bakas-Ioannou-2014-jfm,
Marston-etal-2016, Constantinou-etal-2016, Bouchet-etal-2018}.

In order to formulate the theory for the two-layer flow, we first non-dimensionalize (\ref{eq:brtr})-(\ref{eq:brcl})
by choosing the damping relaxation time scale, $1/\tilde{r}$, as the characteristic time scale and the scale of the excitation,
$1/\tilde{k}_f$, as the characteristic length scale. The non-dimensional variables are: $[\zeta, \eta]=[\tilde{\zeta}, \tilde{\eta}]/\tilde{r}$,
$[\psi, \theta]=[\tilde{\psi}, \tilde{\theta}]\tilde{k}_f^2/\tilde{r}$, $[\xi^\psi, \xi^\theta]=[\tilde{\xi}^\psi, \tilde{\xi}^\theta]/\tilde{r}^2$,
$\varepsilon =\tilde{\varepsilon}\tilde{k}_f^2/\tilde{r}^3$, $\beta=\tilde{\beta}/(\tilde{k}_f\tilde{r})$, $\la=\tilde{\la}/\tilde{k}_f$ and $r=1$.
Thus, the non-dimensional version of (\ref{eq:brtr})-(\ref{eq:brcl}) lacks all tildes and has $r=1$. We also approximate the ring forcing
spectrum (\ref{eq:power}) with a forcing covariance that is isotropic and injects energy in a delta-ring in wavenumber space of
radius $1$:
\begin{equation}
\hat{\Xi}(\kv)=\frac{\varepsilon(1+2\la^2)^2}{2(1+(2+p)\la^4+3\la^2)}\delta(k-1)\ .\label{eq:ringb}
\end{equation}
Again, the amplitude is chosen to yield an energy injection rate $\varepsilon$.

The first
cumulant of the barotropic and baroclinic streamfunctions are
$\Psi(\xv ,t) \equiv  \langle \psi\rangle$ and $\Theta(\xv ,t)\equiv \langle \theta\rangle$, while the eddy field is
denoted with dashes and defined as  $\psi'(\xv ,t)= \psi-\langle \psi\rangle$, $ \theta'(\xv ,t)
\equiv \theta-\langle\theta\rangle$. In order to derive the S3T system  we first derive from~(\ref{eq:brtr})-(\ref{eq:brcl}) the
equations for the mean and the
eddies:
\begin{align}
%\label{eq:NL}
& \partial_t Z + J(\Psi,Z) + J(\Theta,\Del\Theta) +
\b\partial_x\Psi = - \langle J(\psi',\zeta')+J(\theta',\Del\theta')\bit\rangle -Z\ ,\label{eq:EQL1}\\
& \partial_t H + J(\Psi,H) +
J(\Theta,Z) + \b\partial_x\Theta =- \langle J(\psi ',\eta')+J(\theta ',\zeta')\bit\rangle -H\ ,\label{eq:EQL2}\\
& \partial_t \zeta' + J(\Psi,\zeta') + J(\psi',Z) + J(\Theta,\Del\theta') + J(\theta',\Del\Theta) +
\b\partial_x\psi' =\nonumber\\
&\qquad =\underbrace{\langle J(\psi',\zeta')+J(\theta',\Del\theta')\bit\rangle  - \(J(\psi',\zeta')+J(\theta',\Del\theta') \bit\)}_{N_\psi}-\zeta' +\xi^\psi~,\label{eq:psi'}\\
& \partial_t \eta' + J(\Psi,\eta') + J(\psi',H)+ J(\Theta,\zeta')+ J(\theta',Z) +
\b\partial_x\theta' =\nonumber\\
&\qquad = \underbrace{
\langle J(\psi ',\eta')+J(\theta ',\zeta')\bit\rangle - \(J(\psi ',\eta')+J(\theta ',\zeta') \bit\)}_{N_\theta}-
\eta'+\xi^\theta~.\label{eq:theta'}
%& &\(\partial_t + \U^\psi\bcdot\bnabla  \bit\) \Del \psi' - (\Del\U^\psi\bcdot\bnabla) \psi' +   (\U^\theta\bcdot\bnabla) \Del\theta'
%- (\Del\U^\theta\bcdot\bnabla) \theta' + \b\partial_x\psi' =\nonumber\\
%& &\hspace{2cm}=\underbrace{\langle J(\psi',\Del\psi')+J(\theta',\Del\theta')\bit\rangle_a  - \(J(\psi',\Del\psi')+J(\theta',\Del\theta') \bit\)}_{N_\psi}-\Del\psi'  +\xi^\psi~,\label{eq:psi'}\\
%& &\(\partial_t + \U^\psi\bcdot\bnabla  \bit\) \Del_\la \theta'  - (\Del\U^\psi\bcdot\bnabla)  \theta'+   (\U^\theta\bcdot\bnabla) \Del\psi' - (\Del_\la\U^\theta\bcdot\bnabla)  \psi'+\b\partial_x\theta' =\nonumber\\
%& &\hspace{2cm}= \underbrace{
%\langle J(\psi ',\Del_\la\theta')+J(\theta ',\Del\psi ')\bit\rangle_a - \(J(\psi ',\Del_\la\theta')+J(\theta ',\Del\psi ') \bit\)}_{N_\theta}-\Del_\la\theta'+\xi^\theta~,\label{eq:theta'}
\end{align}
Neglecting the eddy--eddy nonlinear terms  $N_\psi$ and $N_\theta$ in  (\ref{eq:psi'})-(\ref{eq:theta'}),
we can obtain the quasi-linear approximation for the eddies:
\begin{equation}
\partial_t[\z', \h']^T  = \A [\z', \h']^T + [\xi^{\psi}, \xi^{\theta}]^T~,
\label{eq:EQL3}
\end{equation}
where
%\begin{equation}
%\A=\(
%\begin{array}{cc}
%   -\(\U^\psi\bcdot\bnabla\) + \(\Del \U^\psi\bcdot\bnabla\)\Del^{-1} - \beta\partial_x\Del^{-1} -1  &
%   -\(\U^\theta\bcdot\bnabla\)\Del\Del_\la^{-1} + \(\Del \U^\theta\bcdot\bnabla\) \Del_\la^{-1}  \\
%   -\(\U^\theta\bcdot\bnabla\) + \(\Del_\la \U^\theta\bcdot\bnabla\)\Del^{-1}   &
%   -\(\U^\psi\bcdot\bnabla\)+\(\Del\U^\psi\bcdot\bnabla\)\Del_\la^{-1}-\b\partial_x\Del_\la^{-1} -1
%\end{array}
%\),
%\end{equation}
\begin{equation}
\A\equiv\(
\begin{array}{cc}
   \(\Del \U^\psi\Del^{-1}-\U^\psi\)\bcdot\bnabla- \beta\partial_x\Del^{-1} -1  &
   -\U^\theta\bcdot\bnabla\Del\Del_\la^{-1} + \Del \U^\theta\bcdot\bnabla \Del_\la^{-1}  \\
   -\U^\theta\bcdot\bnabla + \Del_\la \U^\theta\bcdot\bnabla\Del^{-1}   &
   \(\Del\U^\psi\Del_\la^{-1}-\U^\psi\)\bcdot\bnabla-\b\partial_x\Del_\la^{-1} -1
\end{array}
\),
\end{equation}
is the linear operator governing the evolution of the eddies about the instantaneous mean flow
$[\U ^\psi, \U ^\theta]^T$, $T$ denotes the matrix transpose and $\Del^{-1}$, $\Del_\la^{-1}$ are
the integral operators that
invert $\z '$ and $\h '$ into the barotropic and baroclinic streamfunction fields.
%{\color{blue} Limiting the eddy dynamics to their quasi-linear component us crucial, because
%this is tantamount to obtaining a second-order closure \cite{Herring-1963}.}
%in which the eddy--eddy nonlinear terms in the rhs of (\ref{eq:psi'})-(\ref{eq:theta'}) are neglected. This EQL system can be written:}
Equations (\ref{eq:EQL1})-(\ref{eq:EQL2}) and (\ref{eq:EQL3}) constitute the ensemble quasi-linear (EQL) approximation
to the fully non-linear dynamics.

The S3T system is then obtained by first expressing the eddy fluxes forcing the barotropic and baroclinic
flow in terms of the second cumulant
\begin{equation}
\C(\xv _a, \xv _b) \equiv \C_{ab}=
\(
\begin{array}{cc}
   \< \z'_a\z'_b\>  &    \<\z'_a\h'_b \>  \\
   \<\h'_a \z'_b \>  &    \< \h'_a\h'_b\>
\end{array}
\) =\(
\begin{array}{cc}
   C^{\z\z}_{ab} &  C^{\z\h}_{ab} \\
     C^{\h\z}_{ab} &  C^{\h\h}_{ab}
\end{array}
\),\label{eq:Cdef}
\end{equation}
that is the covariance matrix  of the eddy fields $\z'$, $\h'$. In~(\ref{eq:Cdef}) we use the shorthand
$\z'_a=\zeta'(\xv_a, t)$, to refer to the value of the relative barotropic
vorticity at the specific point $\xv_a$ (and similarly for $\h'$). Defining the streamfunction
covariance matrix
\begin{equation}
\S(\xv _a, \xv _b) \equiv \S_{ab} =
\(
\begin{array}{cc}
   S^{\psi\psi}_{ab} &  S^{\psi\theta}_{ab} \\
   S^{\theta\psi}_{ab} &  S^{\theta\theta}_{ab}
\end{array}
\),
\end{equation}
that is related to the eddy covariance as:
\begin{equation}
\C_{ab} =\(
\begin{array}{cc}
   \Del_a\Del_b\,   S^{\psi\psi}_{ab} &  \Del_\a\Del_{\la,b}\,S^{\psi\theta}_{ab} \\
    \Del_{\la,a}\Del_\b\, S^{\theta\psi}_{ab} &  \Del_{\la,a}\Del_{\la,b}\,S^{\theta\theta}_{ab}
\end{array}
\),
\end{equation}
we can express the Reynolds stress divergences as:
\begin{eqnarray}
R^\psi(\C)&\equiv&\langle J(\psi',\zeta')+J(\theta',\Del\theta')\bit\rangle=\partial_x \[\frac1{2}
\(-\partial_{y_a}\Del_b-\partial_{y_b}\Del_a\) \(S^{\psi\psi}_{ab}
+S^{\theta\theta}_{ab}\) \]_{\x_a=\x_b}\nonumber\\
%&=&\partial_x \[\frac1{2}  \(-\partial_{y_a}\Del_b-\partial_{y_b}\Del_a\) \(S^{\psi\psi}_{ab}
%+S^{\theta\theta}_{ab}\) \]_{\x_a=\x_b}+\nonumber\\
&+&\partial_y  \[ \frac1{2} \(\partial_{x_a}\Del_b+\partial_{x_b}\Del_a\) \(S^{\psi\psi}_{ab}+S^{\theta\theta}_{ab}\) \]_{\x_a=\x_b}~,\\
R^\theta(\C)&\equiv&\langle J(\psi',\eta')+J(\theta',\zeta')\bit\rangle =\nonumber\\
&=&\partial_x \[ \frac{1}{2}\(-\partial_{y_a} \Del_{\la,b}-\partial_{y_b}\Del_{a}\) S^{\psi\theta}_{ab} + \frac{1}{2}\(-\partial_{y_a} \Del_{b}-\partial_{y_b} \Del_{\la,a}\) S^{\theta\psi}_{ab} \]_{\x_a=\x_b}+\nonumber\\
  &+&\partial_y \[ \frac{1}{2}\(\partial_{x_a} \Del_{\la,b}+\partial_{x_b}\Del_{a}\) S^{\psi\theta}_{ab} + \frac{1}{2}\(\partial_{x_a} \Del_{b}+\partial_{x_b} \Del_{\la,a}\) S^{\theta\psi}_{ab} \]_{\x_a=\x_b}~,
\end{eqnarray}
where the subscripts $a$ denote the action of the differential operators on $\x_a$,  and $\x_a=\x_b$
denotes that the function of the two points $\x_a$ and $\x_b$ is to be evaluated  at the same point. The first cumulant therefore evolves as:
\begin{eqnarray}
\(\partial_t + \U^\psi\bcdot\bnabla \) Z  +  (\U^\theta\bcdot\bnabla) \Del {\Theta}+
\b {V^\psi} &=& R^\psi(\C)-Z\ ,\label{eq:first_cumZ}\\
\(\partial_t + \U^\psi\bcdot\bnabla \) H + (\U^\theta\bcdot\bnabla)Z  +
\b V^\theta &=& R^\theta(\C)-H\ ,\label{eq:first_cumH}
\end{eqnarray}
and in the limit of an infinite ensemble, the second cumulant $C$ evolves according to the free of
fluctuations time-dependent Lyapunov equation:
\begin{equation}
\partial_t \C_{ab} = \A_a\, \C_{ab} + \( \A_b\,\C_{ab}^T\)^T + \,\Q_{ab}~,\label{eq:dCdt}
\end{equation}
with
%\begin{equation}
%\label{eq:QQ}
%\Q_{ab} =
%\(
%\begin{array}{cc}
%  \left<\xi_a^{\psi}\, \xi_b^{\psi}\right>  & 0  \\
%   0   &   \left<\xi_a^{\theta}\, \xi_b^{\theta}\right>
%\end{array}
%\)=\(
%\begin{array}{cc}
%  (2-p)\Xi(\x_a-\x_b)  & 0  \\
%   0   &   p\Xi(\x_a-\x_b)
%\end{array}
%\)\ ,
%\end{equation}
\begin{equation}
\label{eq:QQ}
\Q_{ab} =
\(
\begin{array}{cc}
  \left<\xi_a^{\psi}\, \xi_b^{\psi}\right>  & 0  \\
   0   &   \left<\xi_a^{\theta}\, \xi_b^{\theta}\right>
\end{array}
\)=\(
\begin{array}{cc}
  (1+p)\Xi(\x_a-\x_b)  & 0  \\
   0   &   (1-p)\Xi(\x_a-\x_b)
\end{array}
\)\ ,
\end{equation}
the homogeneous spatial covariance of the stochastic forcing. The subscript
$a$ in $\A$ denotes that the differential
operators in $\A$ act on the variables $\x_a$ of $\C_{ab}$
and also that the functions in $\A$ are evaluated at $\x_a$.
%The parameter $p$ specifies the forcing
%covariance $\Q$. Excitation~(\ref{eq:forc_prop}) is obtained by setting $p=1$ and
%excitation
%(\ref{eq:forc_no_brclnc}) is obtained with $p=0$.
Equations (\ref{eq:first_cumZ})-(\ref{eq:dCdt}) form the S3T dynamical system which governs the joint
evolution of the large-scale flow field described by $[Z, H]^T$ and  the covariance $\C$ of the eddy field
and is a second order closure of the SSD, as the quasi-linear approximation to the dynamics amounts to
ignoring the third and higher order cumulants in the SSD.

\section{S3T stability of the homogeneous equilibrium}

The statistical stationary states of the turbulent flow are manifestations of the fixed points
of the SSD. If the fixed points of the SSD are stable then the turbulent flow
is expected to remain close to this stationary state. If the fixed points are unstable, then
the instability manifests as a transition in the turbulent flow and the new
fixed points that result from the equilibration of the SSD instability correspond to the new
attractor in the turbulent flow. Therefore the regime transitions from the homogeneous turbulent
flow that are observed as the energy input rate increases should correspond to the instability of a
homogeneous fixed point of the SSD and the dominant structures should be manifestations of the
new fixed points of the SSD.

Under homogeneous excitation, the  state with no  mean flow $Z=H=0$ and homogeneous covariance
$\C_E= \Q /2$, with  $\Q$ the excitation covariance~(\ref{eq:QQ}), is
a fixed point of the S3T system (\ref{eq:first_cumZ})-(\ref{eq:dCdt}). This is the homogeneous statistical equilibrium
of the two-layer fluid. We study the stability of this equilibrium  as the energy input rate of the excitation, $\varepsilon$,
is increased. The linear stability is addressed by performing an
eigenanalysis of the S3T system linearized about this equilibrium. The eigenfunctions have
both a mean flow component and a perturbation covariance component. The homogeneity of the equilibrium
implies  that the  mean flow component of the  eigenfunctions consists of sinusoidal functions:
\begin{equation}
\label{eq:mn}
[\delta \Psi,\delta\Theta]^T=[a_\psi, a_\theta]^T\,e^{\ij \nv \bcdot \xv} e^{\sigma t}\ ,
\end{equation}
and the covariance eigenfunction component also consists of sinusoidal functions of the form:
\begin{equation}
\label{eq:cn}
[\delta S^{\psi\psi}, \delta S^{\psi\theta}, \delta S^{\theta\psi}, \delta S^{\theta\theta}]^T=
\frac{e^{\ij\nv \bcdot \frac{\xv_a+\xv_b}{2}+\sigma t}}{2\upi}
\int_{-\infty}^{\infty}\int_{-\infty}^{\infty} [\hat{S}^{\psi\psi}, \hat{S}^{\psi\theta}, \hat{S}^{\theta\psi}, \hat{S}^{\theta\theta}]^T
e^{\ij\kv\bcdot (\xv_a-\xv_b)}\,\df^2\kv ~,
\end{equation}
with $\nv = (n_x, n_y)$  the wavevector of the eigenfunction, and  $n=|\nv |$ the total wavenumber.
The inhomogeneity of the covariance eigenfunction is revealed by its dependence on the mean position $(\xv_a+\xv_b)/2$.
The  eigenvalue  $\sigma$  associated with  $\nv$ governs the stability of  perturbation~(\ref{eq:mn}),~(\ref{eq:cn}) to the homogeneous equilibrium state;
the growth rate of this  perturbation is $\sigma_r \equiv \Real(\sigma)$ and its frequency is $\sigma_i \equiv \Imag(\sigma)$.

It is shown  in  Appendix \ref{sec:appB} that the eigenfunctions are either purely barotropic  ($a_\psi \ne 0$, $a_\theta=0$)
or purely baroclinic   ($a_\psi = 0$, $a_\theta \ne0$). The barotropic modes  satisfy the dispersion relation:
\begin{equation}
\sigma+1-\ij\beta n_x/n^2=f_\psi(\sigma)\ , \label{eq:stab_eq1}
\end{equation}
while the baroclinic modes satisfy:
\begin{equation}
\sigma+1-\ij\beta n_x/n_\lambda^2=f_\theta(\sigma)\ ,\label{eq:stab_eq2}
\end{equation}
with $n_\la\equiv\sqrt{n^2+2\la^2}$. The terms $f_\psi$, $f_\theta$ represent the  eddy acceleration feedback
on the  barotropic and baroclinic mean flow induced upon perturbing the equilibrium by the mean flow eigenfunction.
These feedbacks are  shown in Appendix \ref{sec:appB} to be:
\begin{eqnarray}
f_\psi (\sigma) &=&\frac{1+p}{4\upi n^2}\int_{-\infty}^{\infty}\int_{-\infty}^{\infty} \frac{(n_yk_x-n_xk_y)^2(k_{++}^2-k^2)(1-n^2/k^2)
\hat{\Xi}}{(\sigma+2)k^2k_{++}^2+i\beta(k_xk_{++}^2-k_{x++}k^2)}\,\df^2\kv\nonumber\\ & &\quad+
\frac{1-p}{4\upi n^2}\int_{-\infty}^{\infty}\int_{-\infty}^{\infty} \frac{(n_yk_x-n_xk_y)^2(k_{++}^2-k^2)(1-n^2/k_\lambda^2)
\hat{\Xi}}{(\sigma+2)k_\lambda^2k_{\lambda ++}^2+i\beta(k_xk_{\lambda ++}^2-k_{x++}k_\lambda^2)}\,\df^2\kv,\label{eq:fpsi}
\end{eqnarray}
and
\begin{eqnarray}
f_\theta (\sigma)&=&\frac{1+p}{4\upi n_\lambda^2}\int_{-\infty}^{\infty}\int_{-\infty}^{\infty}\frac{(n_yk_x-n_xk_y)^2(k_{\lambda ++}^2-k^2)(1-n_\lambda^2/k^2)
\hat{\Xi}}{(\sigma+2)k^2k_{\lambda ++}^2+i\beta(k_xk_{\lambda ++}^2-k_{x++}k^2)}\,\df^2\kv\nonumber\\& &\quad+
\frac{1-p}{4\upi n_\lambda^2}\int_{-\infty}^{\infty}\int_{-\infty}^{\infty}\frac{(n_yk_x-n_xk_y)^2(k_{++}^2/k_\lambda^2-1)(k^2-n^2)
\hat{\Xi}}{(\sigma+2)k_\lambda^2k_{++}^2+i\beta(k_xk_{++}^2-k_{x++}k_\lambda^2)}\,\df^2\kv,\label{eq:ftheta}
\end{eqnarray}
with the notation: $k_\la^2=1+2\la^2$, $k_{x++}=k_x+n_x$, $k_{++}^2=|\kv+\nv |^2$ and
$k_{\la ++}^2=|\kv+\nv |^2+2\la^2$.

%Numerical solution of (\ref{eq:stab_eq1})-(\ref{eq:stab_eq2}) under
%the isotropic ring excitation   (\ref{eq:ring}) shows that
Both $f_\psi$ and $f_\theta$ are linear functions of the energy input rate
$\varepsilon$ and  the S3T homogeneous equilibrium becomes  unstable when these feedbacks have positive real part and the
energy input rate, $\varepsilon$, exceeds a critical value $\varepsilon_c$. The critical value, $\varepsilon_c$,  is obtained
by first determining the energy input rate $\varepsilon_t(\nv )$ for which the eigenfunction   $\nv$
is neutral (satisfying $\sigma_r (\nv) =0$),
and then determining the barotropic or baroclinic eigenfunction $\nv$ that achieves neutrality
with the least energy and set
$\varepsilon_c=\mbox{min} \left ( \varepsilon_t (\nv) \right )$.

For a single layer barotropic fluid, the dispersion properties were found to depend on the
value of the non-dimensional planetary vorticity gradient $\beta$. For $\beta\ll1$ or $\beta=\mathcal{O}(1)$ stationary zonal
jets grow at the fastest rate, while for large values of $\beta$ non-zonal westward propagating structures are
more unstable. The same behaviour is observed for the two--layer fluid as well.

\begin{figure}
\centerline{\includegraphics[width=.75\textwidth]{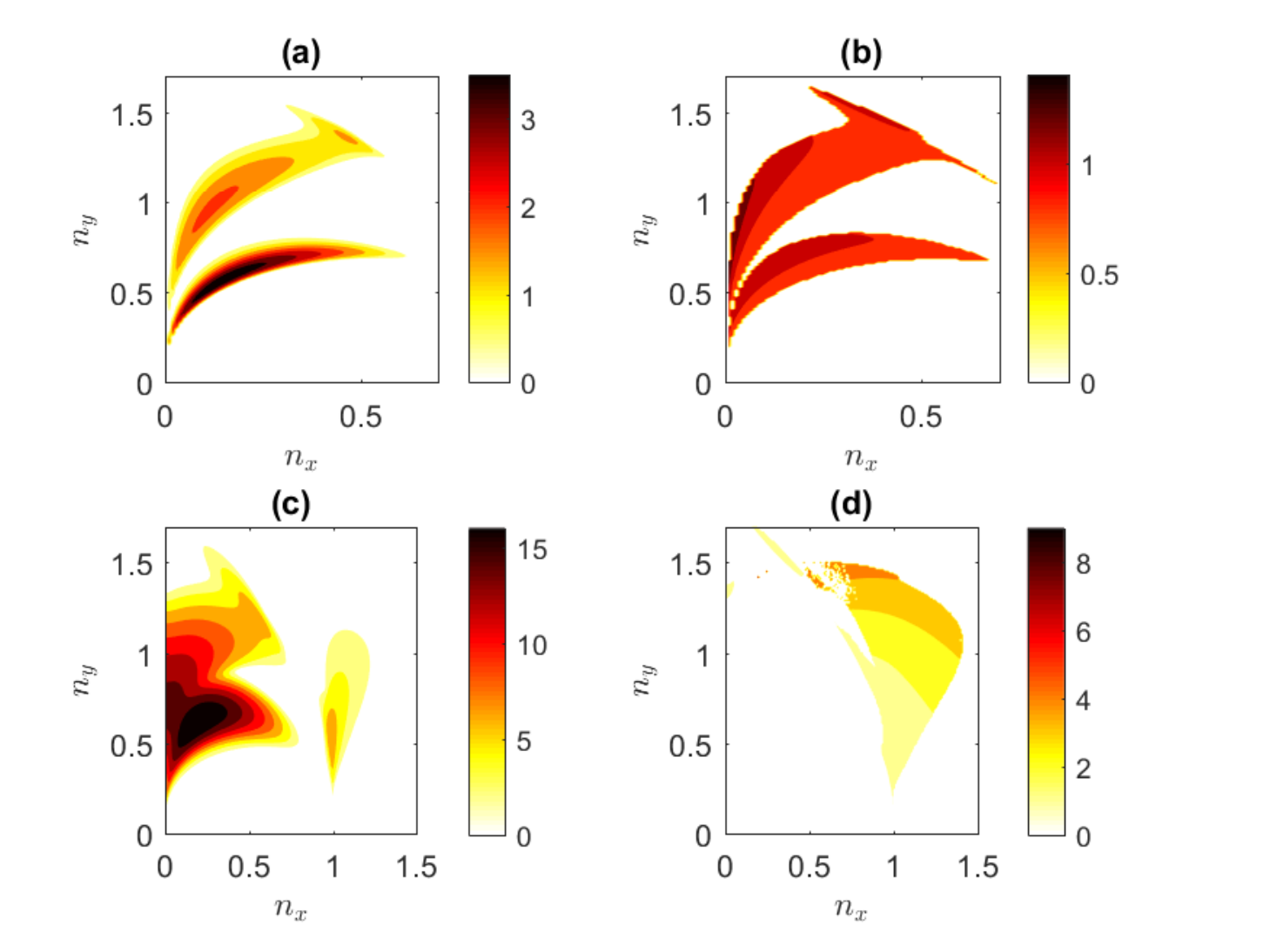}}
\caption{Dispersion properties of the unstable barotropic modes for $\beta=100$, $\lambda=1$.
(a) and (c) Growth rate $\sigma_r$ of the most unstable modes for (a) $\varepsilon=10\varepsilon_c$ and
(c) $\varepsilon=100\varepsilon_c$. (b) and (d) Ratio of the frequency of the most unstable modes over
the corresponding barotropic Rossby wave frequency $R_{ft}=\sigma_i/\sigma_{Rt}$ for (b)
$\varepsilon=10\varepsilon_c$ and (d) $\varepsilon=100\varepsilon_c$. The forcing in the two layers is uncorrelated ($p=0$).}\label{fig:disp3}
\end{figure}

We first consider $\beta=100$   and an uncorrelated forcing between the two
layers ($p=0$). For $\la\geq 0.68$ only barotropic modes become unstable when the energy input rate passes a critical value, while
baroclinic modes are stable for all  values of $\varepsilon$. Therefore the absence of baroclinic structures
when $\tilde{\lambda}=\tilde{k}_f$ ($\lambda=1$) in the NL simulations is due to the stability of the
homogeneous equilibrium to baroclinic perturbations. The growth rate and frequency of the unstable
barotropic modes as a function of the mean flow wavenumber $\nv$ are shown in figure~\ref{fig:disp3} for two values of
$\varepsilon$ ($\varepsilon=10\varepsilon_c$ and $\varepsilon=50\varepsilon_c$) and for $\lambda=1$.
At low supercriticality (cf. figure~\ref{fig:disp3}a) only non-zonal modes ($n_x\neq 0$) are unstable with two branches
of instability: one at meridional scales comparable to the forcing scale
($n_y\simeq 1$) and one at larger meridional scales, with the most unstable modes having larger scales than that of the excitation.
The frequencies of the unstable modes  are very close to the frequencies of the free barotropic Rossby waves $\sigma_{Rt}=\beta n_x/n^2$.
This can be quantified by calculating the ratio $R_{ft}=\sigma_i/\sigma_{Rt}$, which is plotted  in figure~\ref{fig:disp3}b. At larger
supercriticality
(cf. figure~\ref{fig:disp3}b) the two branches merge, stationary ($\sigma_i=0$) barotropic
zonal jet eigenfunctions  ($n_x=0$) become unstable and there is an additional branch of unstable non-zonal modes with
zonal wavenumbers comparable to the excitation wavenumber ($n_x\simeq1$) with the  frequency of these modes
departing substantially from the Rossby wave frequency (cf. figure~\ref{fig:disp3}d). However, the large-scale
non-zonal structures, remain the most unstable modes.

\begin{figure}
\centerline{\includegraphics[width=.75\textwidth]{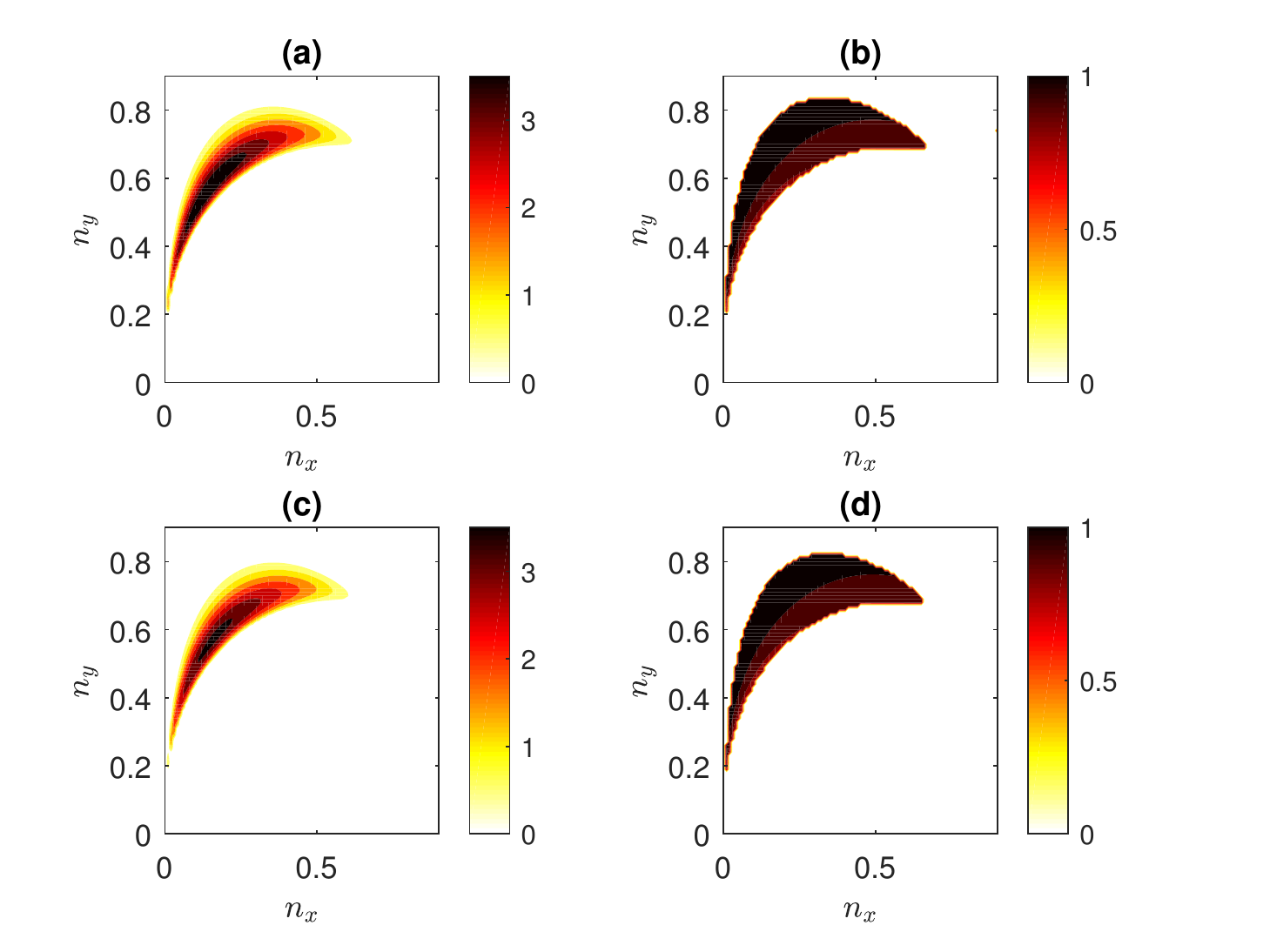}}
\caption{Dispersion properties of the unstable modes for $\beta=100$, $\lambda=0.1$ and $\varepsilon=10\varepsilon_c$.
(a) and (c) Growth rate $\sigma_r$ of the most unstable (a) barotropic and (c) baroclinic modes. (b) and (d)
Ratio of the frequency of the most unstable (b) barotropic modes
over the corresponding barotropic Rossby wave frequency $R_{ft}=\sigma_i/\sigma_{Rt}$ and (d) baroclinic modes
over the corresponding baroclinic Rossby wave frequency $R_{fc}=\sigma_i/\sigma_{Rc}$. The forcing in the two layers is uncorrelated ($p=0$).}\label{fig:disp1}
\end{figure}

\begin{figure}
\centerline{\includegraphics[width=.75\textwidth]{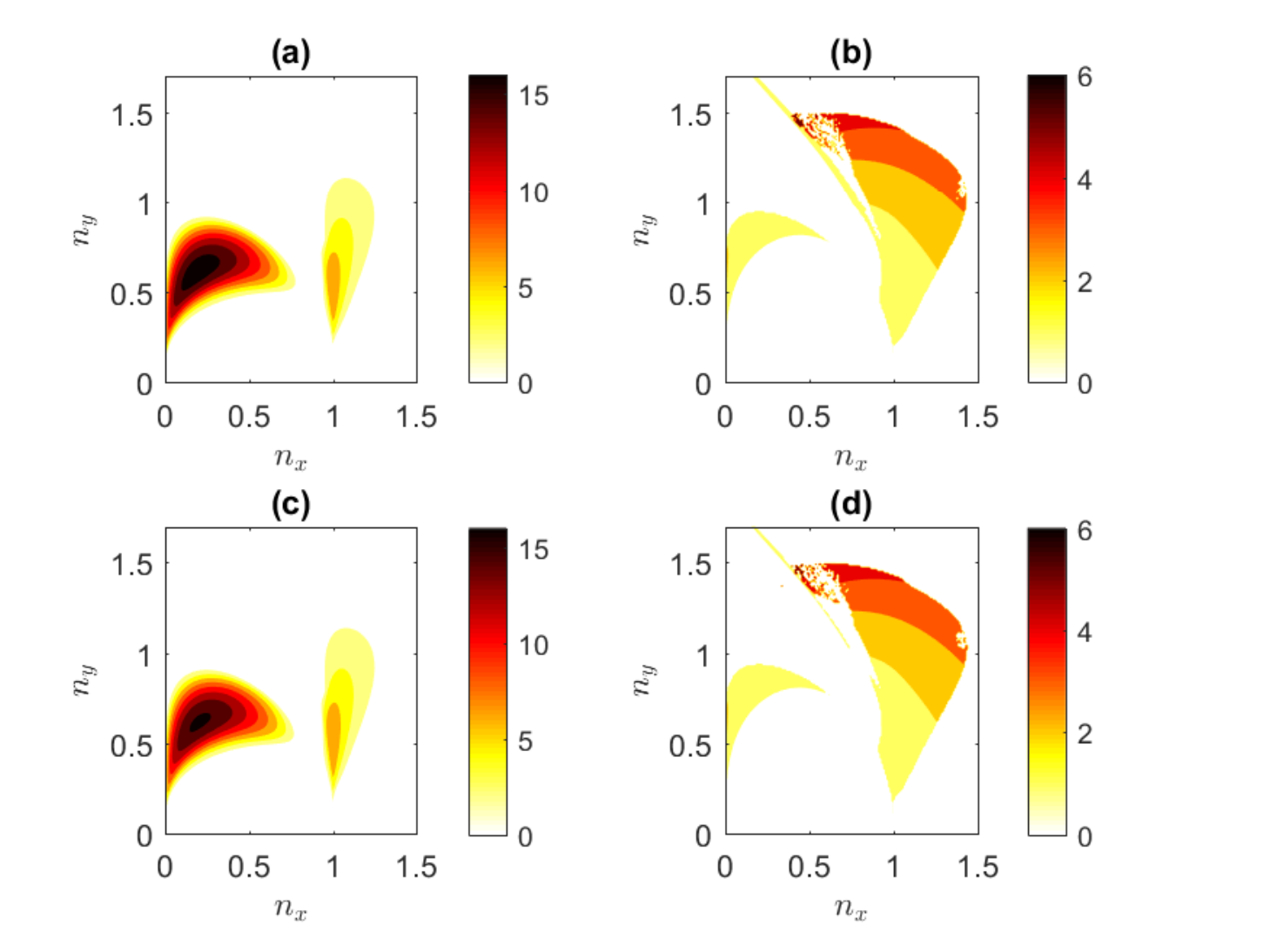}}
\caption{The same as figure~\ref{fig:disp1} but for $\varepsilon=100\varepsilon_c$.}\label{fig:disp2}
\end{figure}

When $\la<0.68$, baroclinic modes become S3T unstable and have for small values of $\lambda$
comparable  growth rate to the barotropic modes (or nearly equal in the case $\lambda\ll 1$). The growth rate and
frequency of the most unstable barotropic and baroclinic modes are shown in figures~\ref{fig:disp1}-\ref{fig:disp2}
for $\lambda=0.1$, that is when the Rossby radius of deformation is  ten times larger than the scale of excitation.
The barotropic and baroclinic modes have similar growth rates as a function of~$\nv$. Their dispersion properties
resemble the large-scale branch of the unstable barotropic modes at larger values of $\lambda$, with the frequencies
of the large-scale modes following the dispersion of the barotropic $\sigma_{Rt}$, and baroclinic $\sigma_{Rc}=\beta n_x/n_\la^2$
Rossby waves as shown in figure~\ref{fig:disp1}c,d. At larger supercriticalities, stationary barotropic and
baroclinic zonal jet eigenfunctions are unstable but the non-zonal modes have larger growth rates (cf. figure~\ref{fig:disp2}).

\begin{figure}
\centerline{\includegraphics[width=.75\textwidth]{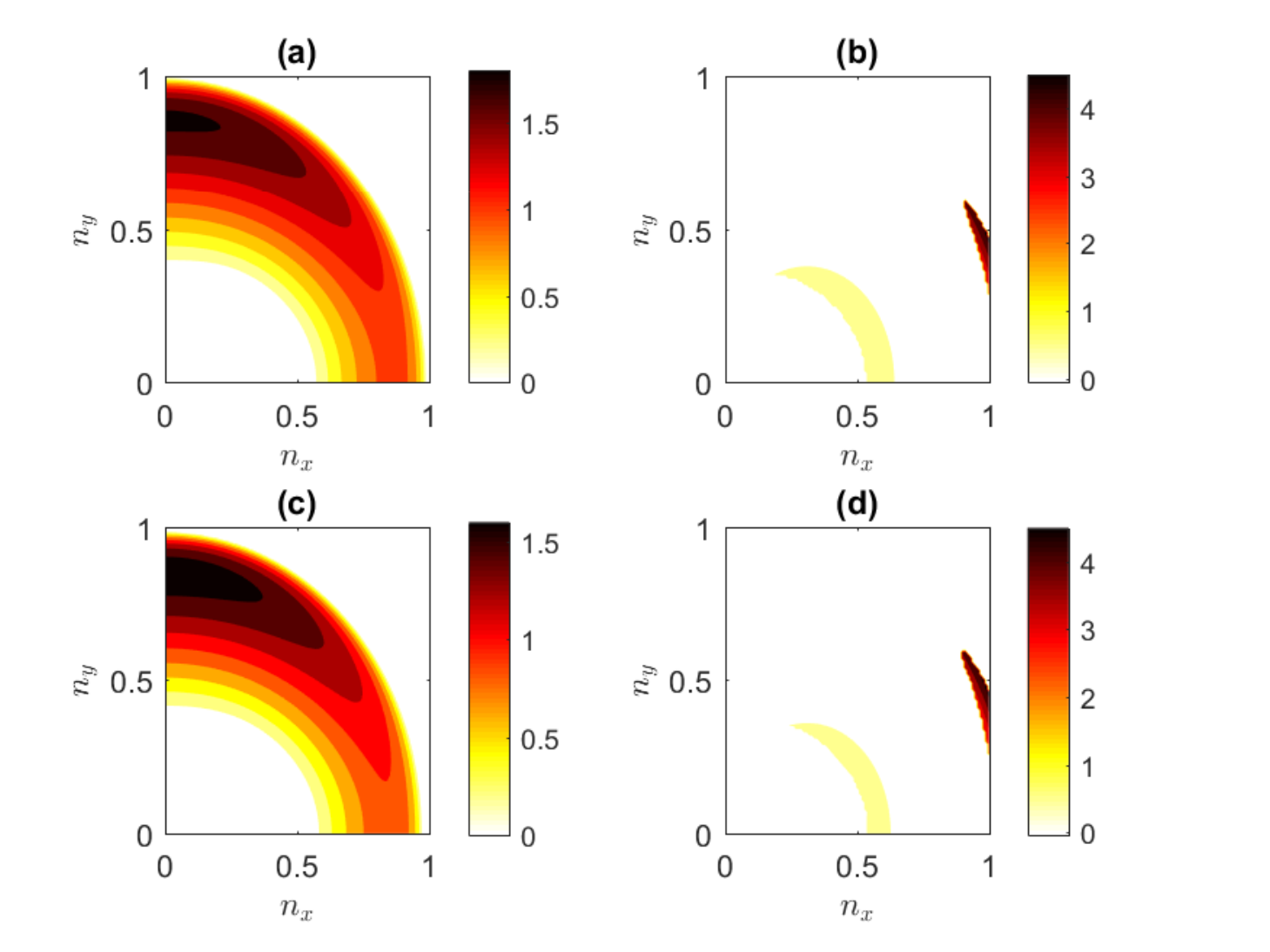}}
\caption{Dispersion properties of the unstable modes for $\beta=1$, $\lambda=0.1$ and $\varepsilon=10\varepsilon_c$.
(a) and (c) Growth rate $\sigma_r$ of the most unstable (a) barotropic and (c) baroclinic modes. (b) and (d)
Frequency $\sigma_i$ of the most unstable (b) barotropic modes and (d) baroclinic modes. The forcing in the two layers is uncorrelated ($p=0$).}\label{fig:disp4}
\end{figure}

We now consider  $\beta=1$. Baroclinic modes are unstable for $\la<0.37$ instead of $\la <0.68$  at $\beta=100$ and
the baroclinic unstable modes have lower growth rates than  the barotropic modes. This is illustrated in figure \ref{fig:disp4} where
the growth rate and frequency
of the unstable barotropic and baroclinic modes are shown for $\lambda=0.1$. As in  a one layer fluid, the most
unstable modes are stationary zonal jets for $\beta=\mathcal{O}(1)$. For $\la>0.37$ the dispersion properties of the barotropic
unstable modes are similar and are not shown.

We finally test the sensitivity of the dispersion properties of the unstable modes to the forcing correlation between the
two layers. Consider first the barotropic forcing ($p=1$). In this case the flux
feedbacks arise only from the first terms in (\ref{eq:fpsi})-(\ref{eq:ftheta}), that is only from the organization of the
barotropic turbulent eddies by the mean flow. For $\lambda~= \mathcal{O}(1)$ the baroclinic
modes are stable and the dispersion properties of the barotropic modes are similar to the ones shown in figure \ref{fig:disp3}
with the only significant difference being the absence of the upper branch. For lower values of $\lambda$, the baroclinic
modes become unstable and have dispersion properties similar to the ones shown in
figure~\ref{fig:disp1}-\ref{fig:disp2} for uncorelated forcing despite the fact that small scale incoherent baroclinic eddies
are not directly excited in this case ($S_E^{\theta\theta}=0$). Therefore the instability of the baroclinic modes when
excitation occurs at much smaller scales than the Rossby radius of deformation is robust regardless of whether incoherent baroclinic
eddies are supported at the homogeneous equilibrium.

\begin{figure}
\centerline{\includegraphics[width=.75\textwidth]{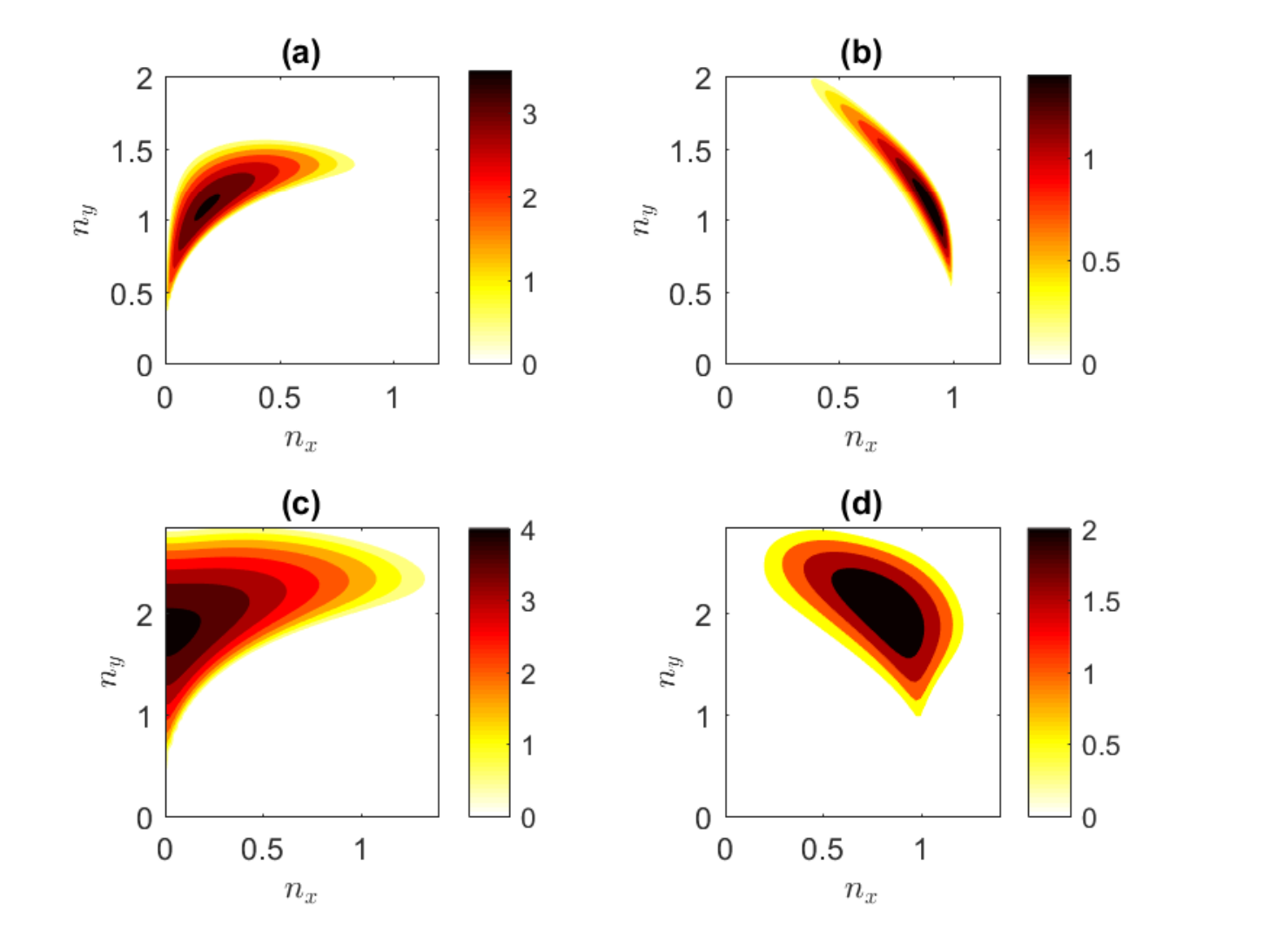}}
\caption{Dispersion properties of the unstable modes for baroclinic forcing ($p=-1$) at
$\beta=100$ and $\varepsilon=10\varepsilon_c$. (a) and (b) Growth rate $\sigma_r$ of the most unstable (a) barotropic and (b)
baroclinic modes for $\lambda=1$. (c) and (d) Growth rate $\sigma_r$ of the most unstable (c) barotropic and (d)
baroclinic modes when $\lambda=2$.}\label{fig:disp5}
\end{figure}

We then consider the baroclinic forcing ($p=-1$), in which case the flux
feedbacks arise only from the organization of the baroclinic turbulent eddies by the mean flow (i.e the second terms in
(\ref{eq:fpsi})-(\ref{eq:ftheta})). For low values of $\lambda$, the dispersion properties
of both the barotropic and the baroclinic modes are similar to the ones obtained for uncorrelated forcing ($p=0$) (shown in
figure~\ref{fig:disp1}) and are not shown. The dispersion properties at $\lambda=1$ are shown in figure \ref{fig:disp5}(a)-(b).
We observe that barotropic modes have faster growth rates
and resemble the upper branch of figure \ref{fig:disp3}, that is they have small zonal scales and meridional scales comparable
to the forcing scale ($n_y\simeq 1$). The most unstable baroclinic modes, that are suboptimal compared to their barotropic counterparts,
have both zonal and meridional scales comparable to the forcing scale. Interestingly, when we force at scales larger than the
Rossby radius of deformation ($\lambda>1$), the instability characteristics change. As illustrated in figure \ref{fig:disp5}(c)-(d) showing
the dispersion properties at $\lambda=2$, the most unstable modes are barotropic zonal jets with scales comparable to
the deformation scale, a characteristic that is also evident at even larger values of $\lambda$ (not shown).

In summary, S3T predicts that the homogeneous statistical equilibrium of a two-layer baroclinic flow becomes unstable
at a critical value of the energy input rate, $\varepsilon_c$, whereupon
large-scale mean flows emerge. Barotropic modes have larger growth rates compared to their baroclinic counterparts for all values
of $\beta$ and
$\lambda$ regardless of the characteristics of the eddies supported at statistical equilibrium. When energy is injected at
a length scale close to the deformation radius, barotropic modes following the Rossby wave dispersion have in general the largest growth
rate for large values of $\beta$ and stationary zonal jets
have the largest growth rate for $\beta=\mathcal{O}(1)$. When energy is injected  at scales much smaller than the deformation
radius, both barotropic and baroclinic large-scale flows
emerge with the barotropic modes having slightly larger growth rates.

\section{Equilibration of the SSD instabilities and comparison to NL simulations}

In this section we examine the equilibration of the SSD instabilities by numerical integrations
of the S3T dynamical system. The goal is twofold. The first is to determine the non-homogeneous
fixed points of the S3T dynamical system with the largest domain of attraction that a random
perturbation will likely end up to. For example, we saw in the previous section that when
the non-dimensional radius of deformation is smaller than a critical value, barotropic and
baroclinic modes have comparable growth rates. Therefore there is the question of whether
the equilibrated state will consist of both barotropic and baroclinic components. The second
is to compare the dominant structures in the turbulent flow with the equilibrated states in
the S3T system and to investigate whether the characteristics of the dominant structures
in the NL simulations are predicted by S3T.

To integrate the S3T dynamical system, we discretize (\ref{eq:first_cumZ})-(\ref{eq:dCdt})
in a doubly periodic channel of size $2\upi \times 2 \upi$ using a pseudo-spectral code and a fourth order time-stepping
Runge-Kutta scheme. Due to the high-dimensionality of the covariance matrix
we use the modest $M=32\times 32$ resolution. To facilitate the comparison with the NL simulations, we
apply the same ring forcing~(\ref{eq:ringb}) and consider the same parameters $\tilde{\beta}=60$, $\tilde{r}=0.1$,
$\tilde{k}_f=6$ and $\Del \tilde{k}_f=1$. Because the S3T simulations are computationally
expensive, we also utilize for comparison purposes simulations of the ensemble quasi-linear
system (EQL) defined in (\ref{eq:EQL1}), (\ref{eq:EQL2}), (\ref{eq:EQL3}). The EQL system is
the finite ensemble version of the deterministic S3T dynamical system and the EQL simulations
converge to the S3T integrations as the number of ensemble members $N_{\textrm{ens}}$ used to
calculate the mean increases. The emergent structures and the statistical turbulent equilibria
predicted by the S3T system manifest in the EQL integrations with the addition of noise from stochastic
forcing fluctuations,  which for any finite $N_{\textrm{ens}}$  do not  average to zero. A disadvantage of
this method is that due to thermal noise from the excitation, only the equilibria with the largest
domain of attraction will emerge. Previous studies have investigated the convergence of the
EQL simulations to the S3T integrations and determined that $N_{ens}=10$ ensemble members
illustrate the relevant dynamics with minimal computational cost \citep{Bakas-Ioannou-2014-jfm,
Constantinou-etal-2016}). We therefore make the same choice.

\begin{figure}
\centerline{\includegraphics[width=.75\textwidth]{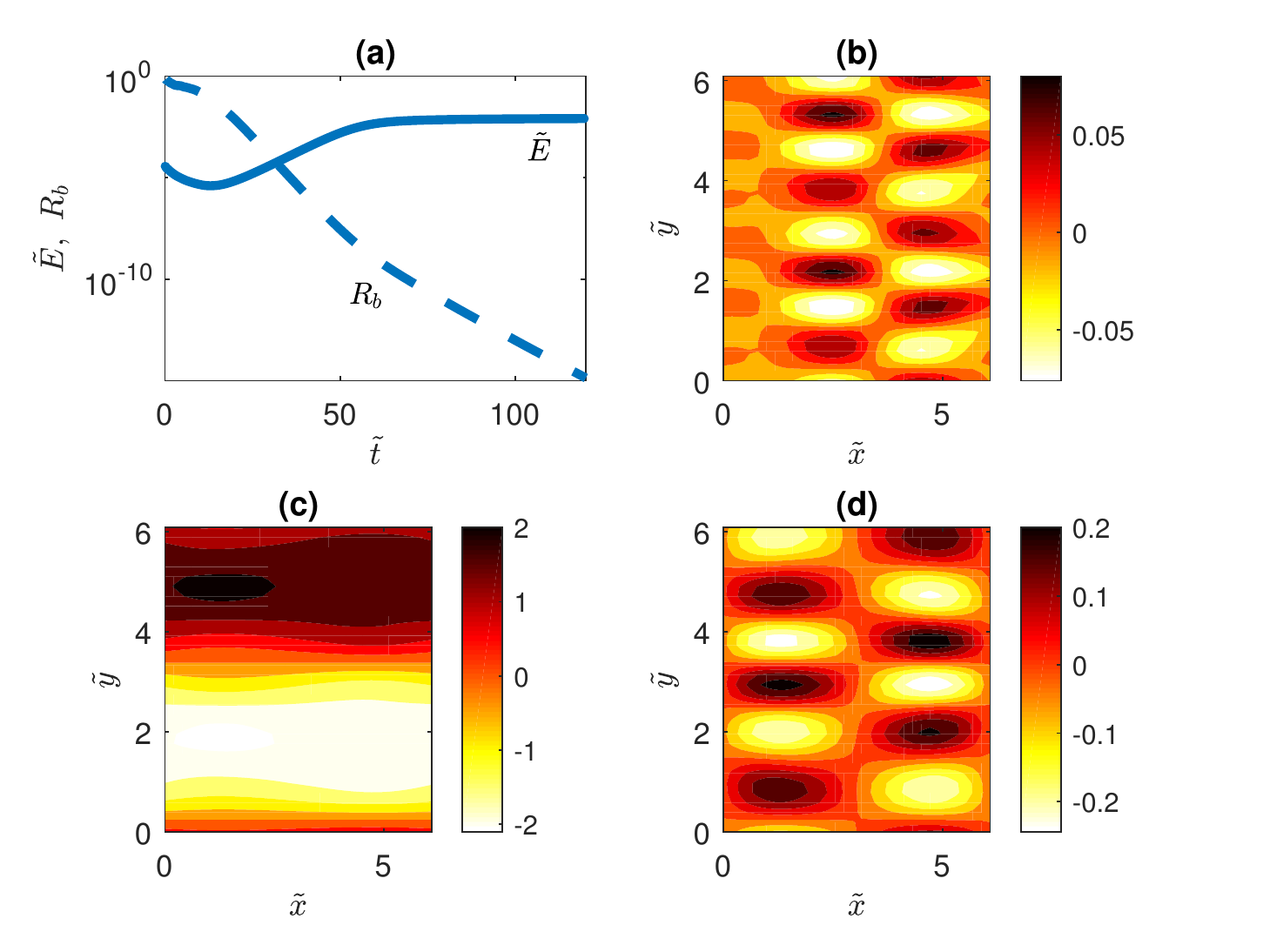}}
\caption{(a) Evolution of the mean flow energy $\tilde{E}$ (solid line) and baroclinicity~$R_b$ defined in~(\ref{eq:brcl_ratio}) (dashed
line) and (b)~contours of mean flow barotropic streamfunction $\Psi$ at equilibrium for
$\tilde{\varepsilon}=2\tilde{\varepsilon}_c$. (c)~Contours of mean flow barotropic streamfunction~$\Psi$ at equilibrium
and (d) non-zonal part of the equilibrium streamfunction when~$\tilde{\varepsilon}=60\tilde{\varepsilon}_c$. The
parameters values are the same as in figure~\ref{fig:NL_L6_bif}.}
\label{fig:PL_L6}
\end{figure}

Consider first the case of uncorrelated forcing between the layers ($p=0$) and $\tilde{\lambda}=\tilde{k}_f$.
%for which the stability analysis predicts the
%emergence of barotropic
%large scale  structures for energy input rates larger than the threshold $\varepsilon_c$.
The critical
energy input rate $\tilde{\varepsilon}_c$  that renders the homogeneous S3T equilibrium unstable is obtained from the discrete
version of equations (\ref{eq:stab_eq1})-(\ref{eq:stab_eq2}). We investigate the equilibration of the SSD instability for
supercritical values of $\tilde{\varepsilon}$ through S3T integrations. For
$\tilde{\varepsilon}=2\tilde{\varepsilon}_c$, the barotropic perturbation with $(\tilde{n}_x, \tilde{n}_y)=(1, 4)$ has the
largest growth rate, while the baroclinic perturbations are S3T stable at this value of $\tilde{\la}$ for all values of
$\tilde{\varepsilon}$. At $t=0$, we introduce
a small random perturbation, with both barotropic and baroclinic mean flow components and
study the evolution of the energy of the  baroclinic and barotropic components through the baroclinicity measure:
\begin{equation}
R_b=\frac{\sum_{\tilde{k}_x,\tilde{k}_y} (\tilde{k}^2+2\tilde{\la}^2)|\hat\Theta(\tilde{\kv})|^2}
{\sum_{\tilde{k}_x,\tilde{k}_y} \tilde{k}^2|\hat\Psi(\tilde{\kv})|^2},\label{eq:brcl_ratio}
\end{equation}
in which $\hat\Psi$ and $\hat\Theta$ are the Fourier components of the barotropic and baroclinic mean flow streamfunction
respectively. As shown in figure~\ref{fig:PL_L6}a,  the baroclinicity, $R_b$, decays exponentially
and the baroclinic part of the flow diminishes as predicted by the  S3T  stability analysis, while
the energy of the barotropic flow, also shown in figure~\ref{fig:PL_L6}a, grows exponentially with the predicted
growth rate, and  the flow finally equilibrates to a  westward propagating wave. The barotropic streamfunction at
equilibrium, shown in figure~\ref{fig:PL_L6}b,
has Fourier components primarily with
wavenumbers $(|\tilde{k}_x|, |\tilde{k}_y|)=(1, 4)$ and secondarily with wavenumbers
$(|\tilde{k}_x|, |\tilde{k}_y|)=(2, 4)$. Investigation of the
propagation properties of the wave shows that the phase speed of each wave component is very close
to the phase speed of the unstable modes that themselves are nearly equal to the Rossby wave
phase speed for each component.

The  finite amplitude state of the SSD consisting of the propagating waves becomes itself secondarily unstable at
larger energy input rates to zonal jet perturbations. This secondary SSD instability has already been discussed in the
context of a single layer  barotropic flow on a $\beta$-plane by \citet{Bakas-Ioannou-2014-jfm} and for the chosen
parameters it occurs at $\tilde{\varepsilon}_{nl}=15\tilde{\varepsilon}_c$. The flow then transitions to a new equilibrium
state such as the one shown in figure~\ref{fig:PL_L6}c,d  for the case $\tilde{\varepsilon}=60\tilde{\varepsilon}_c$.
This state consists of a zonal jet (figure~\ref{fig:PL_L6}c) coexisting with a weak propagating
wave (figure~\ref{fig:PL_L6}d), as in single layer barotropic  flows \citep{Bakas-Ioannou-2014-jfm,Constantinou-etal-2016}.

\begin{figure}
\centerline{\includegraphics[width=.75\textwidth]{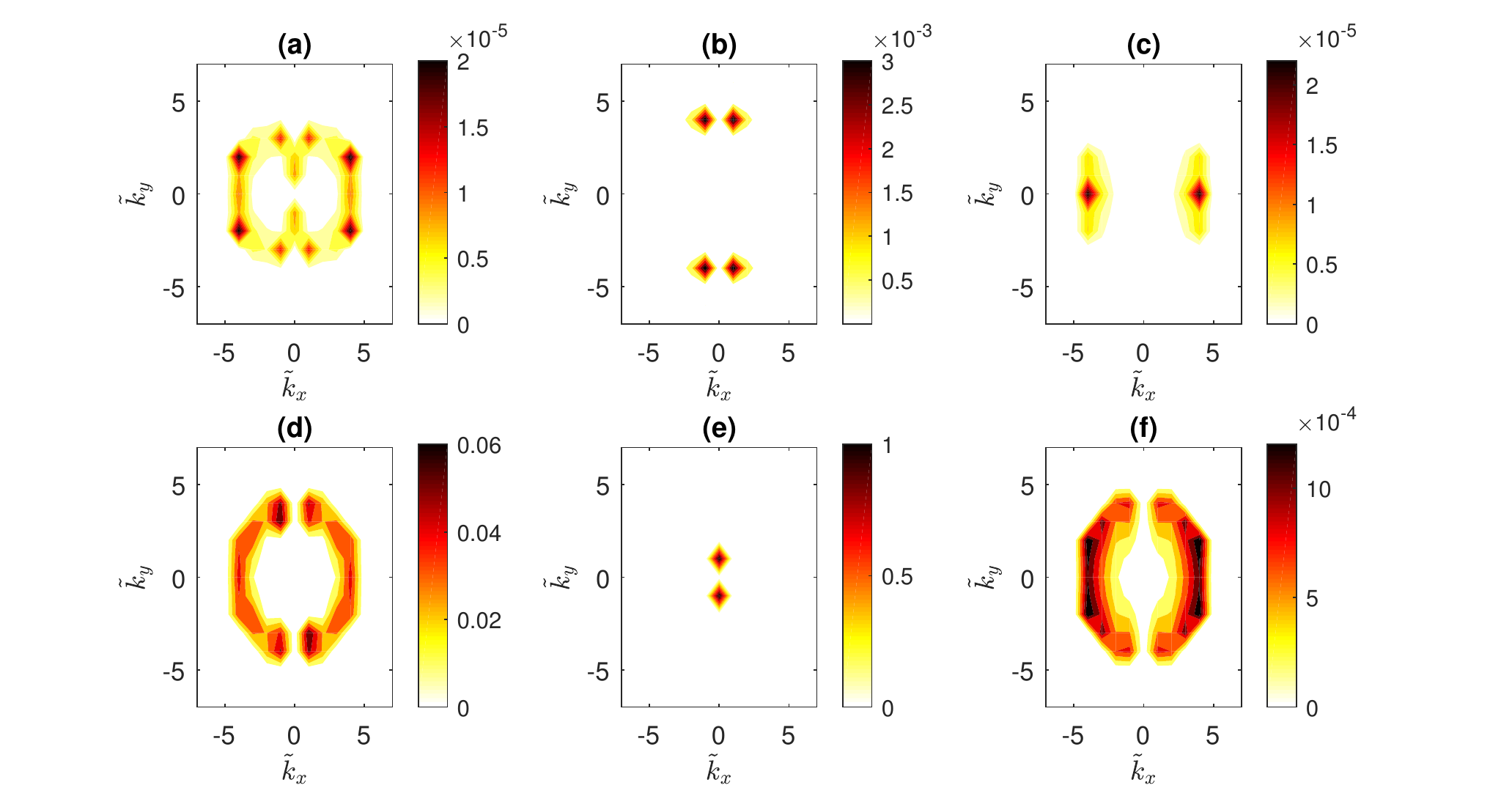}}
\caption{(a) Kinetic energy power spectrum of the barotropic part of the eddy covariance $k^2\hat{S}^{\psi\psi}$, (b)
total kinetic energy power spectrum $\hat{E}_{S3T}^\psi$ of the barotropic part of the flow and (c)
total kinetic energy power spectrum $\hat{E}_{S3T}^\theta$ of the baroclinic part of the flow for
$\tilde{\varepsilon}=2\tilde{\varepsilon}_c$. (d)-(f) The same as in (a)-(c) but for
$\tilde{\varepsilon}=60\tilde{\varepsilon}_c$.}\label{fig:PL_L6_spectra}
\end{figure}

We now investigate how these results relate to the regime transitions and to the characteristics of the dominant
structures in the turbulent flow. The rapid increase of the nzmf index in the NL
simulations when the energy input rate passes the critical value $\tilde{\varepsilon}_c$ calculated from the
stability analysis of the homogeneous fixed point of the SSD shows that
the bifurcation point for the emergence of large scale waves in the turbulent flow is accurately predicted by S3T.
The same holds for the increase of the zmf index observed in figure~\ref{fig:NL_L6_bif} for
$\tilde{\varepsilon}\gtrapprox 12\tilde{\varepsilon}_c$, which shows that zonal jets emerge in the flow approximately
at the stability threshold $\tilde{\varepsilon}_{nl}$ of the finite amplitude traveling wave states. The small
quantitative discrepancy observed should be attributed to both the thermal noise in the NL simulations and in the
quasi-linear approximation of the dynamics. The scales and phase speeds of the dominant large scale structures
in the NL simulations also match the scales and phase speeds of the equilibrated SSD instabilities
in S3T. This is illustrated by comparing the distinct peaks in the kinetic energy power spectra in
the NL simulations and the frequency power spectrum of the wave components shown in figures~\ref{fig:NL_L6_spectra}-\ref{fig:NL_L6_spectra2}
to the scales of the emergent structures in the S3T simulations shown in figure~\ref{fig:PL_L6} and their phase speed.
In order to compare the amplitude of the emergent structures we calculate the equivalent kinetic energy power spectra
\begin{eqnarray}
\hat{E}_{S3T}^\psi({\tilde {\bf k}})&=&\tilde{k}^2\left(|\hat{\Psi}|^2+\hat{S}^{\psi\psi}\right),\\
\hat{E}_{S3T}^\theta({\tilde {\bf k}})&=&\tilde{k}^2\left(|\hat{\Theta}|^2+\hat{S}^{\theta\theta}\right),
\end{eqnarray}
where $\hat{S}^{\psi\psi}$ and $\hat{S}^{\theta\theta}$ are the power spectra of the eddy covariances $S^{\psi\psi}$ and
$S^{\theta\theta}$ respectively. For the barotropic part, the eddy power spectrum $\hat{S}^{\psi\psi}$ shown
in figure \ref{fig:PL_L6_spectra}a and d is two orders of magnitude smaller than $|\hat{\Psi}|^2$, so
the equivalent power spectrum $\hat{E}_{S3T}^\psi$ shown in figure \ref{fig:PL_L6_spectra}b and e is dominated by the
spectrum of the coherent part of the flow. This should come as no surprise, since the dominant structures in
the turbulent flow are phase coherent and should therefore be a manifestation of the coherent part of the
flow $\Psi$. For the baroclinic part, $\Theta=0$, so
$\hat{E}_{S3T}^\theta$ that is shown in figure \ref{fig:PL_L6_spectra}c and e coincides with the eddy power
spectrum  $\hat{S}^{\theta\theta}$. Comparison of the spectra obtained from the NL simulations (figures
\ref{fig:NL_L6_spectra} and \ref{fig:NL_L6_spectra2}) and the equivalent spectrum shown in figure
\ref{fig:PL_L6_spectra} shows that the amplitude of the emergent structures differs by a factor
of two for $\tilde{\varepsilon}=2\tilde{\varepsilon}_c$ and by about 20\% for
$\tilde{\varepsilon}=60\tilde{\varepsilon}_c$. To facilitate comparison of the amplitude of the emerging structures
for a wide range of values for the energy input rate, we utilize the EQL simulations and calculate the zmf and nzmf indices
that are shown in figure \ref{fig:NL_L6_bif}. We observe that apart from the quantitative disagreement
at low supercritical energy input rates for the amplitude of the emergent waves, the EQL dynamics
fairly reproduce the amplitude of the emergent structures. Similar results regarding the comparison
of the S3T prediction with NL simulations were found for barotropic turbulence in \citet{Bakas-Ioannou-2014-jfm},
showing that the quasi-linear dynamics close to the bifurcation point capture the emergence of flows in contrast
to the physical intuition on the importance of the non-linear terms as also discussed in the introduction.

%Note that in these cases, zonal jets emerge only after the inception of the S3T secondary instability of the equilibrated  non-zonal large-scale structure. For this reason, predictions of the jet-forming bifurcation from the stability analysis of the
%homogeneous equilbrium in S3T formulations with the mean taken as the zonal mean do not to correspond to the jet formation bifurcation in fully nonlinear simulations \citep{Srinivasan-Young-2012} and modification of the forcing
%spectrum by the emerged large-scale waves must be included \citep{Constantinou-etal-2014}.

\begin{figure}
\centerline{\includegraphics[width=.75\textwidth]{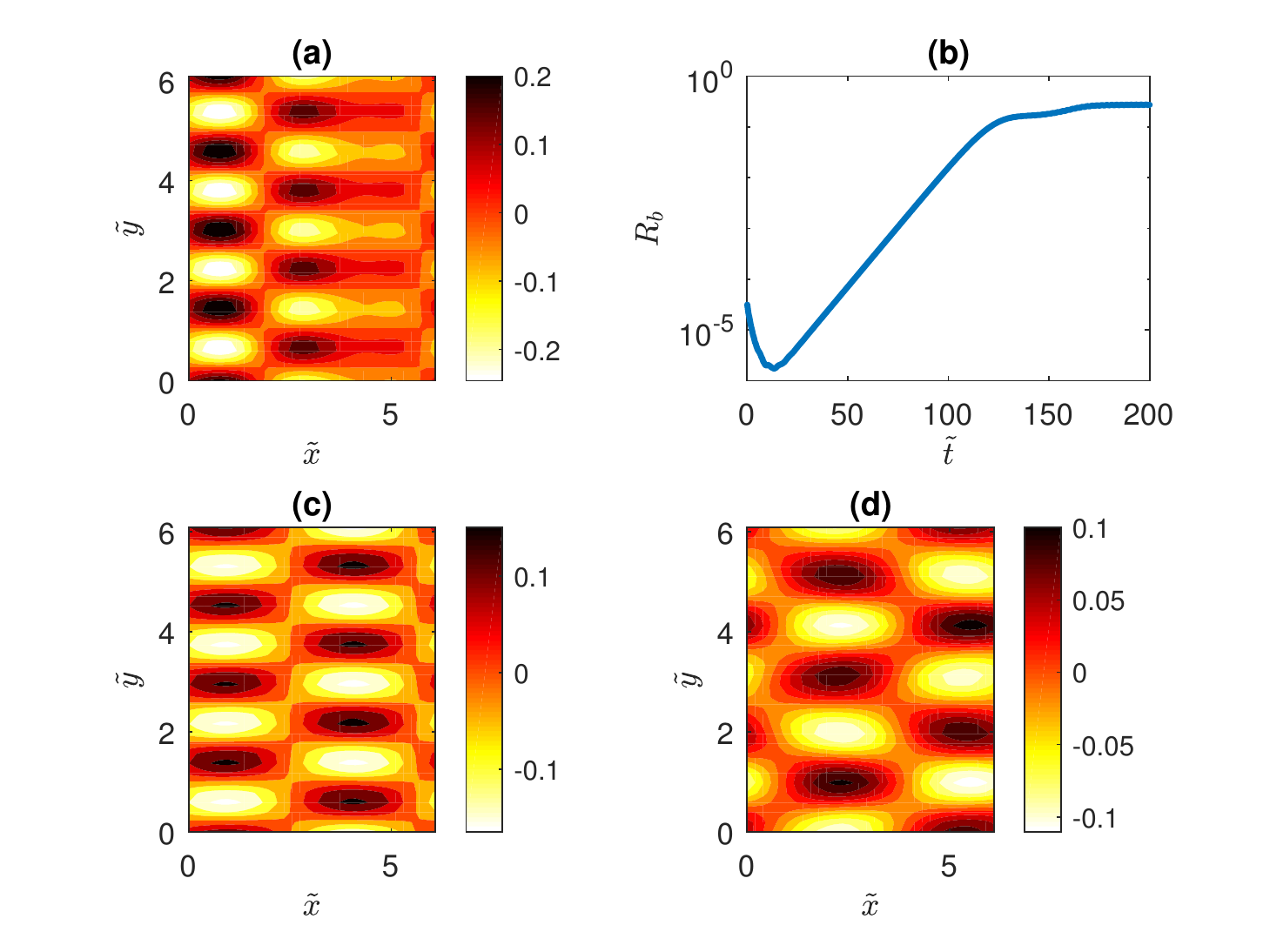}}
\caption{(a) Contours of mean flow barotropic streamfunction $\Psi$ for the pure barotropic equilibrated
state. (b) Evolution of the baroclinicity ratio after inserting a baroclinic mean flow perturbation to
the barotropic equilibrium shown in (a). (c)-(d) Contours of mean flow (c) barotropic $\Psi$ and (d) baroclinic
$\Theta$ streamfunction for the mixed barotropic-baroclinic equilibrium state. The energy input rate is
$\tilde{\varepsilon}=10\tilde{\varepsilon}_c$, $\tilde{\lambda}=\tilde{k}_f/6$ and the rest of the parameters are as
in figure~\ref{fig:PL_L6}.}\label{fig:PL_L1_fig1}
\end{figure}

Consider now $\tilde{\lambda}=\tilde{k}_f/6$ ($\lambda=1/6$) for which the S3T stability analysis predicts the existence
of both barotropic and baroclinic unstable S3T modes. We have identified in this case three attracting equilibrium states with
finite amplitude mean flows. The first is a purely barotropic equilibrium, to which all  purely barotropic perturbations are attracted.
Note that purely barotropic perturbations do not produce any baroclinic fluxes even in their nonlinear stage of evolution and as a result,
the mean flow remains barotropic. An example of such an equilibrated state is shown for $\tilde{\varepsilon}=10\tilde{\varepsilon}_c$ in
figure~\ref{fig:PL_L1_fig1}a, where again the large-scale  flow is a westward propagating wave. However this equilibrium is secondarily
unstable to baroclinic mean flow perturbations. We illustrate this by perturbing the barotropic equilibrium, shown in figure~\ref{fig:PL_L1_fig1}a,
with  a small baroclinic mean flow perturbation. Figure~\ref{fig:PL_L1_fig1}b shows that the baroclinic mean flow component grows
exponentially at first and as soon as it reaches a finite amplitude, the flow transitions to the second non-zonal equilibrium with both
barotropic and baroclinic components and baroclinicity  $R_b\simeq 1/4$. The structure of these components is
shown in figure~\ref{fig:PL_L1_fig1}c,d.

The third equilibrium is also a  westward propagating
wave with both barotropic and baroclinic components and similar scale as the one shown in figure~\ref{fig:PL_L1_fig1}c,d
but is more baroclinic  ($R_b\simeq 2$). However,  this equilibrium  has a very small domain of
attraction and it can be approached only if the initial perturbation is baroclinic\footnote{Specifically, this equilibrium is
approached only if the initial perturbation has a baroclinicity ratio larger than $10^6$.}. As a result, we do not
expect this equilibrium to manifest in noisy simulations of the turbulent flow in which both baroclinic and barotropic perturbations
inevitably arise. Therefore, at low supercriticalities the S3T instabilities equilibrate
to  the finite amplitude mixed barotropic-baroclinic traveling wave shown
in figure~\ref{fig:PL_L1_fig1}c,d with baroclinic to barotropic
streamfunction amplitude ratio of about one half.

\begin{figure}
\centerline{\includegraphics[width=.75\textwidth]{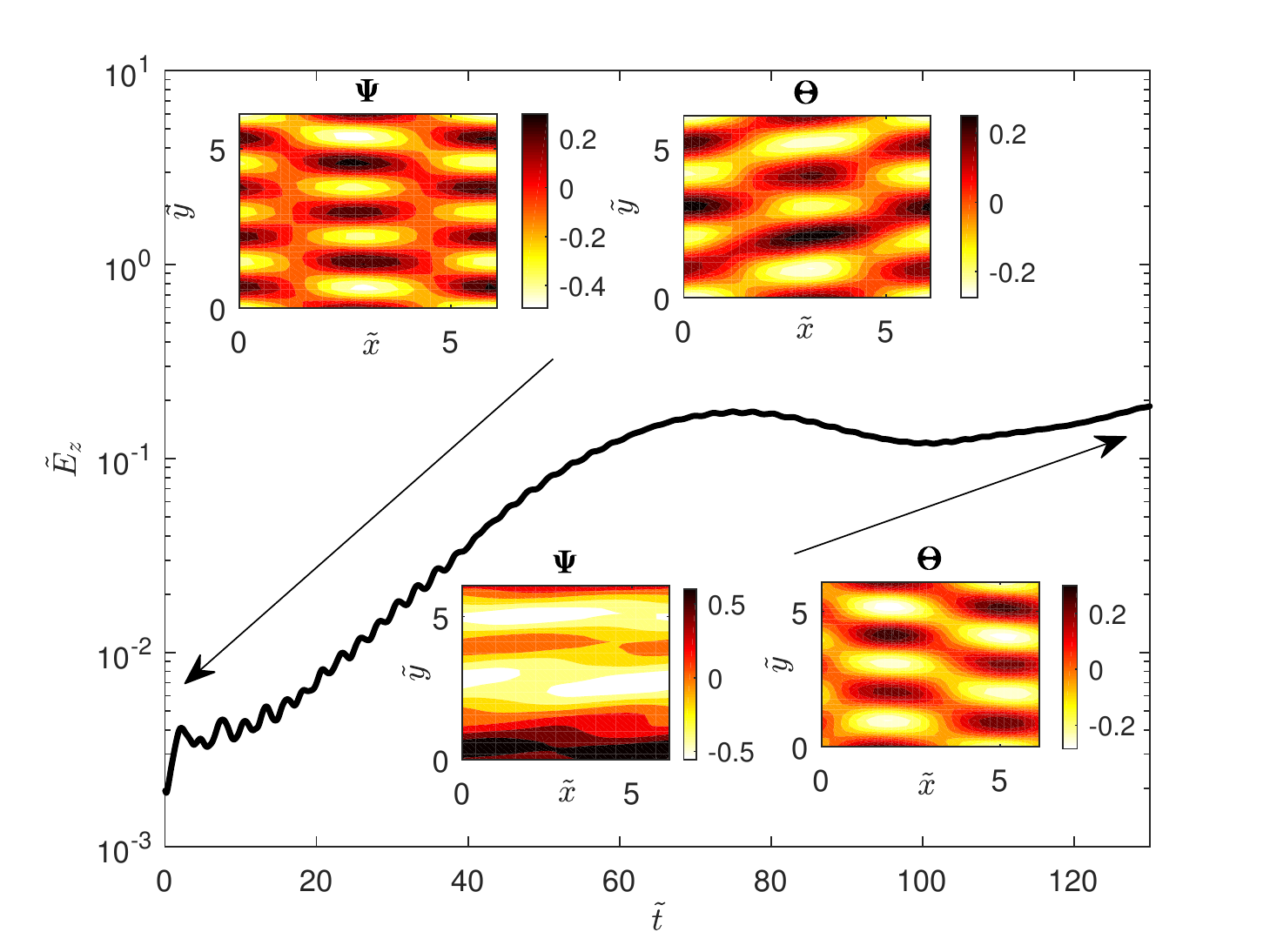}}
\caption{Evolution of the energy of the zonal part of the flow when a zonal mean flow perturbation is imposed on the
finite amplitude equilibrium state shown in the upper insets. The lower insets show the barotropic and baroclinic
streamfunction of the mean flow at $\tilde{t}=130$. The energy input rate is $\tilde{\varepsilon}=40\tilde{\varepsilon}_c$, and the rest of the
parameters are as in figure~\ref{fig:PL_L1_fig1}.}\label{fig:PL_L1_fig2}
\end{figure}

\begin{figure}
\centerline{\includegraphics[width=.75\textwidth]{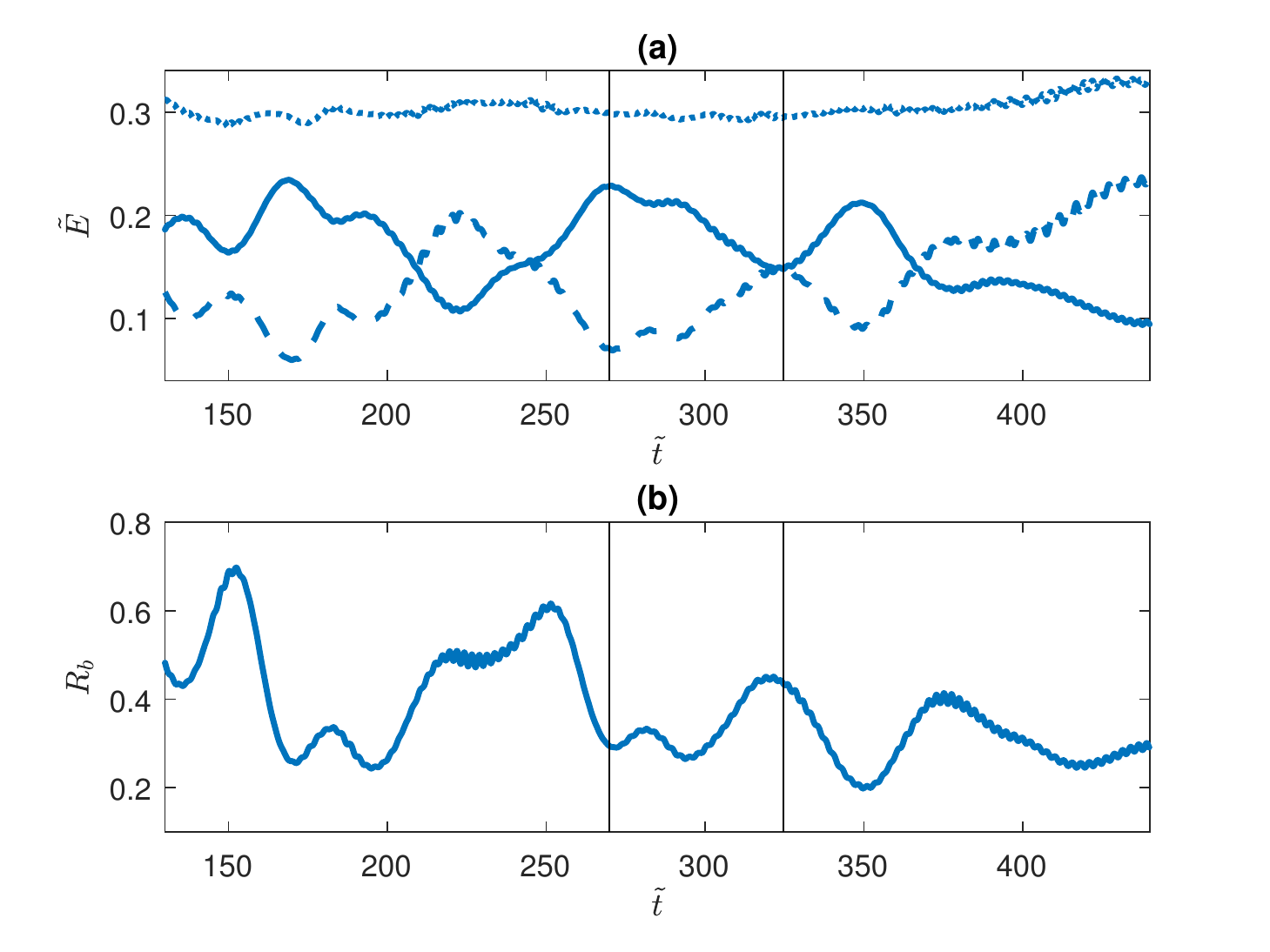}}
\caption{(a) Evolution of the mean flow energy (dotted), the energy of the zonal part (solid) and the energy
of the non-zonal part (dashed) of the flow. (b) Evolution of the baroclinicity $R_b$. The vertical lines
denote the times at which the snapshots shown in figure~\ref{fig:PL_L1_fig4} are taken.}\label{fig:PL_L1_fig3}
\end{figure}

\begin{figure}
\centerline{\includegraphics[width=.75\textwidth]{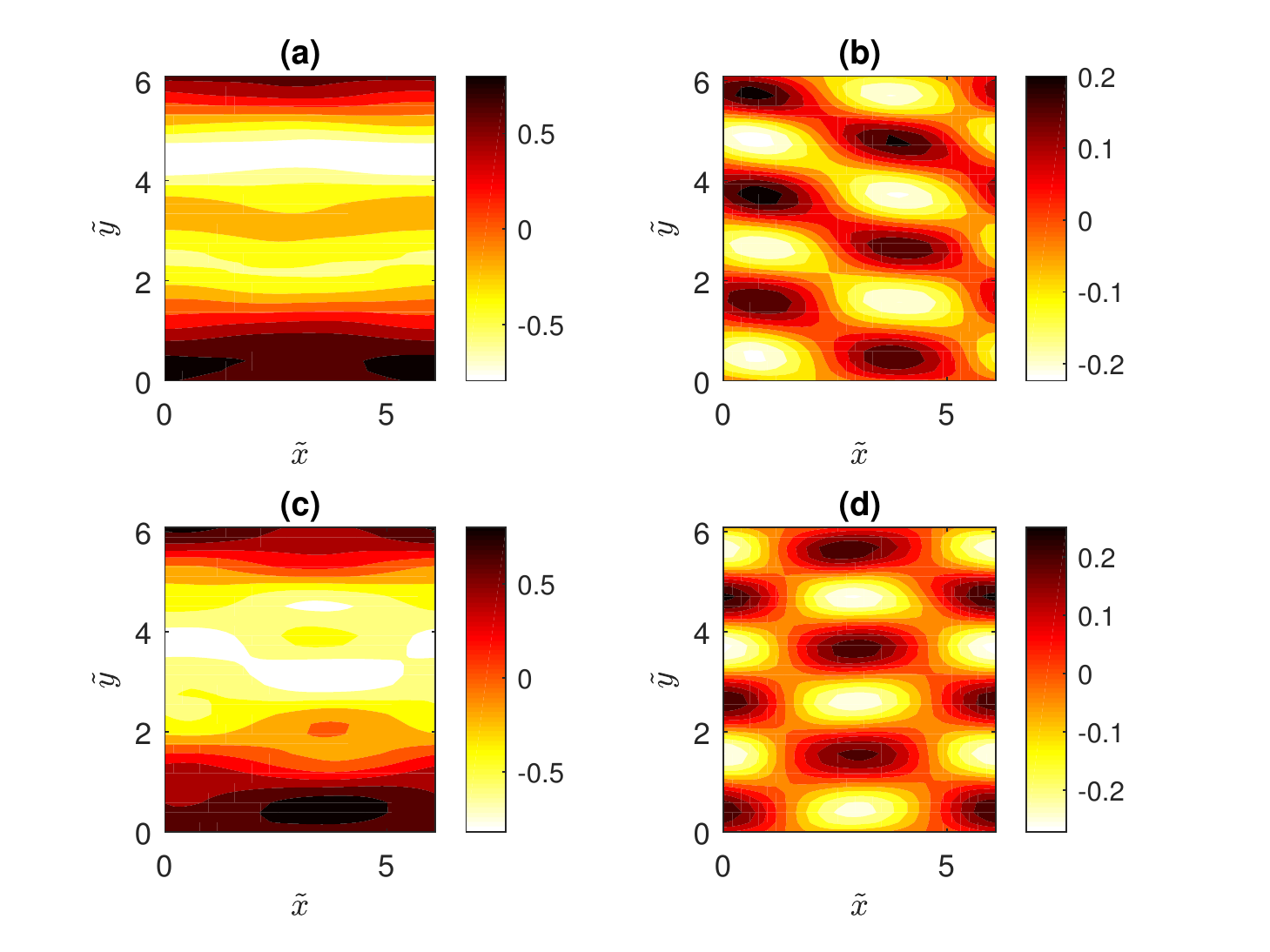}}
\caption{(a)-(b) Snapshot of the mean flow (a) barotropic $\Psi$ and (b) baroclinic $\Theta$ streamfunction
at $\tilde{t}=270$. (c)-(d) The same as in (a)-(b) but for a snapshot at $\tilde{t}=325$.}\label{fig:PL_L1_fig4}
\end{figure}

Again, at higher supercriticality the traveling baroclinic wave states of the SSD
become secondarily unstable to zonal perturbations. However, only the barotropic part of the flow turns into zonal jets. This is shown in the numerical
integration for $\tilde{\varepsilon}=40\tilde{\varepsilon}_c$ illustrated in figure~\ref{fig:PL_L1_fig2}. In this experiment the mixed barotropic-baroclinic traveling wave
state, shown in the upper inset of figure~\ref{fig:PL_L1_fig2}, is perturbed by a zonal mean flow perturbation with both
barotropic and baroclinic components. The energy
of the zonal part of the flow, $\tilde{E}_z$, grows exponentially and the flow transitions to the state  shown at the lower
inset that consists of a zonal barotropic flow and a traveling baroclinic wave. This is a time dependent state
as shown in figure~\ref{fig:PL_L1_fig3}. The energy of the zonal part of the flow and the baroclinicity $R_b$
oscillate and are anti-correlated revealing that the large scale flow vacillates between the state shown in
figure~\ref{fig:PL_L1_fig4}a,b where the barotropic zonal jet is dominant and the state shown in  figure~\ref{fig:PL_L1_fig4}c,d
where the baroclinic wave is dominant. At even higher energy input rates the large scale flow exhibits the
same time dependence but with weaker fluctuations due to the weakening of the baroclinic wave component.

\begin{figure}
\centerline{\includegraphics[width=.75\textwidth]{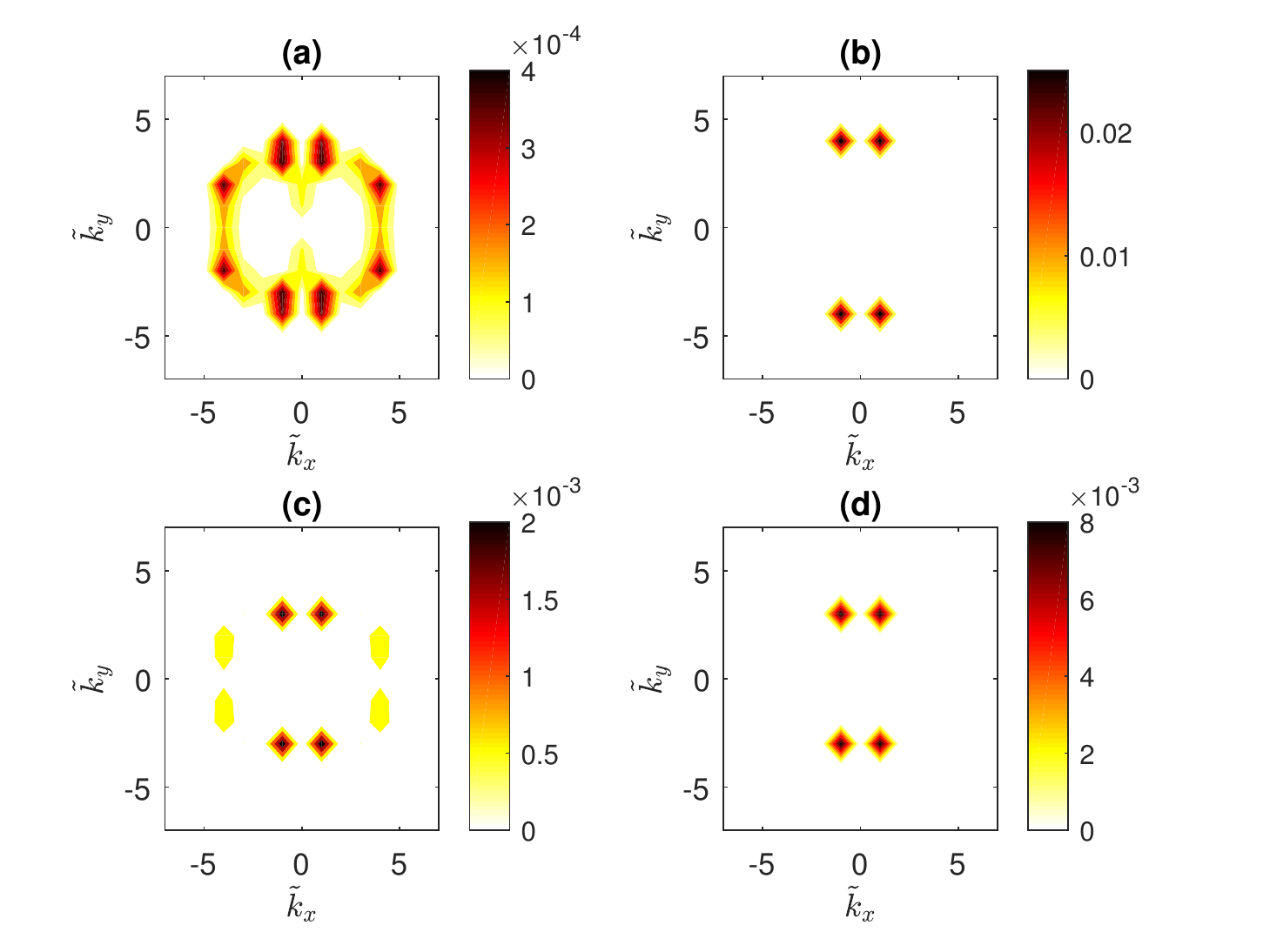}}
\caption{(a) Kinetic energy power spectrum of the barotropic part of the eddy covariance $k^2\hat{S}^{\psi\psi}$. (b)
Equivalent kinetic energy power spectrum $\hat{E}_{S3T}^\psi$ of the barotropic part of the flow. (c)
Kinetic energy power spectrum of the baroclinic part of the eddy covariance $k^2\hat{S}^{\theta\theta}$ and (d)
equivalent kinetic energy power spectrum $\hat{E}_{S3T}^\theta$ of the baroclinic part of the flow for
$\tilde{\varepsilon}=10\tilde{\varepsilon}_c$.}\label{fig:PL_L1_spectra}
\end{figure}

Comparison of the dominant structures in the NL simulations (c.f figures \ref{fig:NL_L1_spectra}-\ref{fig:NL_L1_40ec_trans})
to the equilibrated states in S3T with the largest domain of attraction (c.f figures \ref{fig:PL_L1_fig1}-\ref{fig:PL_L1_fig4})
demonstrates that the scales of the phase coherent large scale waves in the turbulent flow are predicted by S3T.
Figure \ref{fig:PL_L1_spectra} shows the equivalent spectra which are dominated in this case as well
by the coherent part of the flow. Comparison to the spectra obtained from the NL simulations (cf. figure \ref{fig:NL_L1_spectra})
shows quantitative differences of the order of 20\% for the amplitude of the emergent structures. Differences on the
exact points of transition from the wave attractor to the jet attractor are found in this case as well. But even nuances of the dynamics
such as the long time variability of the
turbulent flow at $\tilde{\varepsilon}=40\tilde{\varepsilon}_c$ with a vacillation between a state with stronger barotropic jet/weaker baroclinic wave
and a state with a stronger baroclinic wave/weaker barotropic jet are captured in S3T with approximately the same time scale
as revealed by comparison of figures \ref{fig:NL_L1_40ec_trans} and \ref{fig:PL_L1_fig3}.

We now consider the equilibration of the S3T instabilities for the other two cases of forcing correlation, that is exciting
only the barotropic ($p=1$) and the baroclinic ($p=-1$) eddies. For the case of barotropic forcing, the instability characteristics
as well as their equilibration are similar to the case of uncrorrelated forcing between the two layers and are not shown. For
the case of baroclinic forcing, structures with small meridional scales are predicted to initially emerge and the
low resolution of the S3T calculations is not adequate to resolve these scales. For this reason we choose to study the
equilibration of the instabilities with ensemble quasi-linear simulations (EQL) governed by~(\ref{eq:EQL1})-(\ref{eq:EQL2})
and~(\ref{eq:EQL3}) at higher resolution ($64\times 64$).

\begin{figure}
\centerline{\includegraphics[width=.75\textwidth]{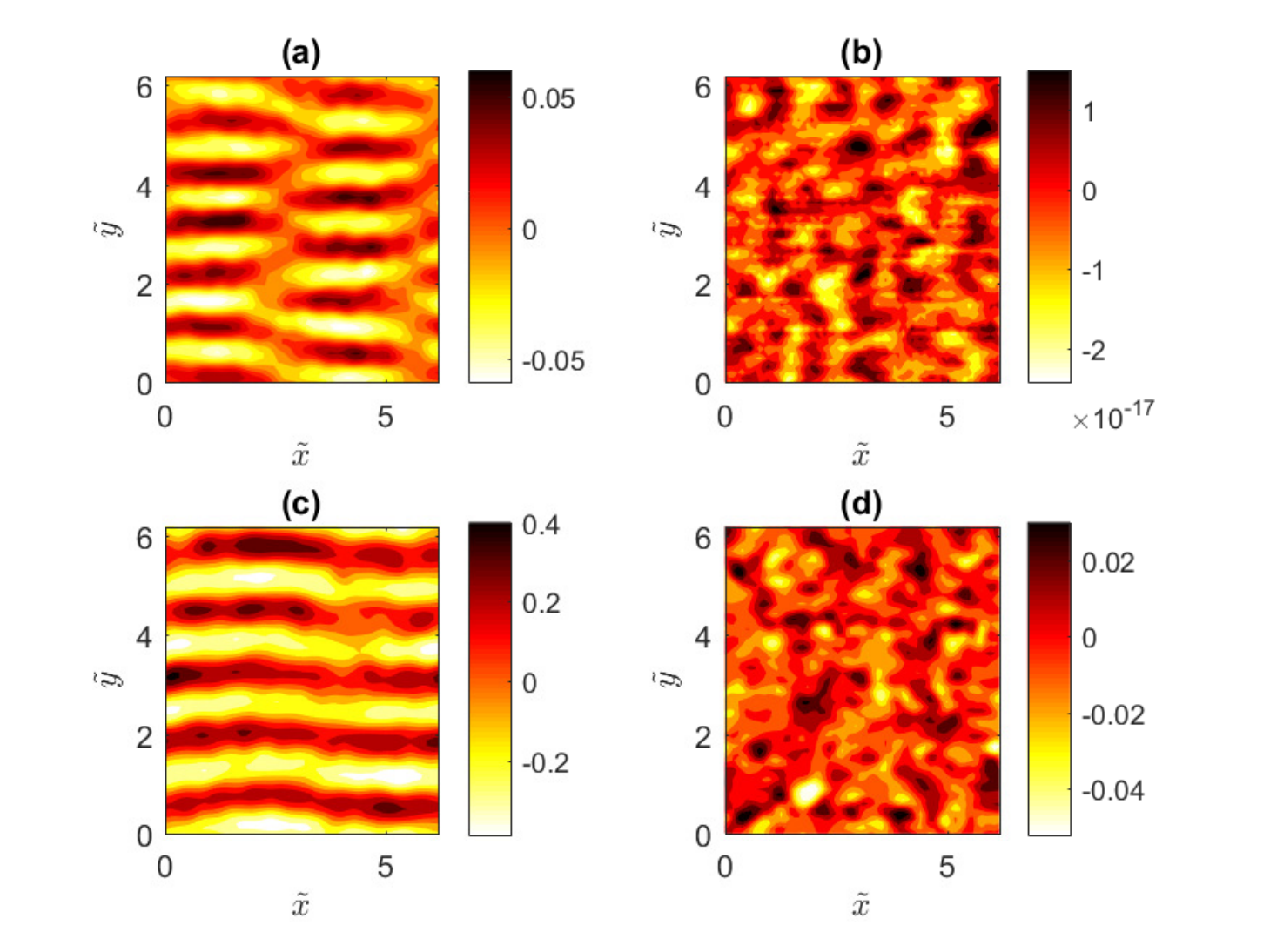}}
\caption{(a)-(b) Contours of the (a) barotropic $\Psi$ and (b) baroclinic $\Theta$ streamfunction of the equilibrated
state for a baroclinic forcing ($p=-1$) at $\tilde{\varepsilon}=4\tilde{\varepsilon}_c$ and $\tilde{\lambda}=\tilde{k}_f$. (c)-(d) The same as in (a) and (b)
but for $\tilde{\varepsilon}=60\tilde{\varepsilon}_c$. The flows are obtained from EQL integrations at a $64\times 64$ resolution and
$N_{ens}=10$ ensemble members.}\label{fig:EQL_L6}
\end{figure}

\begin{figure}
\centerline{\includegraphics[width=.75\textwidth]{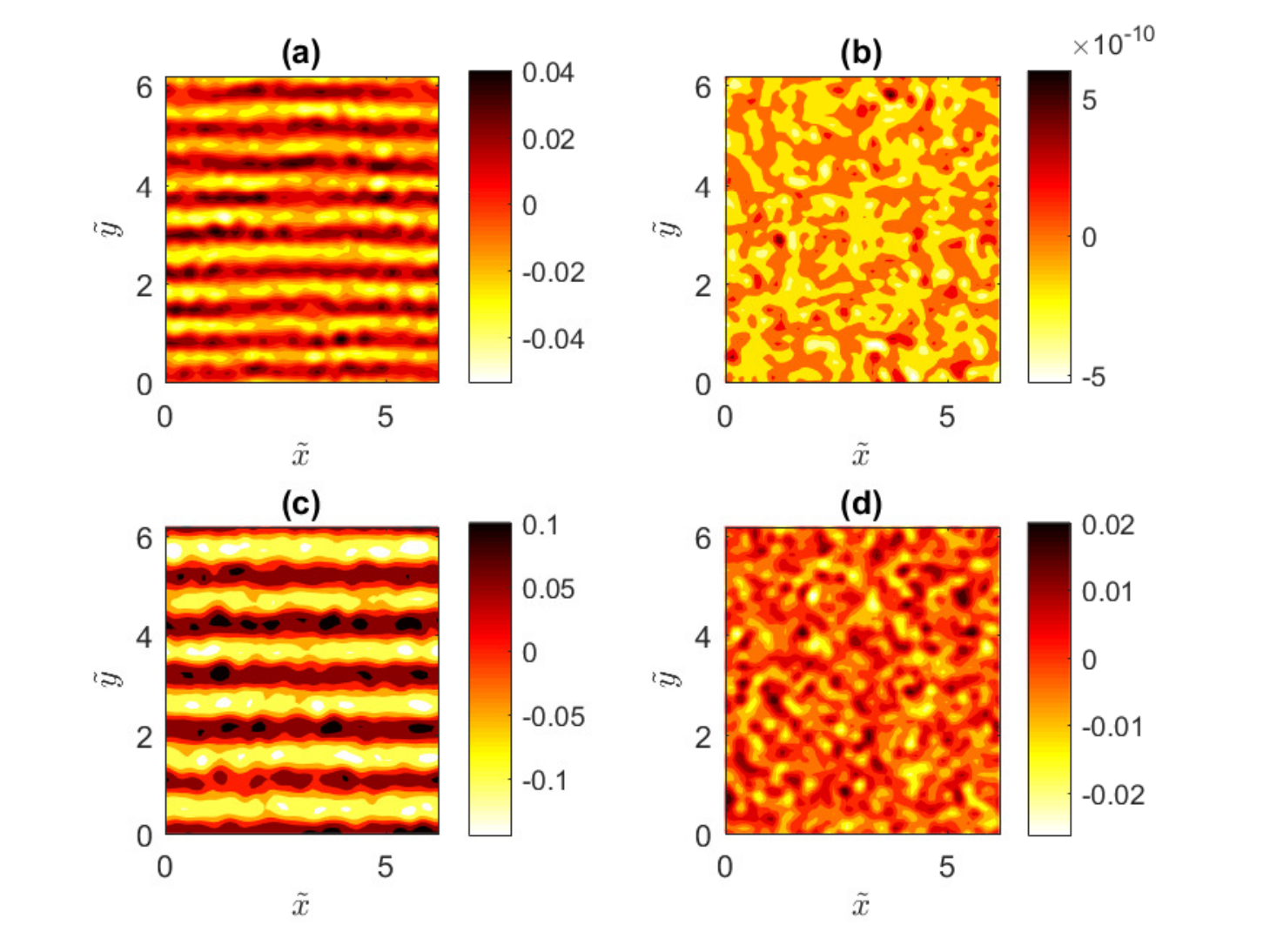}}
\caption{(a)-(b) Contours of the (a) barotropic $\Psi$ and (b) baroclinic $\Theta$ streamfunction of the equilibrated
state for a baroclinic forcing ($p=-1$) at $\tilde{\varepsilon}=10\tilde{\varepsilon}_c$ and $\tilde{\lambda}=2\tilde{k}_f$.
(c)-(d) The same as in (a) and (b)
but for $\tilde{\varepsilon}=60\tilde{\varepsilon}_c$. The flows are obtained from EQL integrations at a $64\times 64$ resolution and
$N_{ens}=10$ ensemble members.}\label{fig:EQL_L12}
\end{figure}

Consider first $\tilde{\lambda}=\tilde{k}_f$ ($\lambda=1$). At low supercriticality ($\tilde{\varepsilon}<5\tilde{\varepsilon}_c$), the
baroclinic modes are stable and barotropic Rossby waves that have the largest growth rate equilibrate at finite amplitude as in the
case of uncorrelated forcing. The only difference in this case is that these waves have meridional scales comparable to the
forcing scale (the most unstable mode has $(|\tilde{n}_x|, |\tilde{n}_y|)=(1, 6)$). As a result the flow equilibrates to this
barotropic traveling wave state. This is shown in figure \ref{fig:EQL_L6}(a)-(b) in which a snapshot of the barotropic and
the baroclinic streamfunction are shown. For higher supercriticality ($\tilde{\varepsilon}>5\tilde{\varepsilon}_c$), baroclinic modes
are rendered unstable as well. However, the finite amplitude equilibria have very weak baroclinicity. This is illustrated in figure
\ref{fig:EQL_L6}(c)-(d), showing that the baroclinic streamfunction is one order of magnitude smaller than the barotropic part of the
flow that has been zonated through the secondary SSD instability.
Consider now the larger value $\tilde{\lambda}=2\tilde{k}_f$ ($\lambda=2$) for which linear stability analysis predicts that
barotropic stationary zonal jets with scales close to the Rossby radius of deformation have the largest growth rate. The equilibrated
structure for $\tilde{\varepsilon}=10\tilde{\varepsilon}_c$ is shown in figure \ref{fig:EQL_L12}(a)-(b). Indeed it consists of
a small scale barotropic jet with a scale comparable to the deformation radius ($\tilde{n}_y=3\tilde{\lambda} /4$) and a
very weak baroclinic flow. At larger supercriticality, the jets assume larger scales (c.f. \ref{fig:EQL_L12}(c)-(d))
and baroclinicity remains weak. Finally we note that at $\tilde{\lambda}=\tilde{k}_f/6$ the equilibrated structures are similar
to the case of uncorrelated forcing and are not shown.

Comparison of the coherent structures in figures \ref{fig:EQL_L6}-\ref{fig:EQL_L12} to the emergent structures in the
NL simulations in figures \ref{fig:NL_pm1_L6_spectra}-\ref{fig:NL_pm1_L12_spectra} reveals that again the
scales of the emergent flows are accurately captured by S3T, while their amplitude is slightly underestimated.

% Finally, consider now the case of a very low $\lambda=0.06$ (non-dimensional $\lambda=0.01$).
%The barotropic and baroclinic streamfunction are shown in
%figure~\ref{fig:EQL_L0p1} for two values of the energy input rate. At low supercriticality we observe the
%barotropic and baroclinic waves with similar amplitude (cf. figure~\ref{fig:EQL_L0p1}a-b). However in contrast
%to the case $\lambda=0.1$, zonation occurs at large supercriticality for both the barotropic and baroclinic
%flows and four strong jets emerge (cf. figure~\ref{fig:EQL_L0p1}c-d). To summarize, when the forcing scale
%is comparable to the deformation scale, barotropic structures emerge in the EQL simulations when the
%energy input rate passes the threshold predicted by the stability analysis of the homogeneous equilibrium.
%For low supercriticality the flow is dominated by propagating waves, while at larger energy input rates
%the waves become S3T unstable to zonal perturbations and barotropic zonal jets emerge. When the forcing
%scale is smaller than the deformation scale, both baroclinic and barotropic waves of comparable amplitude
%emerge with zonation occurring at higher supercriticalities only for the barotropic flow. Finally when the
%forcing scale is much smaller than the deformation scale, zonation occurs for both the barotropic and
%the baroclinic flow and robust zonal jets with both a barotropic and a baroclinic component dominate
%the turbulent flow.

\section{Conclusions}

The emergence of coherent structures in stratified turbulent flows with turbulence supported
by external sources was investigated in this work. A two-layer model on a $\beta$-plane channel was
considered under the influence of homogeneous and isotropic stochastic forcing and linear damping of
potential vorticity. No mean thermal gradient was imposed in the flow. The goal was to extend the
analysis of previous studies which found that coherent flows emerge out of a background of homogeneous
turbulence as a bifurcation when the turbulence intensity increases and that these turbulent bifurcations
can be attributed to a new type of collective instability of the statistical state dynamics of the turbulent
flow. The questions addressed were the following: in the absence of temperature gradients can externally forced
turbulence in a stratified flow support turbulent equilibria in which large scale baroclinic coherent flows are
sustained at finite amplitude? Or are the large scale flows that emerge necessarily barotropic?

Direct non-linear simulations show two major flow transitions as the energy input rate of the forcing
increases. In the first transition large scale, Rossby waves that remain phase coherent over long
time scales emerge in the flow. As the energy input rate increases, the scales of these waves as well as
their energy increases as well. In the second transition, zonal jets emerge instead while the energy in the
large scale waves decreases. Regarding the vertical structure of the emergent flows, the large scale waves
and jets are barotropic when energy is injected at scales comparable to the Rossby radius of deformation. When the
energy is injected at smaller scales, the waves that emerge have comparable barotropic and baroclinic streamfunctions.
After the second transition, the barotropic part of the flow is zonated and the turbulent flow exhibits variability
on long times scales with time periods of stronger barotropic jet and time periods of stronger baroclinic waves.
For larger energy input rates the barotropic jet dominates. These results were found to be independent
of the correlation of the excitation between the two layers as two extreme cases (forcing only baroclinic eddies and forcing
only barotropic eddies) produced qualitatively similar results to the case of uncorrelated forcing between the two layers.
That is, even though the forcing might inject
barotropic eddies only in the flow, these eddies are organized as to yield baroclinic fluxes to
support the baroclinic waves. Similarly, in the case in which we inject baroclinic eddies only in the flow,
the prevalence of barotropic coherent structures remains.

We then developed a theory to explain the emergence of coherent flows and predict their characteristics.
The theory is based on a second order closure of the statistical state dynamics (SSD) of the turbulent flow (S3T).
%closed at second order (S3T) with the
%interpretation {\color{blue} of the ensemble average as a Reynolds average over the phase incoherent turbulent eddies}
%that allows for the study of both zonal and non-zonal coherent structures.
The fixed points of the S3T dynamical system for the joint evolution of the coherent flow and the eddy
covariance define equilibria, the manifestations of which are the states of the turbulent flow at statistical equilibrium.
Instability of these fixed points and equilibration into new steady states of the SSD also
manifest as regime transitions in the turbulent flow with the emergence of new attractors.

The linear stability of the homogeneous equilibrium with no coherent structures was examined analytically
for a wide range of forcing scales $1/\tilde{k}_f$ relative to the deformation radius $1/\tilde{\lambda}$
and a wide range of values for the non-dimensional planetary vorticity
gradient $\beta=\tilde{\beta} \tilde{r}/\tilde{k}_f$, where $1/\tilde{r}$ is the linear dissipation time scale. The
equilibrium was found to be unstable for all $\lambda$ when the energy input rate of the forcing exceeds a
critical value. When turbulence is injected close to the Rossby radius of deformation or at larger scales, baroclinic flows
are either stable or grow at a lower rate compared to barotropic flows. For
large values of $\beta$,
westward propagating barotropic waves that follow the barotropic Rossby wave dispersion grow the most, while for
values of $\beta=\mathcal{O}(1)$ or lower stationary zonal jets have faster growth rates. The equilibration of the incipient
instabilities was then studied for large values of $\beta$ through numerical integrations of the S3T dynamical system. The
flow was found to equilibrate for low supercriticality at finite amplitude traveling wave states that approximately have the
same dispersion properties as the unstable modes. For larger supercriticality, the traveling wave states become S3T unstable
with respect to zonal jet perturbations. The flow then equilibrates at mixed jet-traveling wave states with lower amplitude
traveling waves embedded in strong zonal jets. This dynamics is similar to barotropic turbulence organization that was
investigated in earlier studies \citep{Bakas-Ioannou-2014-jfm,Constantinou-etal-2016}.

When turbulence is injected at scales much smaller than the deformation radius, barotropic and baroclinic modes have
comparable growth rates with barotropic modes growing at slightly faster rates and having dispersion properties similar to
the ones for larger values of the deformation radius. For low supercriticalities,
the flow equilibrates at mixed barotropic-baroclinic traveling wave states with the barotropic streamfunction having a larger
amplitude compared to the baroclinic streamfunction. For higher supercriticalities, a secondary S3T instability of the
traveling wave states similar to the one discussed above zonates the barotropic part of the flow. As a result the flow
transitions to a mixed barotropic-baroclinic jet-wave state where a weaker baroclinic wave is embedded in a stronger barotropic
jet. This state is time dependent with the amplitudes of the barotropic jet and the baroclinic wave oscillating in time as the jet
intensifies the wave for half a cycle and the wave intensifies the jet in the next part of the cycle giving the excess
energy back to the jet. For highly supercritical regimes, the baroclinicity and this vacillation are much weaker. Finally, it was
found that the characteristics of the emergent structures are not sensitive to the correlation of the forcing in the two layers
(with small exceptions in the case of baroclinic forcing).

The predictions of the theory regarding the SSD were then compared to the results obtained in
direct numerical nonlinear simulations. First of all, the absence of baroclinic structures
in the turbulent flow when the energy is injected at scales comparable to the deformation radius, is attributed to
the stability of the homogeneous equilibrium to baroclinic modes. The critical threshold above which non-zonal structures are unstable
according to the stability analysis of the S3T system was found to be in excellent agreement with the critical value above
which large scale, phase coherent waves acquire significant power in the nonlinear simulations. The scale and phase
speed of the dominant structures in the nonlinear simulations were also found to correspond to the coherent structures
predicted by S3T including the existence or not of a baroclinic component for the flow. However, the amplitude of the
emerging waves is underestimated. In addition, the critical turbulence
intensity threshold for the emergence of zonal jets, which is identified in S3T as the energy input rate at which the secondary
instability of the finite amplitude traveling wave states appears, was also found to roughly match the corresponding threshold
for jet formation in the nonlinear simulations, with the emerging jet scale being accurately obtained using
S3T. As in the wave regime, the jet amplitude is underestimated by roughly 20 \%. However, it is surprising that even nuances
of the statistical state dynamics such as the time dependent mixed barotropic-baroclinic jet-wave state have a manifestation
in the fully turbulent flow.

In summary, in stratified flows with no mean thermal gradient imposed, there is a large bias towards the emergence of
barotropic flows even though both barotropic and baroclinic incoherent turbulent eddies can be supported at a homogeneous
turbulent equilibrium. When turbulence is injected at the deformation radius (or larger) baroclinic mean flow perturbations
to the homogeneous equilibrium do not emerge. When turbulence is injected at much smaller
scales compared to the deformation radius, flows with baroclinic components emerge but the baroclinicity of the flow is rather small
and the baroclinic component of the flow remains non-zonal even though the barotropic part is zonated when the turbulence intensity
is above a critical threshold. It is also important to note that similar to the organization of barotropic turbulence,
a second-order closure of the SSD is very accurate in capturing the transitions in the turbulent flow with the emergence of
coherent structures even in this regime in which the shear rate of the emerging mean flows is infinitesimally small. We can
therefore conclude that the coherent structures emerge as instabilities of the SSD of the flow, an analytical treatment of which
is only amenable in the second-order closure framework of the SSD.

\vspace{1em}

Acknowledgements. The authors would like to thank Navid Constantinou for useful discussions. Nikos Bakas is supported
by the AXA Research Fund.

\appendix
%\section{Bound for the ratio of injected potential to kinetic energy}
%\label{sec:appA}
%
%Consider the temporally uncorrelated and spatially homogeneous excitation of the two layers $\xi_1$ and
%$\xi_2$ with forcing covariance given by (\ref{eq:forc_hom})-(\ref{eq:hom_elem}). This excitation injects kinetic energy with spectrum:
%\begin{equation}
%\hat{E}_K=2\frac{\[\lambda^4+(k^2+\lambda^2)^2\]\hat{\Xi}+(k^2+\lambda^2)\lambda^2\hat{\Xi}_{12}}{k^2k_\la^4}
%\end{equation}
%and potential energy with spectrum:
%\begin{equation}
%\hat{E}_P=\frac{2\lambda^2}{k_\la^4}(\hat{\Xi}-\hat{\Xi}_{12}),
%\end{equation}
%where $\hat{\Xi}$ and $\hat{\Xi}_{12}$ are the Fourier transforms of $\Xi$ and $\Xi_{12}$.
%For a given ratio of potential to kinetic energy $\a$, the cross-correlation $\hat{\Xi}_{12}$ is:
%\begin{equation}
%\hat{\Xi}_{12}=\frac{(1-2\a)\la^2k^2-\a(k^4+2\la^4)}{(\a+1)k^2\la^2+\a\la^4}\hat{\Xi}.
%\label{eq:A1}
%\end{equation}
%The homogeneity of $\Xi_{12}$ implies that it should be non-negative definite \citep{Constantinou-2015-phd}, which requires
%that $\hat \Xi_{12} \ge 0$,  and consequently the ratio $a$ should
%satisfy:
%\begin{equation}
%\a\leq \frac{\la^2k^2}{\la^4+(k^2+\la^2)^2}~.
%\end{equation}
%The maximum value of the ratio $\a$ is seen from~(\ref{eq:A1}) to be achieved when $\hat{\Xi}_{12}=0$. For the minimum value,
%$\a=0$, equation~(\ref{eq:A1}) implies  that $\hat{\Xi}_{12}=\hat{\Xi}$.

\section{Stability equation}
\label{sec:appB}
It can be readily shown as in the barotropic case \citep{Bakas-Ioannou-2014-jfm}  that due to the homogeneity of
the forcing covariance, the state with zero
coherent flow $Z=H=0$ and a homogeneous covariance $\C_E=\Q/2$ is a fixed point of the
S3T dynamical system. The linear stability of the homogeneous equilibrium is assessed by considering the joint linear evolution
of  barotropic and baroclinic mean flow perturbations $[\delta\Psi, \delta\Theta]^T$
and covariance  perturbations $\delta \C$. The linearized equations governing the evolution of the coherent mean flow
perturbations are:
\begin{eqnarray}
& &\partial_t\Del {\delta \Psi} + \b\partial_x\delta \Psi=\delta f_\psi-\Del\delta\Psi~,\\
& &\partial_t\Del_\la \delta\Theta+ \b\partial_x\delta\Theta=\delta f_\theta -\Del_\la\delta\Theta~,
\label{eq:stres1}
\end{eqnarray}
where $\delta f_\psi=R^\psi\(\delta\C\)$ and $\delta f_\theta=R^\theta\(\delta\C\)$ are the perturbation eddy stress divergences which depend linearly
on the components of $\delta \C$. The
components of $\delta \C$ evolve according to :
\begin{eqnarray}
\partial_t \delta C^{\z\z}&=&-2\delta C^{\z\z}-\beta\left[\partial_{\overline{x}}\(\tilde{\Del}+\frac{1}{4}\overline{\Del}\)-2\partial_{\tilde{x}}\Gamma\right]\delta S^{\psi\psi}-
 \( \delta\U_a^\psi  \nonumber -\delta\U_b^\psi\)\bcdot\tilde{\bnabla}\tilde{\Del}^2 S_E^{\psi\psi}\nonumber  \\
& &  -\( \Del_a\delta\U_a^\psi-\Del_b\delta\U_b^\psi\)\bcdot\tilde{\bnabla}\tilde{\Del}  S_E^{\psi\psi}~,\label{eq:dC1}\\
\partial_t \delta C^{\z\eta}&=&-\left[\(\delta\U_a^\theta\bcdot\tilde{\bnabla}\)\tilde{\Del}\tilde{\Del}_\la-\(\Del_{a}\delta\U_a^\theta\bcdot\tilde{\bnabla}\)
\tilde{\Del}_\la\right]S_E^{\theta\theta}+
\left[\(\delta\U_b^\theta\bcdot\tilde{\bnabla}\)\tilde{\Del}^2-\(\Del_{\la,b}\delta\U_b^\theta\bcdot\tilde{\bnabla}\)\tilde{\Del}\right]
S_E^{\psi\psi}\nonumber\\
& & - 2\delta C^{\z\eta}-\beta\left[\frac{1}{2}\partial_{\overline{x}}\left(\tilde{\Del}+\tilde{\Del}_\la+\frac{1}{2}\overline{\Del}\right)-
\partial_{\tilde{x}}\left(\tilde{\Del}-\tilde{\Del}_\la+2\Gamma\right)\right]\delta S^{\psi\theta}~,\\
\partial_t \delta C^{\eta\z}&=& \left[\(\delta\U_b^\theta\bcdot\tilde{\bnabla}\)\tilde{\Del}\tilde{\Del}_\la-\(\Del_{b}\delta\U_b^\theta\bcdot\tilde{\bnabla}\)\tilde{\Del}_\la\right]S_E^{\theta\theta}
-\left[\(\delta\U_a^\theta\bcdot\tilde{\bnabla}\)\tilde{\Del}^2-\(\Del_{\la,a}\delta\U_a^\theta\bcdot\tilde{\bnabla}\)\tilde{\Del}\right]S_E^{\psi\psi}\nonumber\\
& &-\delta C^{\eta\z}-\beta\left[\frac{1}{2}\partial_{\overline{x}}\left(\tilde{\Del}+\tilde{\Del}_\la+\frac{1}{2}\overline{\Del}\right)-
\partial_{\tilde{x}}\left(\tilde{\Del}_\la-\tilde{\Del}+2\Gamma\right)\right]\delta S^{\theta\psi}~,\\
\partial_t\delta C^{\eta\eta}&=& -\left[\(\delta\U_a^\psi-\delta\U_b^\psi\)\bcdot\tilde{\bnabla}\tilde{\Del}_\la^2-
\(\Del_a\delta\U_a^\psi-\Del_b\delta\U_b^\psi\)\bcdot\tilde{\bnabla}\tilde{\Del}_\la\right]S_E^{\theta\theta}\nonumber\\
& &-\beta\left[\partial_{\overline{x}}\left(\tilde{\Del}_\la+\frac{1}{4}\overline{\Del}\right)-2\partial_{\tilde{x}}\Gamma\right]
\delta S^{\theta\theta}-2\left[\left(\tilde{\Del}_\la+\frac{1}{4}\overline{\Del}\right)\left(\tilde{\Del}+\frac{1}{4}\overline{\Del}\right)-\Gamma^2\right]
\delta S^{\theta\theta},\label{eq:dC4}
\end{eqnarray}
where $\Gamma\equiv\partial_{~\tilde{x}~\overline{x}}^2+\partial_{~\tilde{y}~\overline{y}}^2$, and operators with tildes indicate
differentiation with respect to $\tilde{\xv }=\xv _a-\xv _b$ while operators with overbars indicate  differentiation with
respect to $\overline{\xv }=(\xv _a+\xv _b)/2$.

Because of the statistical homogeneity of the equilibrium state, the eigenfunctions are sinusoidal  and the
coherent flow
component of the eigenfunctions   are  $[\delta \Psi,\delta\Theta]=[a_\psi, a_\theta]e^{\ij\nv \bcdot\xv }
e^{\sigma t}$, with  covariance components:
\begin{equation}
[\delta S^{\psi\psi}, \delta S^{\psi\theta}, \delta S^{\theta\psi}, \delta S^{\theta\theta}]^T=
{e^{\ij\nv \bcdot\overline{\xv }}e^{\sigma t}\over 2\upi}
\int_{-\infty}^{\infty}\int_{-\infty}^{\infty} [\hat{S}^{\psi\psi}, \hat{S}^{\psi\theta}, \hat{S}^{\theta\psi}, \hat{S}^{\theta\theta}]^T
e^{\ij\kv\bcdot\tilde{\xv }}\,\df^2\kv~.
\end{equation}
Inserting the eigenfunction  into (\ref{eq:dC1})-(\ref{eq:dC4}), we determine the perturbation covariance amplitudes in terms of the amplitudes of the perturbation
mean flow. They are given  by:\begin{subequations}
\begin{equation}
\hat{S}^{\psi\psi}= \frac{a_\psi\hat{\boldsymbol{z}}\bcdot (\kv_+\times \nv )k_+^2(k_+^2-N^2)\hat{S}_{E+}^{\psi\psi}}{\left(\sigma+2\right)k_+^2k_-^2+i\beta\left(k_{x-}k_+^2-k_{x+}k_-^2\right)}
- \frac{a_\psi\hat{\boldsymbol{z}}\bcdot (\kv_-\times \nv )k_-^2(k_-^2-N^2)\hat{S}_{E-}^{\psi\psi}}{\left(\sigma+2\right)
k_+^2k_-^2+i\beta\left(k_{x-}k_+^2-k_{x+}k_-^2\right)},
\end{equation}
%\begin{equation}
%\hat{S}^{\psi\theta}=a_\theta \frac{(n_yk_{x+}-n_xk_{y+})k_+^2(k_+^2-n_\la^2)\hat{S}_{E+}^{\psi\psi}}{\left(\sigma+2\right)k_+^2k_{\la -}^2+i\beta\left(k_{x-}k_+^2-k_{x+}k_{\la %-}^2\right)}-a_\theta \frac{(n_yk_{x-}-n_xk_{y-})k_{\la -}^2(k_-^2-n^2)
%\hat{S}_{E-}^{\theta\theta}}{\left(\sigma+2\right)k_+^2k_{\la -}^2+i\beta\left(k_{x-}k_+^2-k_{x+}k_{\la -}^2\right)},\label{eq:cross1}
%\end{equation}
\begin{equation}
\hat{S}^{\psi\theta}= \frac{a_\theta\hat{\boldsymbol{z}}\bcdot (\kv_+\times \nv )k_+^2(k_+^2-n_\la^2)\hat{S}_{E+}^{\psi\psi}}{\left(\sigma+2\right)k_+^2k_{\la -}^2+i\beta\left(k_{x-}k_+^2-k_{x+}k_{\la -}^2\right)}- \frac{a_\theta\hat{\boldsymbol{z}}\bcdot (\kv_-\times \nv )k_{\la -}^2(k_-^2-n^2)
\hat{S}_{E-}^{\theta\theta}}{\left(\sigma+2\right)k_+^2k_{\la -}^2+i\beta\left(k_{x-}k_+^2-k_{x+}k_{\la -}^2\right)},\label{eq:cross1}
\end{equation}
\begin{equation}
\hat{S}^{\theta\psi}=
\frac{a_\theta\hat{\boldsymbol{z}}\bcdot (\kv_+\times \nv )k_{\la +}^2(k_+^2-n^2)\hat{S}_{E+}^{\theta\theta}}{\left(\sigma+2\right)k_{\la +}^2k_-^2+i\beta\left(k_{x-}k_{\la +}^2-k_{x+}k_-^2\right)}-\frac{a_\theta\hat{\boldsymbol{z}}\bcdot (\kv_-\times \nv )k_-^2(k_-^2-n_\la^2)
\hat{S}_{E-}^{\psi\psi}}{\left(\sigma+2\right)k_{\la +}^2k_-^2+i\beta\left(k_{x-}k_{\la +}^2-k_{x+}k_-^2\right)},\label{eq:cross2}
\end{equation}
\begin{equation}
\hat{S}^{\theta\theta}=\frac{a_\psi\hat{\boldsymbol{z}}\bcdot (\kv_+\times \nv )k_{\la +}^2(k_{\la +}^2-n^2)\hat{S}_{E+}^{\theta\theta}}
{\left(\sigma+2\right)k_{\la +}^2k_{\la -}^2+i\beta\left(k_{x-}k_{\la +}^2-k_{x+}k_{\la -}^2\right)}
-\frac{a_\psi \hat{\boldsymbol{z}}\bcdot (\kv_-\times \nv )k_{\la -}^2(k_{\la -}^2-n^2)
\hat{S}_{E-}^{\theta\theta}}{\left(\sigma+2\right)k_{\la +}^2k_{\la -}^2+i\beta\left(k_{x-}k_{\la +}^2-k_{x+}k_{\la -}^2\right)},
\end{equation}
\end{subequations}
with the notation $\kv_\pm=\kv\pm \nv /2$, $k_\pm^2=|\kv_\pm|^2$, $k_{\la\pm}^2=|\kv_\pm|^2+2\la^2$,
$n_\la=n^2+2\la^2$, $\hat{S}_{E\pm}^{\psi\psi}(\kv)=(1+p)\hat{\Xi}(\kv_\pm)/2k_\pm^4$,
$\hat{S}_{E\pm}^{\theta\theta}(\kv)=(1-p)\hat{\Xi}(\kv_\pm)/2k_{\la \pm}^4$ and $\hat{\boldsymbol{z}}$ the unit vector
representing the vertical direction.

The perturbation eddy stress divergences  can then be determined as a function of the perturbation coherent flow and the spectrum of the excitation.
They are:
\begin{equation}
[\delta f_\psi, \delta f_{\theta}]^T=\[a_\psi\(f_{\psi\psi +}-f_{\psi\psi -}\),
a_\theta \(f_{\theta\theta +}-f_{\theta\theta -}\)\]^T e^{\ij\nv \bcdot \xv }~,\label{eq:eddy_str1}
\end{equation}
where
\begin{eqnarray}
f_{\psi\psi\pm}&=&\frac{1}{4\upi}\int_{-\infty}^{\infty}\int_{-\infty}^{\infty}\frac{\hat{\boldsymbol{z}}\bcdot\left[k_+^2(\kv_-\times \nv )-k_-^2(\kv_+\times \nv )\right]\hat{\boldsymbol{z}}\bcdot(\kv_\pm\times \nv )k_{\la \pm}^2(k_{\la \pm}^2-n^2)\hat{S}_{E \pm}^{\theta\theta}}{\left(\sigma+2\right)k_{\la +}^2k_{\la -}^2+i\beta (k_{x-}k_{\la +}^2-k_{x+}k_{\la -}^2 )}\,\df^2\kv\nonumber\\
&+&\frac{1}{4\upi}\int_{-\infty}^{\infty}\int_{-\infty}^{\infty}\frac{\hat{\boldsymbol{z}}\bcdot\left[k_+^2(\kv_-\times \nv )-k_-^2(\kv_+\times \nv )\right] \hat{\boldsymbol{z}}\bcdot(\kv_\pm\times \nv )k_\pm^2(k_\pm^2-N^2)\hat{S}_{E \pm}^{\psi\psi}}{\left(\sigma+2\right)k_+^2k_-^2+i\beta (k_{x-}k_+^2-k_{x+}k_-^2 )} \,\df^2\kv\ ,
\end{eqnarray}
%\begin{eqnarray}
%f_{\psi\psi\pm}&=&\frac{1}{4\upi}\int_{-\infty}^\infty\int_{-\infty}^\infty\frac{\( -(n_yk_{x+}-n_xk_{y+})k_-^2+(n_yk_{x-}-n_xk_{y-})k_+^2\)(n_yk_{x\pm}-n_xk_{y\pm})k_{\la \pm}^2(k_{\la \pm}^2-n^2)\hat{S}_{E \pm}^{\theta\theta}}{\left(\sigma+2\right)k_{\la +}^2k_{\la -}^2+i\beta\left(k_{x-}k_{\la +}^2-k_{x+}k_{\la -}^2\right)}\,\df^2\kv\nonumber\\
%&+&\frac{1}{4\upi}\int_{-\infty}^\infty\int_{-\infty}^\infty\frac{\( -(n_yk_{x+}-n_xk_{y+})K_-^2+(n_yk_{x-}-n_xk_{y-})K_+^2\)(n_yk_{x\pm}-n_xk_{y\pm})K_\pm^2(K_\pm^2-N^2)\hat{S}_{E \pm}^{\psi\psi}}{\left(\sigma+2\right)k_+^2k_-^2+i\beta\left(k_{x-}k_+^2-k_{x+}k_-^2\right)} \,\df^2\kv,
%\end{eqnarray}
\begin{eqnarray}
f_{\theta\theta\pm}&=&\frac{1}{4\upi}\int_{-\infty}^{\infty}\int_{-\infty}^{\infty}
\frac{\hat{\boldsymbol{z}}\bcdot\left[ -k_{\la \mp}^2(\kv_\pm\times \nv )+k_\pm^2(\kv_\mp\times \nv )\right] \hat{\boldsymbol{z}}\bcdot(\kv_\pm\times \nv )
k_\pm^2(k_\pm^2-n_\la^2)\hat{S}_{E \pm}^{\psi\psi}}{\left(\sigma+2\right)k_+^2k_{\la -}^2+i\beta (k_{x-}k_+^2-k_{x+}k_{\la -}^2 )}\,\df^2\kv\nonumber\\
&+&\frac{1}{4\upi}\int_{-\infty}^{\infty}\int_{-\infty}^{\infty}
\frac{\hat{\boldsymbol{z}}\bcdot\left[ -k_\mp^2(\kv_\pm\times \nv )+k_{\la \pm}^2(\kv_\mp\times \nv )\right]\hat{\boldsymbol{z}}\bcdot(\kv_\pm\times \nv )k_{\la \pm}^2(k_\pm^2-n^2)\hat{S}_{E\pm}^{\theta\theta}}{\left(\sigma+2\right)k_{\la +}^2k_-^2+i\beta (k_{x-}k_{\la +}^2-k_{x+}k_-^2 )}\,\df^2\kv\ .
\end{eqnarray}
Note that when there is no correlation of the barotropic and the baroclinic streamfunction at equilibrium
($S_E^{\psi\theta}=S_E^{\theta\psi}=0$), perturbations to the homogeneous equilibrium
with no temperature difference across the channel have the
property that barotropic mean flow perturbations
induce  only barotropic mean flow accelerations and baroclinic  mean flow perturbations
induce only baroclinic mean flow accelerations decoupling the barotropic  and baroclinic  mean flow tendencies.

A further reduction in the above expressions can be achieved,  as  noted by \cite{Srinivasan-Young-2012},
by utilizing the exchange symmetry of the covariance $\C(\xv _a, \xv _b) = \C(\xv _b, \xv _a)^T$, by
changing the sign of $\kv$ in the integrals  to obtain
\begin{equation}
f_{\psi\psi +}=-f_{\psi\psi -}\ ,\ \ f_{\theta\theta +}=-f_{\theta\theta -}\ .\label{eq:symm}
\end{equation}
Introduction of~(\ref{eq:symm}) into (\ref{eq:eddy_str1}) and  the change of variables  $\kv\rightarrow \kv+\nv /2$,
in the integrals yields the more compact representation of the eddy stress divergences:
\begin{equation}
[\delta f_\psi, \delta f_{\theta}]^T=\[a_\psi f_{\psi}(\sigma),  a_\theta f_{\theta}(\sigma)\]^Te^{\ij\nv \bcdot\xv }\ ,\label{eq:stres_fin}
\end{equation}
with
\begin{eqnarray}
f_\psi(\sigma)&=&\frac{1}{2\upi n^2}\int_{-\infty}^{\infty}\int_{-\infty}^{\infty} \frac{(n_yk_x-n_xk_y)^2(k_{++}^2-k^2)k^2(k^2-n^2)
\hat{S}_E^{\psi\psi}}{(\sigma+2)k^2k_{++}^2+i\beta(k_xk_{++}^2-k_{x++}k^2)}\,\df^2\kv\nonumber\\
&+&
\frac{1}{2\upi n^2}\int_{-\infty}^{\infty}\int_{-\infty}^{\infty} \frac{(n_yk_x-n_xk_y)^2(k_{++}^2-k^2)k_\lambda^2(k_\lambda^2-n^2)
\hat{S}_E^{\theta\theta}}{(\sigma+2)k_\lambda^2k_{\lambda ++}^2+i\beta(k_xk_{\lambda ++}^2-k_{x++}k_\lambda^2)}\,\df^2\kv\ ,
\end{eqnarray}
and
\begin{eqnarray}
f_\theta (\sigma)&=&\frac{1}{2\upi n_\lambda^2}\int_{-\infty}^{\infty}\int_{-\infty}^{\infty} \frac{(n_yk_x-n_xk_y)^2(k_{\lambda ++}^2-k^2)k^2(k^2-n_\lambda^2)
\hat{S}_E^{\psi\psi}}{(\sigma+2)k^2k_{\lambda ++}^2+i\beta(k_xk_{\lambda ++}^2-k_{x++}k^2)} \df^2\kv\nonumber\\
&+& \frac{1}{2\upi n_\lambda^2}\int_{-\infty}^{\infty}\int_{-\infty}^{\infty}\frac{(n_yk_x-n_xk_y)^2(k_{++}^2-k_\lambda^2)k_\lambda^2(k^2-n^2)
\hat{S}_E^{\theta\theta}}{(\sigma+2)k_\lambda^2k_{++}^2+i\beta(k_xk_{++}^2-k_{x++}k_\lambda^2)}\,\df^2\kv\ ,
\end{eqnarray}
with the notation:  $\kv_{++}=\kv+\nv $, $k_{++}^2=|\kv_{++}|^2$ and $k_{\la ++}^2=|\kv_{++}|^2+2\lambda^2$.

Because the barotropic and baroclinic perturbation components decouple
upon substitution of~(\ref{eq:stres_fin}) in~(\ref{eq:stres1}) we obtain that either
$\sigma$ satisfies:
\begin{equation}
\sigma+1-\ij\beta n_x/n^2=f_\psi(\sigma)\ ,
\end{equation}
and the eigenfunction is purely barotropic with $a_\psi \ne 0$ and $a_\theta=0$, or that $\sigma$ satisfies:
\begin{equation}
\sigma+1-\ij\beta n_x/n_\lambda^2=f_\theta(\sigma)\ ,
\end{equation}
and the eigenfunction is purely baroclinic with $a_\psi = 0$ and $a_\theta \ne 0$.

\end{document}